\documentclass[12pt,a4paper]{article}
\usepackage{jheppub}
\usepackage[dvipsnames]{xcolor}
\usepackage{amscd,amsmath,amssymb,amsfonts,xspace,mathrsfs,mathtools,amsthm}
\usepackage[utf8]{inputenc}
\usepackage{bbm}
\usepackage{graphicx}
\usepackage{tabu} 
\usepackage[color,matrix,arrow]{xy}
\usepackage{wasysym} 
\usepackage{stmaryrd}
\usepackage[shortlabels]{enumitem}
\usepackage{array,amsmath,booktabs}
\usepackage{slashed}
\usepackage{tensor}
\usepackage{caption}
\usepackage{subcaption}
\usepackage{enumitem}
\usepackage[textsize=tiny]{todonotes}

\newcommand{\tuv}{\tau_{\text{\tiny{UV}}}}
\newcommand{\uv}{\tau_{\text{\tiny{UV}}}}
\newcommand{\vdim}{\text{vdim}}

\newcommand{\bp}{{\rm bp}}

\newcommand{\nres}[1]{\underset{#1}{\text{nRes}}}

\newcommand{\mad}{m_{\text{AD}}}

\newcommand{\tad}{\tau_{\text{AD}}}

\newcommand{\spinc}{\text{Spin}^c} 
\newcommand{\rk}{\text{rk}}

\newcommand{\psl}{\text{PSL}(2,\mathbb Z)} 
\newcommand{\slz}{\text{SL}(2,\mathbb Z)}  
\newcommand{\spin}{\text{Spin}} 
 
\newcommand{\nstar}{\CN=2^*}

\newcommand{\jt}{\vartheta}
\newcommand{\tn}{\tau_0} 
\newcommand{\luv}{\lambda(\tuv)} 
\newcommand{\lau}{\lambda(\tau)}

\newcommand{\be}{\begin{equation}} 
\newcommand{\ee}{\end{equation}} 
\newcommand{\bes}{\begin{equation*}}
\newcommand{\ees}{\end{equation*}}

\newcommand{\im}{i}

\newcommand{\CB}{\mathcal{B}}

\newcommand{\CE}{\mathcal{E}}  
\newcommand{\CF}{\mathcal{F}}

\newcommand{\CI}{\mathcal{I}}
\newcommand{\CJ}{\mathcal{J}}
\newcommand{\CK}{\mathcal{K}}
\newcommand{\CL}{\mathcal{L}} 
\newcommand{\CM}{\mathcal{M}}  
\newcommand{\CN}{\mathcal{N}}
\newcommand{\CO}{\mathcal{O}} 

\newcommand{\CQ}{\mathcal{Q}}

\newcommand{\CS}{\mathcal{S}}

\newcommand{\BZ}{\mathbb{Z}}

\newcommand{\bfb}{{\boldsymbol b}}

\newcommand{\bfell}{{\boldsymbol \ell}} 
\newcommand{\bfm}{{\boldsymbol m}}

\newcommand{\bfv}{{\boldsymbol v}} 
\newcommand{\bfx}{{\boldsymbol x}}

\newcommand{\bfw}{{\boldsymbol w}}

\newcommand{\bfnu}{{\boldsymbol \nu}}

\newcommand{\bfl}{{\boldsymbol l}}
\newcommand{\bfk}{{\boldsymbol k}}
\newcommand{\bfz}{{\boldsymbol z}}

\newcommand{\bfmu}{{\boldsymbol \mu}}

\title{Topological twists of massive SQCD, Part I}

\author{Johannes Aspman, Elias Furrer, Jan Manschot \\
{\it School of Mathematics, Trinity College, Dublin 2, Ireland\\
\it Hamilton Mathematical Institute, Trinity College, Dublin 2 \vspace{20pt}
}}
   
\abstract{
We consider topological twists of four-dimensional $\mathcal{N}=2$ supersymmetric QCD with gauge group SU$(2)$ and $N_f\leq 3$ fundamental hypermultiplets. The twists are labelled by a choice of background fluxes for the flavour group, which provides an infinite family of topological partition functions. In this Part I, we demonstrate that in the presence of such fluxes the theories can be formulated for arbitrary gauge bundles on a compact four-manifold. Moreover, we consider arbitrary masses for the hypermultiplets, which introduce new intricacies for the evaluation of the low-energy path integral on the Coulomb branch. We develop techniques for the evaluation of these path integrals. In the forthcoming Part II, we will deal with the explicit evaluation.

}

\preprint{}

\begin{document}
\maketitle

\section{Introduction}

Correlations functions of topologically twisted quantum field theories provide many insights into non-perturbative aspects of quantum field theory as well as the geometry of four-manifolds \cite{Witten:1988ze,Witten:1994ev,Witten:1994cg}. We consider topologically twisted $\CN=2$ supersymmetric Yang-Mills theories with additional matter multiplets on a compact four-manifold, which were introduced in  \cite{YAMRON1988325,anselmi1993,anselmi1994,ANSELMI1995247,ALVAREZ1993251,Alvarez:1994ii}. After the work by Seiberg and Witten on the full non-perturbative solution \cite{Seiberg:1994rs,Seiberg:1994aj,Witten:1994cg}, these theories have received much attention in physics \cite{Moore:1997pc,LoNeSha,labastida1995,Labastida:1995zj,Hyun:1995mb,Hyun:1995hz,laba1998,Dijkgraaf:1997ce,Kanno:1998qj,Marino:1998uy,mm1998, Moore:2017cmm, Dedushenko:2017tdw, Manschot:2021qqe, Edelstein:2000wg} and mathematics \cite{gttsche2010,Feehan:1997gj,Feehan:1997rt,Feehan:1997vp,Feehan:2001jc,Gorodentsev:1996cu,Nakajima:2005fg,nakajima2011perverse,marcolli_2011,bryan1996,pidstrigach1995localisation,okonek1996recent}.\footnote{For reviews, see for example \cite{Laba05,Marino:2000hx,Nakajima:2003uh,moore2017}.} Their path integrals can in many cases  be explicitly evaluated after topologically twisting. The study of the analytical structure of partition functions as function of the parameters such as the masses and UV couplings allows to study  the limits in parameters space, as well as relations between different field theories. Moreover, the $Q$-fixed equations give rise to topological invariants of the underlying space-time geometry. The path integral derivation relates distinct notions of invariants for the same space-time geometry, and improves the understanding of such invariants \cite{Taubes1994,Bershadsky:1995vm,Gottsche:2006,Nakajima:2005fg,Harvey_1995,donaldson1995floer,kim2023,Gadde:2013sca,Dedushenko:2017tdw,Manschot:2021qqe,Marino:1998uy}.

We consider in this article topological twists of $\CN=2$ QCD with gauge group SU(2) and matter multiplets in the fundamental representation of the gauge group. By including background fluxes for the flavour group, we obtain an infinite family of topological theories \cite{Hyun:1995mb}. The choice of a background flux makes it possible to formulate topologically twists for $\CN=2$ SQCD for arbitrary 't Hooft fluxes, or first Chern classes of the gauge bundle. This is similar to the topological twist of $\CN=2^*$ SU(2) gauge theory, which requires a non-vanishing background flux on a non-spin four-manifold \cite{Manschot:2021qqe}. We moreover develop techniques to determine correlation functions for arbitrary values of the masses of the hypermultiplets.

The starting point of our approach is the low-energy effective field theory on the Coulomb branch. This phase of the theory contributes for a compact four-manifold $X$  with the topological condition that $b_2^+(X)=1$ \cite{Moore:1997pc}. In this way, the classical Donaldson invariants can be derived starting from the Seiberg-Witten (SW) solution to $\CN=2$ supersymmetric Yang-Mills theory with gauge group SU(2). 
The Coulomb branch integral (or $u$-plane integral) reduces to an integral over zero modes \cite{Moore:1997pc}, and reads schematically
\be 
\label{PhiIntro}
\Phi=\int_{\CB} da\wedge d\bar a\,\rho(a) \,   \Psi(a,\bar a), 
\ee 
where $\CB$ is the Coulomb branch with local coordinates $a$ and $\bar a$, $\rho(a)$ contains the  couplings to the background and $ \Psi(a,\bar a)$ is a sum over fluxes of the unbroken U(1) gauge group. For simplicity, we have suppressed the dependence on the metric and not included observables here. For the pure SU(2) theory, the Coulomb branch integral can be formulated and evaluated for arbitrary four-manifolds, without a requirement for K\"ahler or toric properties.

Recently, progress has been made on evaluating these $u$-plane integrals using a change of variables from $a$ to the running coupling $\tau$. As a result, the integration domain becomes a fundamental domain $\CF\subseteq \mathbb H$  in the upper half-plane $\mathbb H$ for the running coupling \cite{Malmendier:2008db,Griffin:2012kw,Malmendier:2012zz,Malmendier:2010ss,Korpas:2017qdo,Moore:2017cmm,Korpas:2019ava,Korpas:2019cwg,Manschot:2021qqe,Aspman:2021kfp,Korpas:2022tij}.
The integral then takes the  form
\be \label{phi_mu_schematic}
\Phi=\int_{\CF} d\tau\wedge d\bar \tau\,\nu(\tau)\,\Psi(\tau,\bar \tau),
\ee 
where the measure factor $\nu(\tau)$ further contains the Jacobian for the change of variables from $a$ to $\tau$.
The domain $\CF$ is a modular fundamental domain in previous analyses, corresponding to the duality group $\Gamma^0(4)$  for the pure SU(2) theory \cite{Moore:1997pc, Korpas:2017qdo, Korpas:2019cwg},  $\Gamma(2)$ for the $\nstar$ theory \cite{Manschot:2021qqe} and similarly $\Gamma(2)$ and $\Gamma_0(4)$ for the theories with two and three massless flavours \cite{Malmendier:2008db}.

As mentioned above, we aim to apply this approach to $\CN=2$ supersymmetric SU(2) theories with $N_f\leq 3$ hypermultiplets in the fundamental representation. Topological correlators of these asymptotically free theories have been considered in various papers before, in particular the formulation of the low energy path integral in \cite{Moore:1997pc, LoNeSha}, SW contributions for four-manifolds with $b_2^+>1$ \cite{Moore:1997pc, Kanno:1998qj,Marino:1998uy,Dedushenko:2017tdw}, the $u$-plane integral for $\mathbb{P}^2$ \cite{Malmendier:2008db}, and the calculation of the partition function of the AD theory within the $N_f=1$ theory \cite{Moore:2017cmm}. Since no background fluxes are included in these works, the 't Hooft flux necessarily matches the second Stiefel-Whitney class of the four-manifold, $w_2(E)=w_2(X)$, since the twisted hypermultiplets are not well-defined otherwise.

Extending to generic 't Hooft fluxes, and application of the above approach (\ref{phi_mu_schematic}) to fundamental hypermultiplets with generic masses,  gives rise to several new aspects. In particular:
\begin{enumerate}
\item The fundamental domain of the effective coupling constant becomes more intricate for massive theories, and does for generic masses not correspond to a modular fundamental domain for a  subgroup of $\psl$. The domain contains generically a set of branch points, and branch cuts starting from these points \cite{Aspman:2021vhs, Aspman:2020lmf,Aspman:2021evt}. These aspects have to be dealt with appropriately.
\item We couple the hypermultiplets to background fluxes $\bfk_j$ for the flavour group to formulate the theories for arbitrary 't Hooft fluxes.  This gives rise to additional couplings in \eqref{PhiIntro} and \eqref{phi_mu_schematic}, 
\be 
\prod_{j,k=1}^{N_f} \exp\!\left(-2\pi i \frac{\partial^2 F}{\partial m_j\partial m_k}B(\bfk_j,\bfk_k)\right), 
\ee 
where $F$ is the prepotential of the massive theory, and $B(\cdot,\cdot)$ is the quadratic form associated to the intersection form on the middle homology $H_2(X,\mathbb{Z})$ of $X$.  Such couplings were suggested by  Shapere and Tachikawa \cite{Shapere:2008zf}, and are also essential for the formulation of the $\CN=2^*$ Yang-Mills theory on a non-spin four-manifold \cite{Manschot:2021qqe}. Similarly to \cite{Manschot:2021qqe}, we also deduce a non-holomorphic coupling to $\bfk_j$. Moreover, for arriving at a single-valued integrand, we fix an ambiguity in the quadratic terms of the prepotential. These terms have appeared earlier in the literature in the context of singularities of the SW differential and winding numbers \cite{Seiberg:1994aj,Ohta_1997}.
\item Special points on the Coulomb branch give rise to superconformal theories, such as the Argyres-Douglas (AD) theories \cite{Argyres:1995jj, Argyres:1995xn} and the massless $N_f=4$ theory \cite{Seiberg:1994aj}. Their topological partition functions and correlators can be found by considering them in certain mass deformations. The case of $N_f=1$ is analysed in \cite{moore2017}.
\end{enumerate}

The paper is organised as follows. In Section \ref{sec:SWtheories}, we present the Seiberg-Witten solution of SU(2) $\CN=2$ SQCD in flat space, focusing on the fundamental domains for the effective coupling, which we  illustrate  in several interesting examples. In Section \ref{sec:UVtheory}, we formulate the topological twist by coupling the hypermultiplets to external fluxes, such that the topological field theory is well-defined for arbitrary 't Hooft flux and non-spin manifolds. The topological low-energy effective theory coupled to $N_f$ background fluxes is then modelled in Section \ref{sec:eff_theories} as a $\mathrm{SU}(2)\times \mathrm{U}(1)^{N_f}$ theory, with the matter fields corresponding to frozen U(1) factors. This allows to compute the path integral explicitly as an integral over the $u$-plane. In Section \ref{sec:uplane_integral}, we formulate the $u$-plane integral as an integral over the fundamental domains. We prove that the single-valuedness under monodromies holds for a specific choice of magnetic winding numbers. 
Finally, in Section \ref{sec:integrationFD} we demonstrate that such integrals may be evaluated using mock modular forms, and we show that they localise at the cusps, elliptic points and interior singularities of the fundamental domains. 

A following paper, Part II, will be dedicated to explicit analyses and computations for specific examples. There, we will give a detailed discussion on the contribution from the AD points for each type, as briefly mentioned in point 3 above, as well as calculating the $u$-plane integrals for a number of examples.

\section{Special geometry and SW theories}\label{sec:SWtheories}
In this Section, we review aspects of the non-perturbative solution for the low energy effective theory of $\CN=2$ SQCD with gauge group SU(2) and $0\leq N_f\leq 3$ fundamental hypermultiplets \cite{Seiberg:1994rs, Seiberg:1994aj}. See \cite{Klemm:1997gg} for a review.  Throughout, we let $\Lambda_{N_f}$ denote the scale of the theory with $N_f$ hypermultiplets having masses $m_j$, $j=1,\dots,N_f$, and $a$ the mass of the W-boson on the Coulomb branch. 

\subsection{Field content}
The $\CN=2$ theories we consider contain a vector multiplet and $N_f\leq 3$ hypermultiplets. The fields in these multiplets form representations of ${\rm Spin}(4)=\text{SU}(2)_+\times \text{SU}(2)_-$ and $\text{SU}(2)_R$, which we denote by $({\boldsymbol k},{\boldsymbol l}, {\boldsymbol m})$, with ${\boldsymbol k},{\boldsymbol l}$ and ${\boldsymbol m}$ dimensions of the representations.

The vector multiplet consists of a gauge field $A_\mu$, complex scalar field $\phi$, and a pair of Weyl fermions $\Psi^I_\alpha$, $\bar \Psi^I_{\dot \alpha}$. This multiplet transforms under the adjoint representation of the gauge group $G$. The representation of $\text{SU}(2)_+\times \text{SU}(2)_- \times \text{SU}(2)_R$ formed by the bosonic fields is,
\be 
\label{repvecbos}
({\bf 2}, {\bf 2},{\bf 1})\oplus ({\bf 1}, {\bf 1},{\bf 1}) \oplus ({\bf 1}, {\bf 1},{\bf 1}),
\ee 
while the representation for the fermions is
\be 
\label{repvecferm}
({\bf 1}, {\bf 2},{\bf 2})\oplus ({\bf 2}, {\bf 1},{\bf 2}).
\ee 

The hypermultiplet consist of a pair of complex scalar fields, $q$ and $\tilde q$, and Weyl fermions, $\lambda_\alpha$, $\bar \lambda_{\dot \alpha}$, $\chi_\alpha$ and $\bar \chi_{\dot \alpha}$. We fix the gauge group $G=\text{SU}(2)$, and let the hypermultiplets transform under the fundamental representation of this group. With the same notation as above, the bosonic fields of this multiplet form the representation,
\be 
\label{rephypbos}
({\bf 1}, {\bf 1},{\bf 2})\oplus ({\bf 1}, {\bf 1},{\bf 2}),
\ee 
while the fermions form the representation
\be 
\label{rephypferm}
({\bf 2}, {\bf 1},{\bf 1})\oplus ({\bf 1}, {\bf 2},{\bf 1})\oplus ({\bf 2}, {\bf 1},{\bf 1})\oplus ({\bf 1}, {\bf 2},{\bf 1}).
\ee

\subsection{Seiberg-Witten geometry} \label{sec:SWgeometry}
The Seiberg-Witten geometry underlies the Coulomb branch of $\CN=2$ gauge theory. The Coulomb branch is the phase of the theory where $\text{SU}(2)$ is broken to $\text{U}(1)$ by a vacuum expectation value (vev) of the vector multiplet scalar $\phi$. The vev is semi-classically parametrised by 
 a complex parameter $a$, 
\be 
\phi=\left(\begin{array}{cc} a & 0 \\ 0 & -a \end{array}\right),
\ee 
up to gauge transformations. In particular, $a\to -a$ is a gauge transformation. The gauge invariant order parameter is the Coulomb branch expectation value of the theory in $\mathbb{R}^4$,
\be 
\label{defu}
u=\frac{1}{16\pi}\left<{\rm Tr}(\phi^2)\right>_{\mathbb{R}^4}.
\ee 

The non-perturbative effective action of $\CN=2$ SQCD is characterised by the prepotential $F(a,\bfm)$, with $\bfm$ the mass vector $\bfm=(m_1,\dots,m_{N_f})$. The semi-classical part of $F$ reads \cite{Ohta:1996hq, Ohta_1997,DHoker_1997, Ne}
\be\label{prepotential}
\begin{split}
&F(a,\bfm)=\frac{2\im}{\pi}  a^2 \left(\log(4a/\Lambda_{N_f})-\frac{3}{2}\right)-\frac{1}{2}\sum_{j=1}^{N_f}\left( n_j\frac{m_j}{\sqrt{2}}\,a+\, C\, (a^2+m_j^2/2)\right)\\
&\quad -\frac{\im }{4\pi} \sum_{j=1}^{N_f} \left(a+\tfrac{m_j}{\sqrt 2}\right)^2\log((a+\tfrac{m_j}{\sqrt 2})/\Lambda_{N_f})+\left(a-\tfrac{m_j}{\sqrt 2}\right)^2\log((a-\tfrac{m_j}{\sqrt 2})/\Lambda_{N_f})\\
&\quad +\dots,
\end{split} 
\ee 
where the $\dots$ indicate further non-perturbative corrections. Here $C=\frac{1}{2}+\frac{i}{2\pi}\log(2)-\frac{3i}{2\pi}$. The classical terms proportional to $a^2$ and $m^2$ on the first line are chosen to facilitate the decoupling of hypermultiplets, which will be discussed later in more detail. 

The $n_j\in \mathbb{Z}$ in \eqref{prepotential} are the \emph{magnetic winding numbers} of the periods $a_D\coloneqq \frac{\partial F}{\partial a}$ dual to $a$
\cite{Ohta:1996hq,Ohta_1997,ohta1999}. These numbers seem to be only rarely discussed in the literature beyond these references.\footnote{Nekrasov's partition function gives a specific choice upon expanding the function $\gamma_{\hbar}(x;\Lambda)$ in the perturbative part \cite{Ne,
  NekOk}.} Generally, the theory allows for $N_f$ electric winding numbers for $a$ and $N_f$ magnetic winding numbers for $a_D$. These appear in the massive $N_f>0$ theories since the Seiberg-Witten differentials now have poles with nonzero residues \cite{Ohta:1996hq}. See also Appendix \ref{sec:winding}.
The choice \eqref{prepotential} of the prepotential corresponds to fixing the electric winding numbers to be zero, or equivalently fixing the monodromy at infinity to map $a\to e^{\pi i}a$. Compare for example with \cite[Eq. (2.17)]{Ohta:1996hq}.  In Section \ref{sec:uplane_integral}, we will discuss that the single-valuedness of the
$u$-plane integral requires $n_j\equiv -1 \mod 4$.

We introduce the period $a_D$ dual to $a$, and the parameters $m_{D,j}$ dual to $m_j$ by
\be\label{dualmass}
a_D=\frac{\partial F}{\partial a},\qquad m_{D,j}=\sqrt{2}\frac{\partial F}{\partial m_j}.
\ee
These parameters are further combined into the $(2+2N_f)-$dimensional vector $\Pi$, 
\be
\Pi=\left( \begin{array}{c} a_D \\ a \\ m_{D,1} \\ \tfrac{m_{1}}{\sqrt{2}} \\ \vdots \\ m_{D,N_f} \\ \tfrac{m_{N_f}}{\sqrt{2}} \end{array}\right).
\ee
This vector forms a local system over the $u$-plane. The elements of the vector form the symplectic form,
\begin{equation}\label{sympForm}
    \omega_{N_f}=da_D\wedge da+ \frac{1}{\sqrt2}\sum_{j=1}^{N_f}dm_{D,j}\wedge d m_j.
\end{equation}

The effective gauge coupling is related to the prepotential through
\be
\label{tauF}
\tau=\frac{\partial^2 F}{\partial a^2}.
\ee
We also introduce the couplings $v_j$ and $w_{jk}$ with $j,k\in 1,\dots, N_f$,\footnote{If we consider $F$ as a function of $(a,\frac{1}{\sqrt2} m_1,\dots, \frac{1}{\sqrt2} m_{N_f})$, then the dual parameters are encoded in the Jacobian $\boldsymbol{J}_F=(a_D,\bfm_D)$, while the couplings are the elements of the Hessian $\boldsymbol{H}_F=\begin{psmallmatrix} \tau & \bfv^T \\ \bfv & \bfw \end{psmallmatrix}$.}
\be
\label{Defvw}
v_j=\sqrt{2}\frac{\partial^2 F}{\partial a \partial m_j}, \qquad w_{jk}=2\frac{\partial^2 F}{\partial m_j\partial m_k}.
\ee
The derivative of the prepotential with respect to the scale $\Lambda_{N_f}$ provides the  order parameter $u$ (\ref{defu}) on the Coulomb branch, 
\be\label{udlambdadF}
u=\frac{4\pi i}{4-N_f} \Lambda_{N_f} \frac{\partial F(a,\bfm)}{\partial \Lambda_{N_f}}+\frac{1}{4-N_f}\sum_j m_j^2.
\ee

Using the different relations introduced above together with the fact that the prepotential satisfies the homogeneity equation \cite{Sonnenschein_1996,Eguchi:1995jh,DHoker:1996yyu}
\begin{equation}\label{homogeneity_F}
	2 F=\Lambda_{N_f} \frac{\partial F}{\partial \Lambda_{N_f}} +\sum_{j=1}^{N_f}m_j\frac{\partial F}{\partial m_j}+a \frac{\partial F}{\partial a},
\end{equation}
we can find some non-trivial relations between the respective quantities. For instance, from the perturbative prepotential (\ref{prepotential}), we deduce that the leading terms of $u$ are
\be 
\label{uwc}
u=2\,a^2+\CO(a^{-2})
\ee
The weak-coupling limit then becomes in our conventions\footnote{Note that this differs slightly from some of the previous literature. However, it is shown in \cite{Aspman:2021vhs} to be the unique limit consistent with the RG flow.}
\be 
\label{decouplingConv}
\left\{ \begin{array}{l} \tau\to +i\infty, \\ a\to -i\infty, \\ u\to -\infty. \end{array} \right.
\ee

The Seiberg-Witten (SW) solution provides a family of elliptic curves parametrised by the order parameter $u$ and the masses $m_i$, whose complex structure corresponds to the running coupling $\tau=\frac{\theta}{\pi}+\frac{8\pi i}{g^2}$. For the theories of interest in this paper, the curves $\mathcal{S}_{N_f}$ are given by \cite{Seiberg:1994aj}
\begin{equation}\label{eq:curves}
	\begin{aligned}  
		N_f=0:\quad y^2&=x^3-ux^2+\frac{1}{4}\Lambda_0^4x, \\
		N_f=1:\quad y^2&=x^2(x-u)+\frac{1}{4}m\Lambda_1^3x-\frac{1}{64}\Lambda_1^6, \\
		N_f=2:\quad y^2&=(x^2-\frac{1}{64}\Lambda_2^4)(x-u)+\frac{1}{4}m_1m_2\Lambda_2^2x-\frac{1}{64}(m_1^2+m_2^2)\Lambda_2^4, \\
		N_f=3:\quad y^2&=
		x^2(x-u)-\frac{1}{64}\Lambda_3^2(x-u)^2-\frac{1}{64}(m_1^2+m_2^2+m_3^2)\Lambda_3^2(x-u)\\&\quad
		+\frac{1}{4}m_1m_2m_3\Lambda_3x-\frac{1}{64}(m_1^2m_2^2+m_2^2m_3^2+m_1^2m_3^2)\Lambda_3^2.
	\end{aligned}
\end{equation} 
The family of SW curves (\ref{eq:curves}) are Jacobian rational elliptic surfaces with singular fibres \cite{Malmendier:2008yj, Caorsi:2018ahl, Closset:2021lhd}. These surfaces are well studied in the mathematical literature  \cite{shioda1972,schuett2009,maier2006}. Rational in this context means that $g_2$ and $g_3$ are polynomials in $u$ of
degree at most $4$ and $6$, respectively \cite{miranda1995}. In Appendix \ref{app:class_S}, we summarise the class $\CS$ representation of the SW curves \cite{Gaiotto:2009hg}. 

The curve for $N_f$ hypermultiplets reduces to the curve for $N_f-1$ hypermultiplets upon decoupling a hypermultiplet in the double
scaling limit
\cite{Eguchi1999}
\begin{equation}\label{decouplimit}
	m_{N_f}\to \infty, \quad \Lambda_{N_f}\to 0, \quad m_{N_f}\Lambda_{N_f}^{4-N_f}=\Lambda_{N_f-1}^{4-(N_f-1)}.
\end{equation} 

The curves (\ref{eq:curves}) provide the exact results for the vevs of the scalar $a$ and its dual $a_D$ as period integrals. We have explicitly,
\be
a=\int_A \lambda,\qquad a_D=\int_B \lambda, 
\ee
with $\lambda$ the SW differential, for which we refer to \cite{Seiberg:1994aj,Ohta_1997,Ohta:1996hq}.

\subsection{Characteristic functions on the Coulomb branch}  \label{sec:order_parameters}
In this subsection, we discuss various characteristic functions on the Coulomb branch, which are instrumental for determining the $u$-plane integral. They are: 
\begin{itemize}
\item the physical discriminant $\Delta_{N_f}$,
\item the order parameter $u$,
\item the period $da/du$, and 
\item the derivatives $du/d\tau$ and $da/d\tau$.
\end{itemize}
\vspace{.2cm}
{\it The physical discriminant $\Delta_{N_f}$}\\
We define the physical discriminant $\Delta_{N_f}$ as the monic polynomial
\begin{equation}
    \Delta_{N_f}=\prod_{j=1}^{N_f+2}(u-u_j),
\end{equation}
where $u_j$ for $j=1,\dots, N_f+2$ are the singular points of the effective theory. We let $j=1,\dots, N_f$ label the singular points where one of the matter hypermultiplets becomes massless; and $j=N_f+1, N_f+2$ denote the strong coupling singularities where a monopole and a dyon, respectively, becomes massless.

The discriminant is a polynomial of degree $2+N_f$ in $u$. It can also be determined directly from the SW curve  \cite{Aspman:2021vhs},
\begin{equation}\label{phys_disc}
    \Delta_{N_f}=(-1)^{N_f}\Lambda_{N_f}^{2N_f-8}(g_2^3-27g_3^2),
\end{equation}
where $g_2$ and $g_3$ are the Weierstra{\ss} invariants of the SW curve. Up to the sign and scale, the physical discriminant $\Delta_{N_f}$ equals the mathematical discriminant $g_2^3-27g_3^2$ for these theories.

As the masses are tuned, some of the singularities on the Coulomb branch can collide. 
If we consider $\Delta_{N_f}$ as a polynomial in $u$, its discriminant $D(\Delta_{N_f})$ vanishes if and only if two roots coincide. It is straightforward to show that for $N_f\leq 3$,
\begin{equation}\label{dDelta}
D(\Delta_{N_f})=\left( D^{\text{AD}}_{N_f}\right)^3\prod_{i<j}(m_i-m_j)^2(m_i+m_j)^2.
\end{equation}
 This factorises the locus in mass space where singularities collide into two orthogonal components: The first component is the Argyres-Douglas (AD) locus given by the polynomial equation $D^{\text{AD}}_{N_f}=0$, where mutually \emph{non-local} singularities collide \cite{Argyres:1995xn,Aspman:2021vhs,EGUCHI1996430}. The other component is characterised by the equations $m_i=\pm m_j$, and one can check that this gives rise to mutually \emph{local} singularities colliding. Here, the flavour symmetry gets enhanced and a Higgs branch opens up \cite{Seiberg:1994aj}. This extends the analysis of \cite{Aspman:2021vhs}. 
 
\begin{figure}[ht]\centering
	\includegraphics[scale=1]{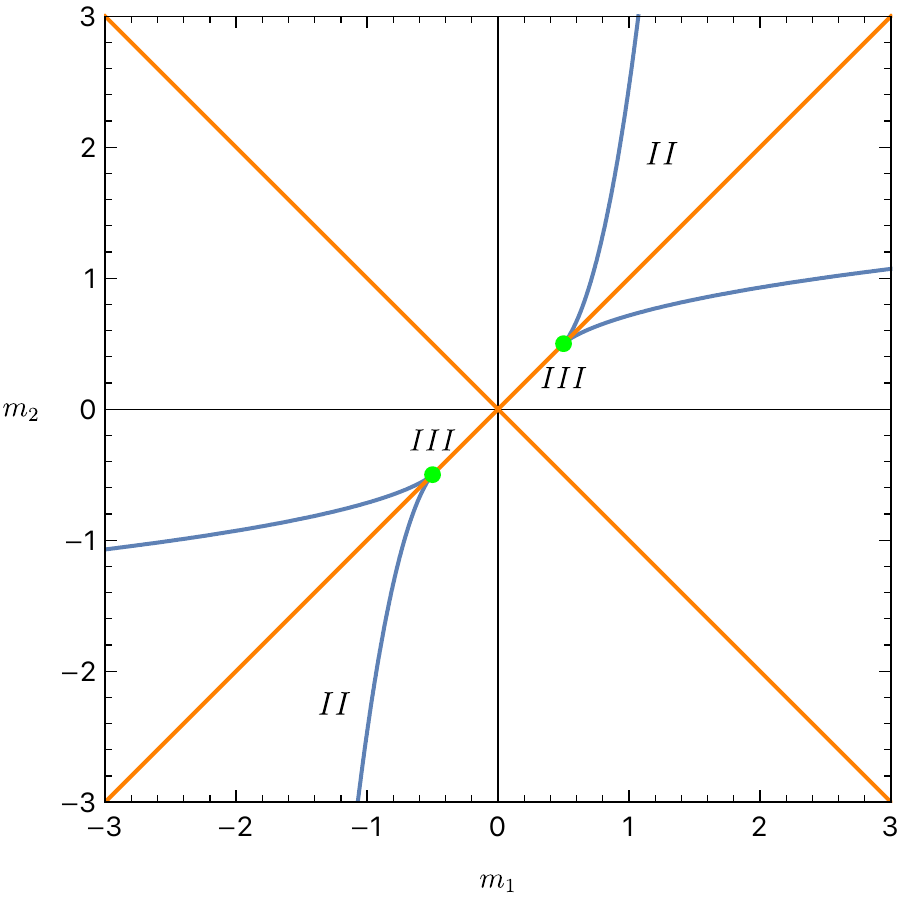}
	\caption{Partitioning of the real $\boldsymbol m=(m_1,m_2)$ plane in $N_f=2$, in units of $\Lambda_2=1$. On the AD component  the Coulomb branch $\mathcal B_{2}(\boldsymbol m)$ contains an AD point of Kodaira type II (blue) or III (green). On the other component (orange), mutually local singularities collide. If $\boldsymbol m$ is varied along a continuous path that does not cross the partitioning $0=D(\Delta_{N_f})$, the weight vector $\boldsymbol k(\boldsymbol m)$ is constant.  }\label{fig:DDlocus2}
\end{figure}

Given a mass configuration $\bfm=(m_1,\dots, m_{N_f})$, we can denote by $k_l$ the weight (or multiplicity) of the $l$-th singularity, and by $\bfk(\bfm)=(k_1,k_2,\dots)$ the vector of those weights. Since the Coulomb branch  $\CB_{N_f}(\bfm)$ contains $2+N_f$ singularities aside from weak coupling $u=\infty$, it is clear that $\bfk(\bfm)$ provides a partition of $2+N_f$. 
This in turn partitions the mass space $\mathbb C^{N_f}\ni\bfm$ into finitely many regions where $\bfk(\bfm)$ is locally constant. 
 As an example,  in Fig. \ref{fig:DDlocus2} we plot  the contours of \eqref{dDelta} for $N_f=2$ in the real $\bfm=(m_1,m_2)$ plane.

The possible singularity structures of the rational elliptic curves \eqref{eq:curves} are classified in Persson's list of allowed configurations of singular fibres \cite{Persson:1990,Miranda:1990}. From Kodaira's classification, it follows that any solution to \eqref{dDelta} gives rise to a singularity on the Coulomb branch of Kodaira type $I_k$, $II$, $III$, or $IV$. As described in \cite{Aspman:2021vhs,Closset:2021lhd}, the solutions to $0=D^{\text{AD}}_{N_f}$ give rise to AD points of Kodaira type $II$, $III$ and $IV$. The second component $0=\prod_{i<j}(m_i-m_j)(m_i+m_j)$ can be studied in more detail. These are $2(N_f-1)$ independent equations. Whenever one of the factors vanishes, the SW surface contains an $I_k$ singularity with $k\geq 2$. For $N_f=2$, the only possibility is $I_2$, while for $N_f=3$ singularities of type $I_2$, $I_3$ and $I_4$ are possible. The point in the Coulomb branch $\CB_{N_f}$ corresponding to an $I_k$ singularity with $k\geq 2$ intersects with a Higgs branch of quaternionic dimension $k-1\geq 1$ \cite{Seiberg:1994aj}. Further merging these $I_k$ singularities with a mutually non-local singularity does not affect the Higgs branch, such that the points with a $III$ or $IV$ singularity also intersect with a Higgs branch of quaternionic dimension one or two, respectively, while the points with AD theories of type $II$ do not intersect with any Higgs branch.
\vspace{.2cm}\\
{\it Order parameter $u$}\\ 
The order parameter $u$, introduced in \eqref{defu}, is invariant under monodromies on the Coulomb branch. By bringing the SW curve to Weierstra{\ss} form,
$u$ can be expressed in terms of the running coupling $\tau\in \mathbb{H}$ (\ref{tauF}) \cite{Nahm:1996di, Brandhuber:1996ng, Huang:2009md,Aspman:2021vhs,Aspman:2021evt}. As in \cite{Aspman:2021vhs}, we restrict the domain for $\tau\in \mathbb{H}$ to $\CF(\bfm)\subseteq \mathbb{H}$, such that $u$ is 1-to-1. 
In specific cases, $u$ can be expressed in terms of modular forms.  We follow the conventions of \cite[Section 2.2]{Aspman:2021vhs} in picking  the solutions.   We discuss the order parameters in more detail in the following Subsection \ref{SecFundDom}.
\vspace{.2cm}\\ 
{\it The period  $da/du$}\\
The period $\frac{da}{du}$ can be related to the complex structure $\tau$ of the curve through \cite{Brandhuber:1996ng}
\begin{equation}\label{dadu_def}
	\frac{da}{du}=\frac{1}{6}\sqrt{\frac{g_2}{g_3}\frac{E_6}{E_4}},
\end{equation}
where the Weierstra{\ss} invariants $g_2$, $g_3$ are polynomials in $u$, and $E_4=E_4(\tau)$, $E_6=E_6(\tau)$ are the Eisenstein series of modular weights four and six respectively, defined in Eq. (\ref{Ek}). The square root in (\ref{dadu_def}) leads to an ambiguity in the sign of $da/du$. This can be resolved by taking a branch of the square root, for example at weak coupling. On the other hand, $da/du$ is not a single-valued function on the $u$-plane. It changes by a sign under monodromies at weak coupling, and transforms with a modular weight under monodromies around strong coupling singularities.
\vspace{.2cm}\\
{\it The derivatives $du/d\tau$ and $da/d\tau$}\\
The derivative $du/d\tau$ is related to the period $\frac{da}{du}$ through \cite{Matone:1995rx, Aspman:2021vhs}
\begin{equation}\label{Matone}
	\frac{du}{d\tau}=-\frac{16\pi i}{4-N_f}\frac{\Delta_{N_f}}{P^{\text{M}}_{N_f}}\left(\frac{da}{du}\right)^2,
\end{equation}
where $P^{\text{M}}_{N_f}$ is a polynomial in $u$. For $N_f\leq 3$, it reads
\begin{equation}
    P^{\text{M}}_{N_f}=\frac{6}{4-N_f}(-1)^{N_f}\Lambda_{N_f}^{2N_f-8}(2g_2g_3'-3g_2'g_3),
\end{equation}
where the prime denotes differentiation with respect to $u$. It appears naturally in differential equations associated with elliptic surfaces (see e.g. \cite{stiller1981}).

When $u$ is known in terms of modular forms, it is also straightforward to determine $\frac{du}{d\tau}$ by differentiating modular forms \cite{Aspman:2021vhs}. For generic masses $\bfm$, we can compute the $q$-series of $u$ and thus $\frac{du}{d\tau}$ to any order. We will not explicitly need $du/d\tau$ for evaluating the $u$-plane integral, but through
\begin{equation}\label{dadtau}
    \frac{da}{d\tau}=\frac{da}{du}\frac{du}{d\tau}
\end{equation}
we can use it to evaluate $da/d\tau$. This derivative appears as the Jacobian for the change of integration variables from the periods $a$ to the couplings $\tau$.

\subsection{Fundamental domains}
\label{SecFundDom}
The recent progress   in computing $u$-plane integrals has been enabled by mapping the $u$-plane to a modular fundamental domain, on which the $u$-plane integrand can be related to mock modular forms and thus be efficiently evaluated \cite{Malmendier:2008db,Griffin:2012kw,Malmendier:2012zz,Malmendier:2010ss,Korpas:2017qdo,Moore:2017cmm,Korpas:2019ava,Korpas:2019cwg,Manschot:2021qqe,Aspman:2021kfp,Korpas:2022tij}. It has been known since the 1990s that the $u$-planes for $\CN=2$ SQCD with $N_f=0,2,3$ massless hypermultiplets are modular and correspond to fundamental domains for congruence subgroups of $\psl$ \cite{Nahm:1996di}. On the other hand, the generic mass case including the peculiar role of massless $N_f=1$ have remained elusive. 

 The order parameter $u$ for a given mass $\bfm=(m_1,\dots, m_{N_f})$ can be considered as a function 
\begin{equation}
    u: \mathbb H\to \CB_{N_f}(\bfm),
\end{equation}
where $\CB_{N_f}(\bfm)$ is the Coulomb branch of the theory with $N_f$ hypermultiplets of mass $\bfm$. In \cite{Aspman:2021vhs}, it was found that for $N_f\geq 1$ and generic masses the duality group does not act on $\tau$ by fractional linear transformations. This prevents the preimage $u^{-1}(\CB_{N_f}(\bfm))$ from being a modular fundamental domain for a subgroup of $\psl$. Instead, we can define a fundamental domain $\CF_{N_f}(\bfm)$  through the equivalence relation $\tau\sim \tau' \Leftrightarrow u(\tau)=u(\tau')$, such that $u: \CF_{N_f}(\bfm)\to \CB_{N_f}(\bfm)$ is bijective. In \cite{Aspman:2021vhs,Aspman:2021evt}, such fundamental domains for the effective gauge coupling  have been found explicitly   for $\CN=2$ SQCD with $N_f\leq 4$ generic masses. They decompose as
\begin{equation}\label{fundamental_domain}
    \CF_{N_f}(\bfm)=\bigcup_{j=1}^{n}\alpha_j\CF,
\end{equation}
where 
$\CF=\psl\backslash \mathbb H$ is the key-hole fundamental domain of $\psl$, and  $\alpha_j\in\psl$ are $n\leq 6$ maps that are locally constant as functions of the masses $\bfm$. More precisely, $\CF_{N_f}(\bfm)$ is constant on any of the finitely many connected components of $\{\bfm\in \mathbb C^{N_f}\, | \, D(\Delta_{N_f})=0\}$, where $D(\Delta_{N_f})$ is given in \eqref{dDelta}.
We also call $n$ the index of the domain $ \CF_{N_f}(\bfm)$.

The geometry of $\CF_{N_f}(\bfm)$ agrees with that of the Coulomb branch $\CB_{N_f}(\bfm)$. Namely, $\CB_{N_f}(\bfm)$ is the complex plane with $2+N_f$ singular points removed.  In \eqref{fundamental_domain}, those singularities are reflected in the cusps $\alpha_j(\infty)$, where the number of $\alpha_j$ with the same $\alpha_j(\infty)$ (giving the \emph{width} of the cusp) agrees with the multiplicity of the singularity $u(\alpha_j(\infty))$.

The domains $ \CF(\bfm)\coloneqq \CF_{N_f}(\bfm)$ are endowed with further data. For generic masses $\bfm$,  there are $N_f$ pairs of branch points on $\CF(\bfm)$ that are connected by branch cuts. The branch points are determined by the non-removable singularities of the rational function $\Delta_{N_f}(u)/P^{\text{M}}_{N_f}(u)$, as it appears in Matone's relation \eqref{Matone}. They correspond to the nontrivial zeros of the discriminant of the sextic equation associated with the SW curve \cite{Aspman:2021vhs}. As the masses are varied, the branch points move on continuous paths in the fundamental domain. 

When two branch points collide, there appears an elliptic point, which is a  further dedicated point  on the domain $\CF(\bfm)$. These elliptic points 
correspond to the superconformal Argyres-Douglas (AD) points \cite{Argyres:1995jj,Argyres:1995xn}. By Kodaira's classification of singular fibres, they are always in the $\psl$ orbit of $i$ or $e^{\pi i/3}$. Their presence is also responsible for the index $n$ of the domain to be reduced. The possible values are $n=2,3,4$ if $\CF(\bfm)$ contains such an elliptic point, and $n=6$ otherwise.

The $N_f=4$ theory furthermore contains a UV coupling $\tuv$. Since $u(\tau)\to \infty$ for $\tau\to \tuv$, it can be viewed as a puncture of the fundamental domain, and has to be excluded from the integration domain.

Let us review the explicit construction of $\CF(\bfm)$ in a few important examples. Many more examples can be found in \cite{Aspman:2021vhs, Aspman:2021evt}. Here, we are mainly considering the masses to be small compared to the period $a$, such that the mass singularities discussed in previous sections can be considered as strong coupling singularities, or in other words as cusps on the real line of the fundamental domains. When the masses are increased it is more natural to think of the mass singularities as weakly coupled. 

\begin{figure}[ht]\centering
	\includegraphics[scale=0.8]{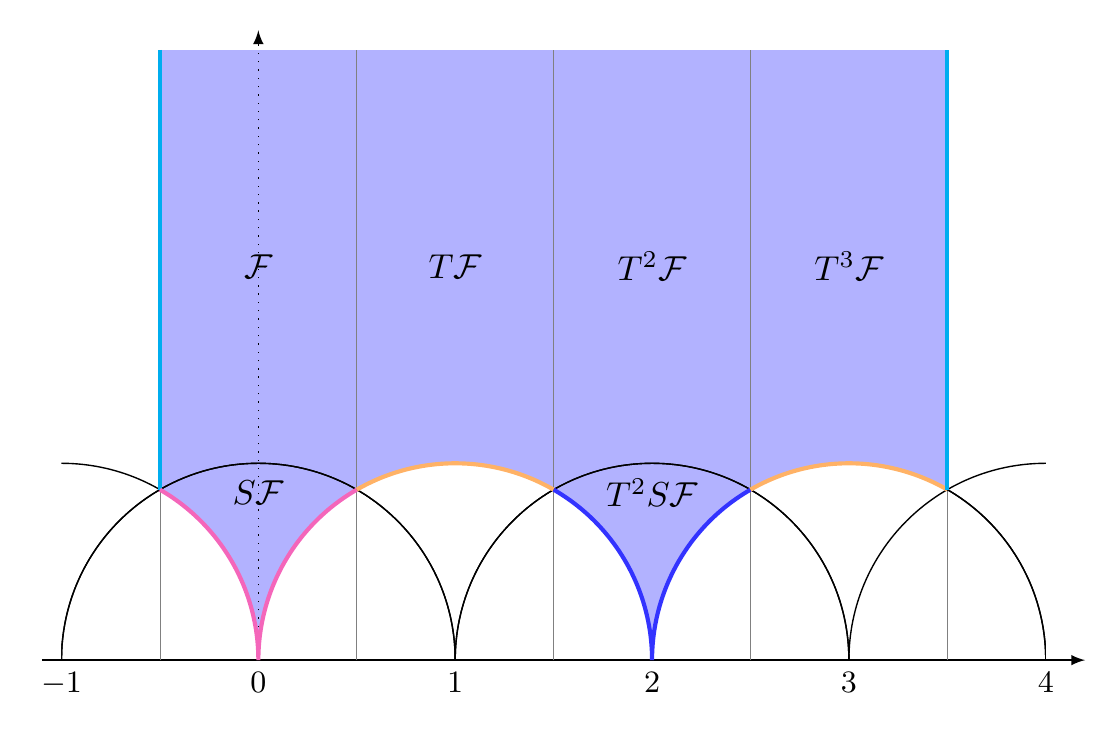}
	\caption{Fundamental domain of $\Gamma^0(4)$. This is the duality group of the pure SU(2) theory. The two cusps on the real line correspond to the strong coupling singularities of the gauge theory, while the cusp at $\tau=i\infty$ corresponds to weak coupling. Boundaries of the same colour are identified.  }\label{fig:fundgamma0(4)}
\end{figure}

\vspace{.2cm}
\noindent {\it The pure SU(2) $N_f=0$ theory}\\
It is well-known that the pure SU(2) Coulomb branch corresponds to the fundamental domain for the congruence subgroup $\Gamma^0(4)$ of $\psl$. The order parameter can be explicitly determined in terms of a modular function for the duality group \cite{matone1996},
\begin{equation}\label{nf0parameter}
	\begin{aligned}
		\frac{u}{\Lambda_0^2}=&-\frac{1}{2}\frac{\vartheta_2^4+\vartheta_3^4}{\vartheta_2^2\vartheta_3^2},
	\end{aligned}
\end{equation}
	where the Jacobi theta functions $\jt_i$ are defined in Appendix \ref{sec:jacobitheta}.
A fundamental domain for the pure SU(2) theory is given in Fig. \ref{fig:fundgamma0(4)}. Since the domain is modular, the branch points are absent.

\begin{figure}[ht]\centering
	\includegraphics[scale=0.8]{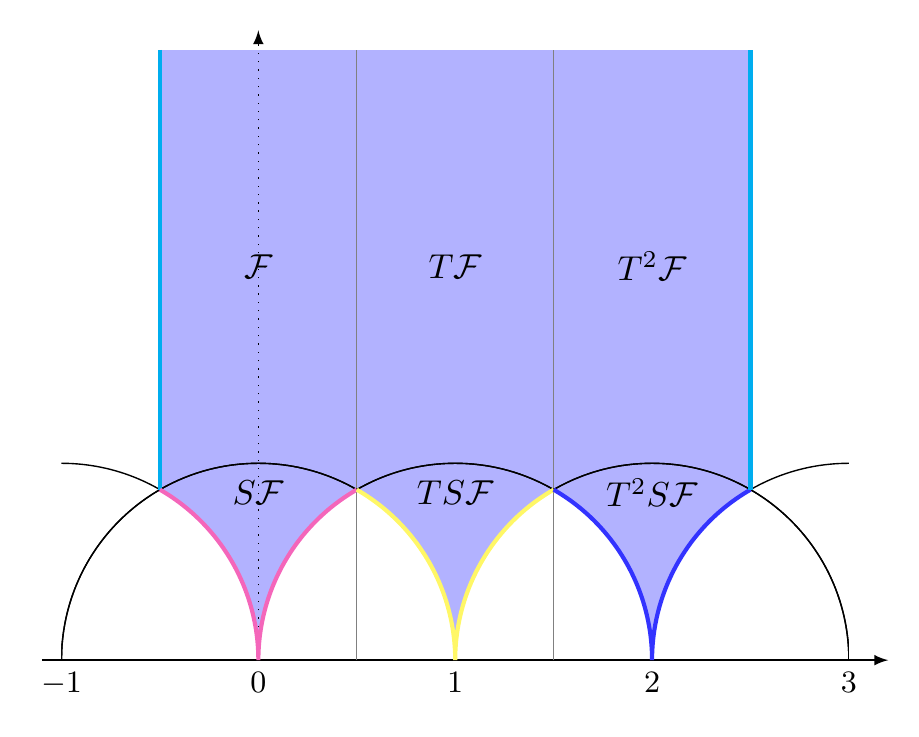}
	\caption{Fundamental domain for massless $N_f=1$.}\label{fig:fundnf1}
\end{figure}

\vspace{.2cm}
\noindent {\it The massless $N_f=1$ theory}\\
From the massless $N_f=1$ SW curve, one directly finds \cite{Nahm:1996di}
\begin{equation}\label{nf1u}
	\begin{aligned}
		\frac{u}{\Lambda_1^2}=&-\frac{3}{2^{\frac73}}\frac{E_4^{\frac12}}{(E_4^{\frac32}-E_6)^{\frac13}},
	\end{aligned}
\end{equation}
where $E_4$ and $E_6$ are the Eisenstein series defined in (\ref{Ek}).  This function also appears as an order parameter in pure SU(3) SW theory \cite{Aspman:2020lmf} as well as in the description of certain elliptically fibred Calabi-Yau spaces \cite{Klemm:2012sx}. Due to the presence of the square root, this function is not modular. The massless $N_f=1$ curve has three distinct strongly coupled singularities, which become the three cusps of the fundamental domain aside from the weakly coupled region $u\to\infty$. The fundamental domain is plotted in Fig. \ref{fig:fundnf1}. Note that, this is not the fundamental domain of any subgroup of $\psl$. 

\begin{figure}[ht]
	\begin{subfigure}{.5\textwidth}
		\centering
		\includegraphics[width=\linewidth]{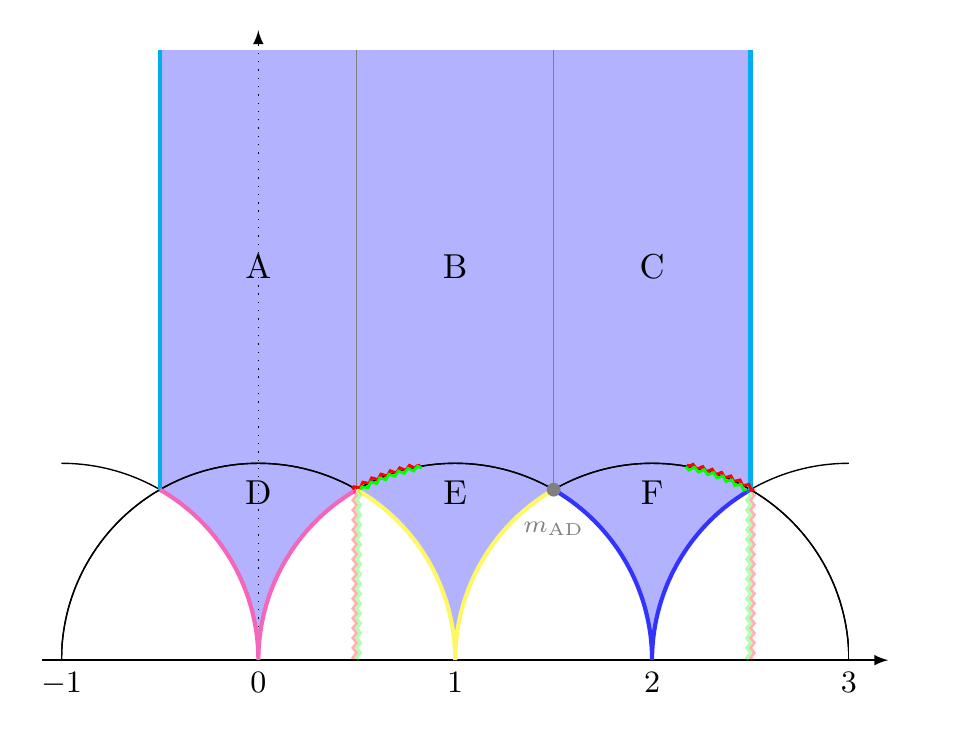}  
		\caption{}
	\end{subfigure}
	\begin{subfigure}{.5\textwidth}
		\centering
		\includegraphics[width=\linewidth]{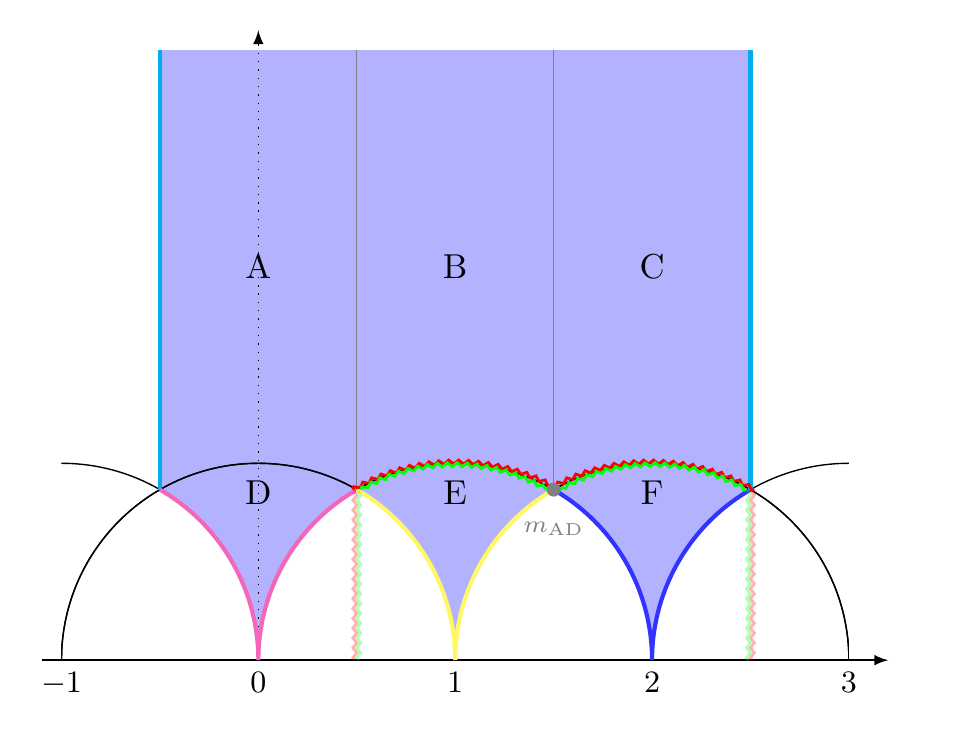} 
		\caption{} 
	\end{subfigure}
\newline
	\begin{subfigure}{.5\textwidth}
		\centering
		\includegraphics[width=\linewidth]{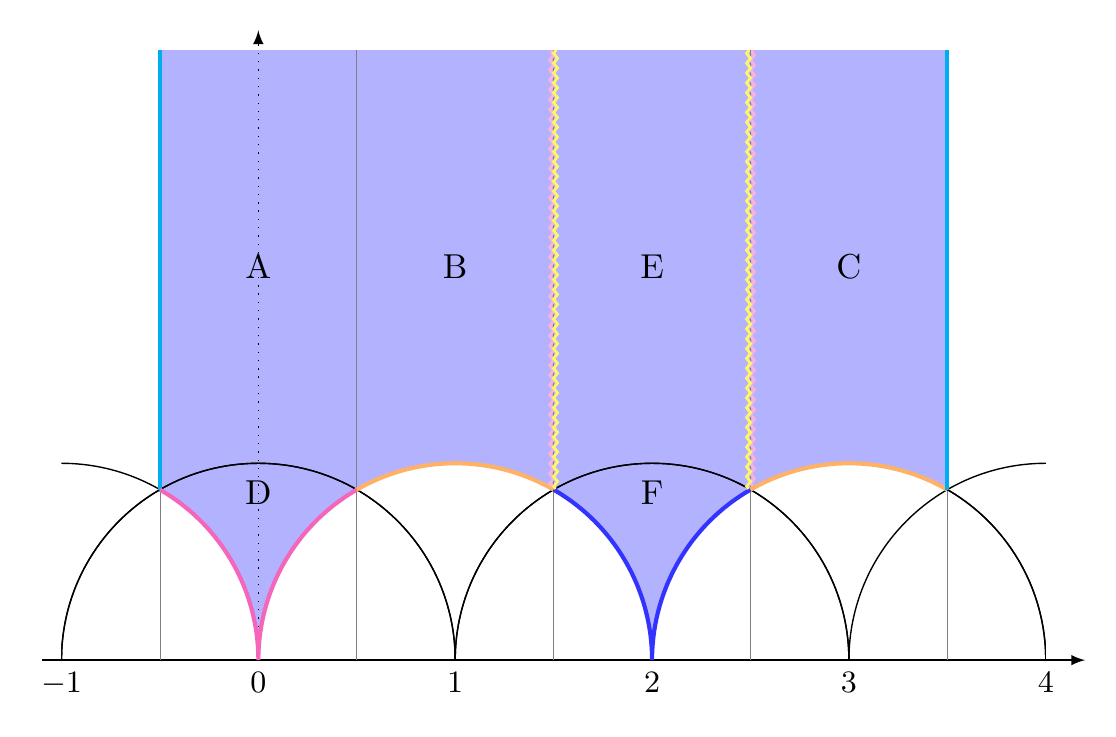}  
		\caption{}
	\end{subfigure}
\begin{subfigure}{.5\textwidth}
	\includegraphics[width=\linewidth]{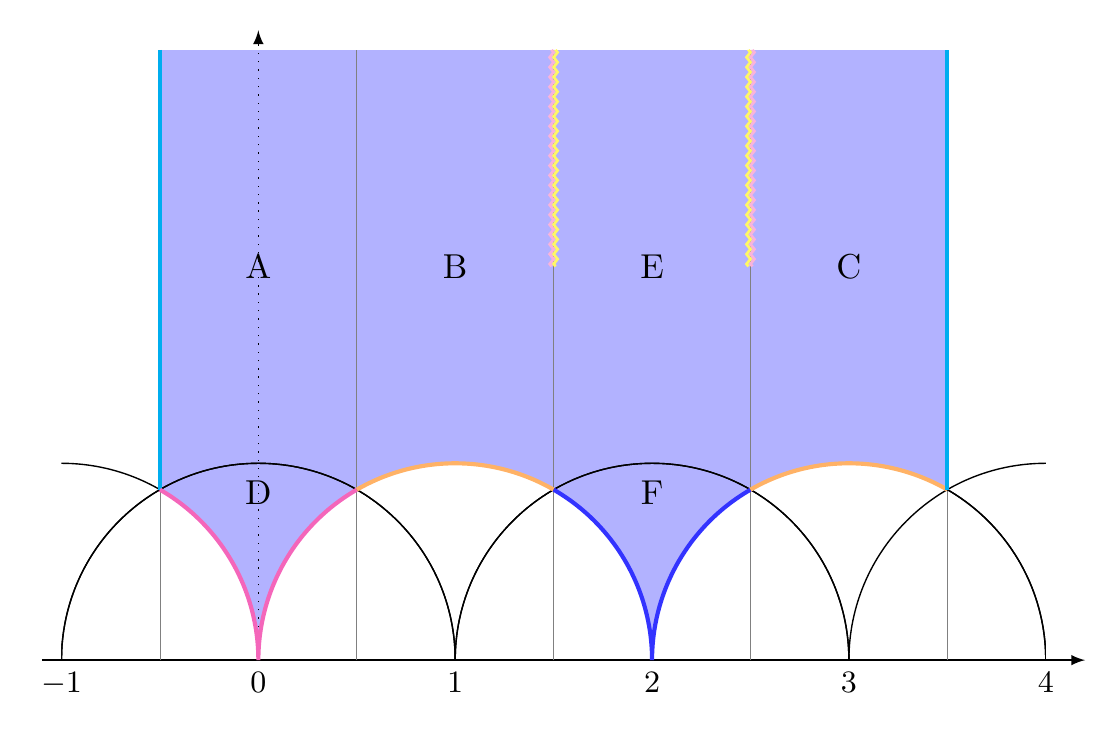}
	\caption{}
\end{subfigure}
	\caption{Choice of branch cuts (zigzag lines) for varying (real) mass in $N_f=1$. Lines with the same color are identified. Starting with a small mass in Figure (a), we cut along the arcs of radius 1 around $\tau=1$ and $\tau=2$. At the AD mass we can use the identifications of the different boundaries to reorganise the domain in Figure (b) to the one of Figure (c). 
	When we increase the mass further the cuts of Figure (c) move upwards as in Figure (d) eventually reaching infinity and disappearing, leaving us with the domain in Fig.  \ref{fig:fundgamma0(4)} of the pure theory.}
	\label{fig:nf1CutsAlt}
\end{figure}

When a positive mass $m$ is turned on, the branch points move in the fundamental domain emerging from those of the massless case. They move as a function of the mass as shown in \cite{Aspman:2021vhs}, and depicted in Fig. \ref{fig:nf1CutsAlt}. This further incorporates the hypermultiplet decoupling on the level of the fundamental domain.

\begin{figure}[ht]\centering
	\includegraphics[scale=0.8]{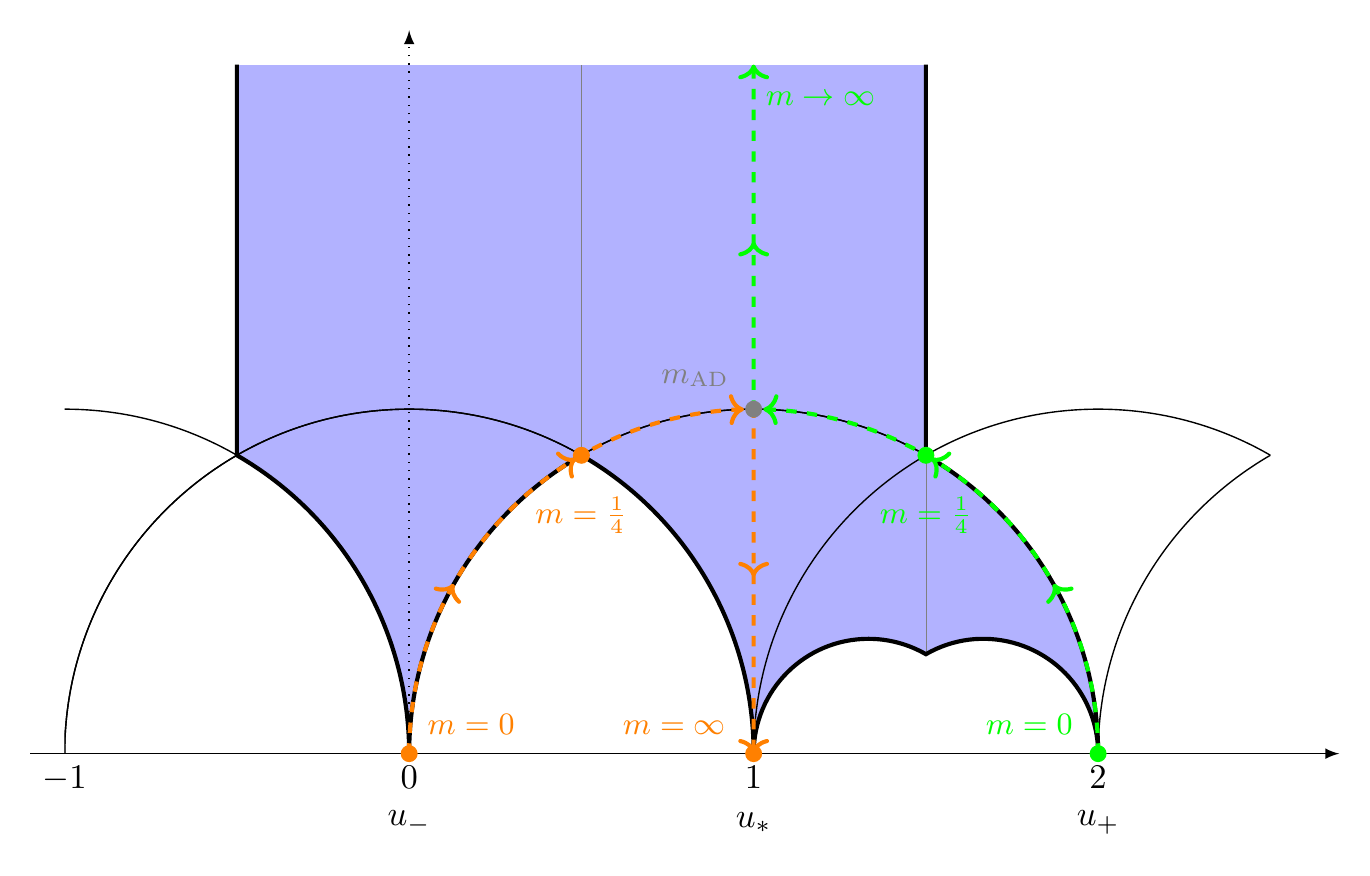}
	\caption{Fundamental domain $\mathcal F_2(m,m)$ of the massive
		$\boldsymbol m=(m,m)$ $N_f=2$ theory. The dashed lines correspond to
		the paths of the branch points from zero to
		infinite mass. For given positive mass $m$, the two branch
		points are identified under $TST^{-1}$, such that there is
		only one branch point
		$u_{\text{bp}}=2m^2-\tfrac{\Lambda_2^2}{8}$ on the
		$u$-plane.}\label{fig:nf2domain} 
\end{figure}

\vspace{.2cm}
\noindent {\it The equal mass $N_f=2$ theory}\\
For $N_f=2$ with $m_1=m_2=m$, the order parameter can be determined explicitly \cite{Aspman:2021vhs},
\begin{equation}\label{umm}
	\frac{u}{\Lambda_2^2}=-\frac{\jt_4^8+\jt_2^4\jt_3^4+(\jt_2^4+\jt_3^4)\sqrt{16\frac{m^2}{\Lambda_2^2}\jt_2^4\jt_3^4+\jt_4^8}}{8\jt_2^4\jt_3^4}.
\end{equation}
The curve has a double singularity at  $u_*=m^2+\frac{\Lambda_2^2}{8}$ and two simple singularities at $u_\pm=-\frac{\Lambda_2^2}{8}\pm m\Lambda_2$. 
When $m$ is varied, it can be determined how the branch point $\tau_{\text{bp}}$ moves in the fundamental domain. When $m$ is not zero and not equal to $\mad=\tfrac{\Lambda_2}{2}$, the singularities $u_*$, $u_+$ and $u_-$ are distinct and the fundamental domain $\CF(m,m)$ is given in Fig. \ref{fig:nf2domain}. For any such mass, there is a pair of branch points in $\CF(m,m)$, which are connected by a branch cut. 
Near a branch point $\tau_\bp$, $u$ reads
\be
u(\tau)=u_{\rm bp}+c(\tau_\bp)\,(\tau-\tau_{\bp})^{1/2},
\ee
for some constant $c(\tau_\bp)$.

\begin{figure}[ht]\centering
	\includegraphics[scale=0.7]{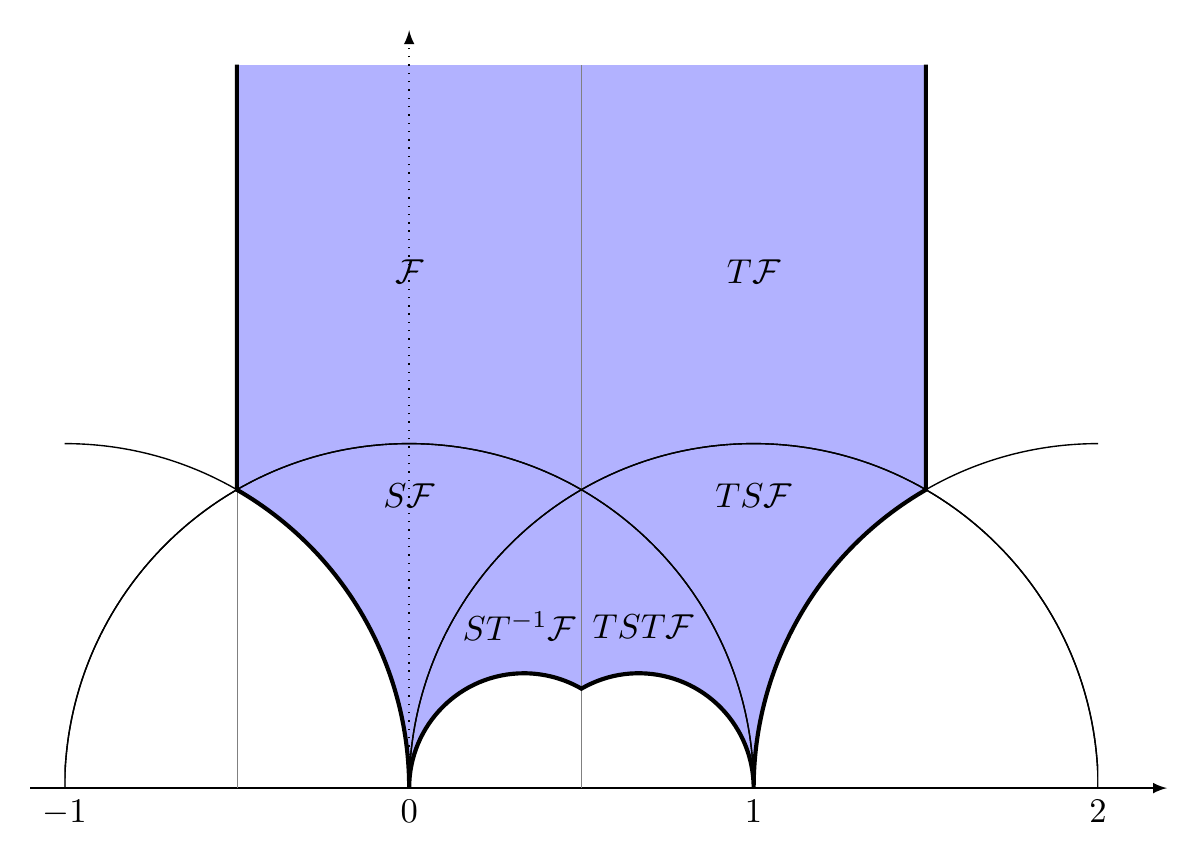}
	\caption{Fundamental domain for $\Gamma(2)$, the duality group of the massless $N_f=2$ theory. }\label{fig:nf2massless} 
\end{figure}

\vspace{.2cm}
\noindent {\it The massless limit in $N_f=2$}\\
In the limit where $m_1=m_2=m\to 0$, one finds from \eqref{umm} that 
\begin{equation}\label{unf2m=0}
	\frac{u}{\Lambda_2^2}=-\frac{1}{8}\frac{\vartheta_3^4+\vartheta_4^4}{\vartheta_2^4}
\end{equation}
This expression is a modular function for the principal congruence subgroup $\Gamma(2)$ of $\psl$. The holomorphy of \eqref{unf2m=0} is explained by the fact that the pair of branch points for $m\neq 0$ collides in the limit $m\to 0$, which annihilates the branch cut. This is because the mutually local singularities $u_\pm\to -\frac{\Lambda_2^2}{8}$ collide, and thus the singularities corresponding to $\tau\to 0$ and $\tau\to 2$ in Fig. \ref{fig:nf2domain} are identified. The fundamental domain of the massless theory is thus the modular fundamental domain for the duality group $\Gamma(2)$, as depicted in  Fig. \ref{fig:nf2massless}.

\begin{figure}[ht]\centering
	\includegraphics[scale=0.7]{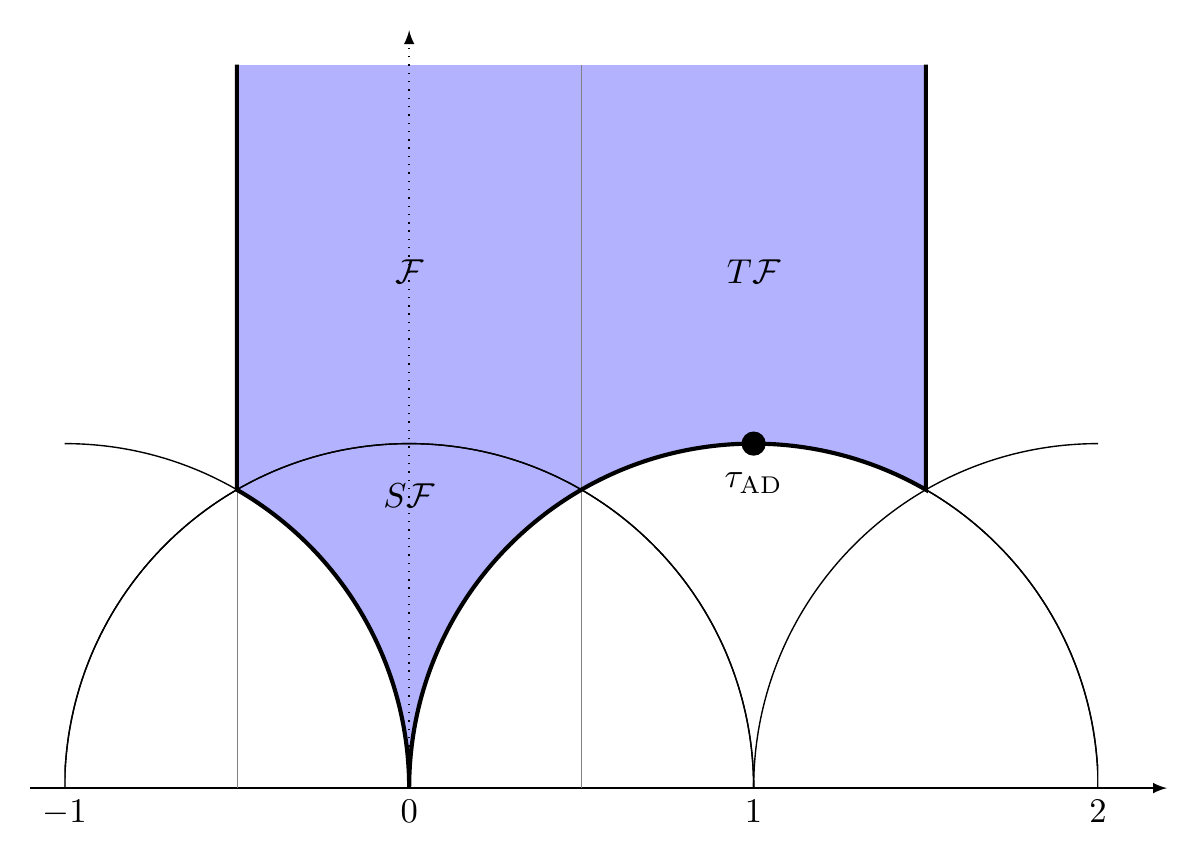}  
	\caption{Fundamental domain of the $N_f=2$ theory with equal AD mass $m_1=m_2=m_{\text{AD}}=\frac{\Lambda_2}{2}$. It is the modular fundamental domain for the congruence subgroup $\Gamma_0(2)$ of $\text{PSL}(2,\mathbb Z)$. The AD point $\tau_{\text{AD}}=1+i$ is the elliptic fixed point of $\Gamma_0(2)$. The three singularities collided in $u_{\text{AD}}$ results in a reduction of the index from 6 to 3.}\label{fig:nf2AD} 
\end{figure}

\vspace{.2cm}
\noindent {\it The AD limit in $N_f=2$}\\
When $m\to\mad=\tfrac{\Lambda_2}{2}$, the singularities $u_*$ and $u_+$ collide and form the superconformal AD point $(A_1,A_3)$ of Kodaira type III. As studied in detail in \cite{Aspman:2021vhs,Closset:2021lhd,Magureanu:2022qym}, this AD point forms an elliptic point of the fundamental domain. 
Since three mutually non-local singularities collide for this mass, the index of the domain $\CF(\mad,\mad)$ is reduced by 3, giving a domain of index $6-3=3$. By taking the limit from \eqref{umm}, one finds that the square root resolves by virtue of the Jacobi identity \eqref{jacobiabstruseidentity}. This restores holomorphy and even modularity. One finds that the order parameter 
\begin{equation}
    \frac{u}{\Lambda_2^2}=-\frac{1}{64}\left(\frac{16\jt_4^8}{\jt_2^4\jt_3^4}+40\right)
\end{equation}
is a modular function for the congruence subgroup $\Gamma_0(2)$ of $\psl$ \cite{Aspman:2021vhs}. Its elliptic fixed point $\tad=1+i$ is precisely the AD point. 
We depict the corresponding fundamental domain  in Fig. \ref{fig:nf2AD}.

\begin{figure}[ht]\centering
	\includegraphics[scale=0.7]{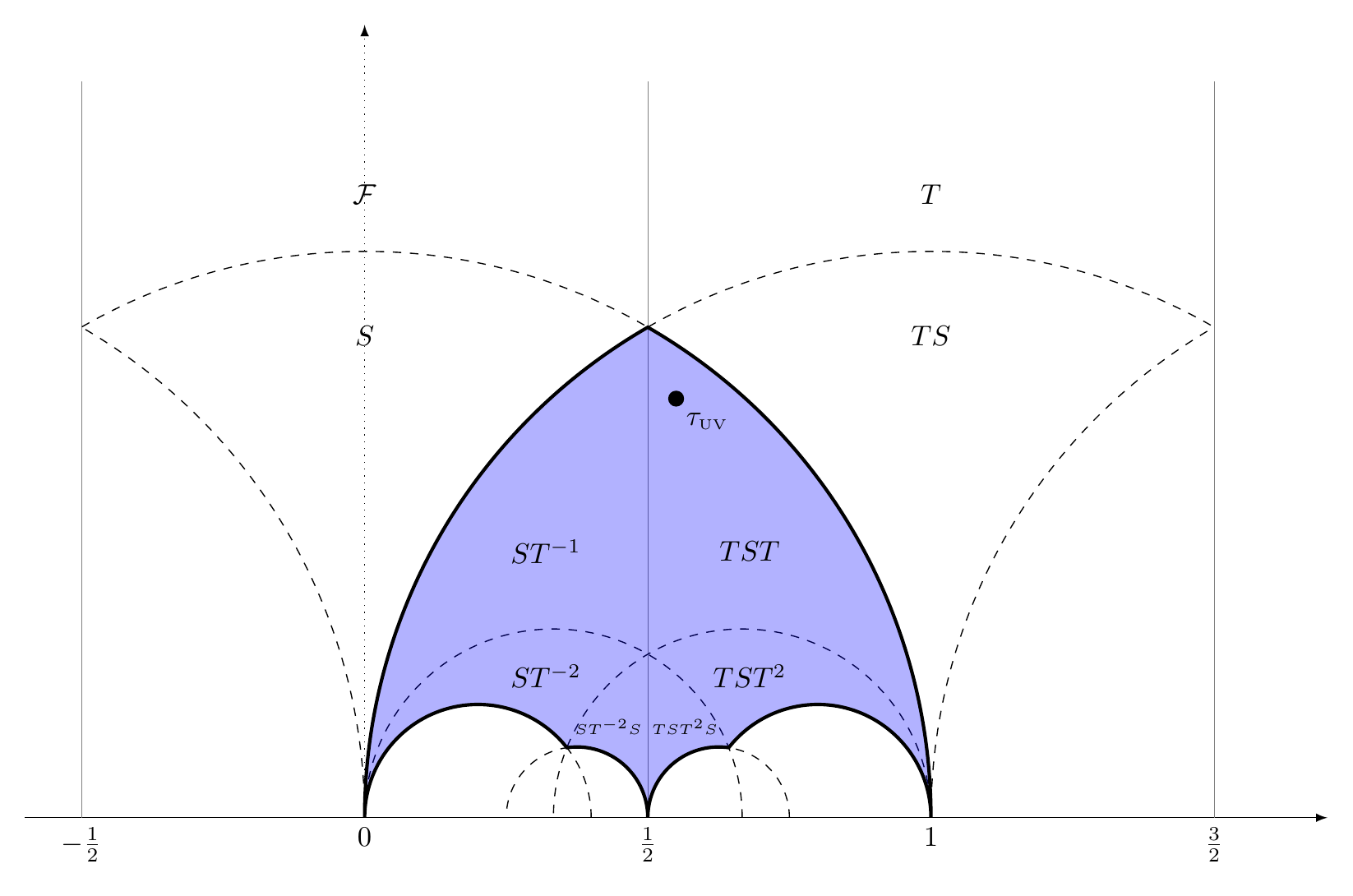}
	\caption{Fundamental domain of the $N_f=4$ theory with mass $\boldsymbol m=(m,m,0,0)$. The six singularities on the Coulomb branch $\mathcal B_4$ are described by the three cusps, each of width $2$. The $I_0^*$ singularity corresponding to $u=\infty$ sits at $\tau=\tau_{\text{\tiny{UV}}}$. }\label{fig:nf4mm00} 
\end{figure}

\vspace{.2cm}
\noindent {\it The $N_f=4$ theory }\\
In $N_f=4$, the fundamental domain for the effective coupling $\tau$ can be found analogously to the asymptotically free cases. One notable new feature is the additional singularity of the theory when $\tau$ approaches the UV coupling $\tuv$. For example, with the mass configuration $\bfm=(m,m,0,0)$ one finds \cite{Aspman:2021evt}
\begin{equation}\label{uASW}
u(\tau,\tuv)=-\frac{m^2}{3}\jt_3(\tuv)^4\frac{\luv^2+2\left(\lau-1\right)\luv-\lau}{\luv-\lau},
\end{equation}
where $\lambda= \frac{\jt_2^4}{\jt_3^4}$. This function is a meromorphic $\Gamma(2)$ modular form of weight $2$ in $\tuv$ for fixed $\tau$, and of weight $0$ in $\tau$ for fixed $\tuv$. Furthermore, it has mixed weight $(0,2)$ under $\psl$ acting on $\tau$ and $\tuv$ simultaneously. As such, it is an example of a \emph{bimodular} form for the triple $(\Gamma(2),\Gamma(2);\psl)$ \cite{Aspman:2021evt, Manschot:2021qqe}.
Since $u(\tau,\tuv)$ has a pole as $\tau\to\tuv$, the fundamental domain $\CF_4(\bfm)$ has a puncture at $\tuv$. This is depicted in Fig. \ref{fig:nf4mm00}. 

\subsection{Monodromies}\label{sec:monodromies}
This section determines the monodromies around the $N_f+2$ monodromies. We leave the winding numbers $n_j$, $j=1,\dots, N_f$, for $a_D$ generic. Starting with the monodromy around infinity, $a\to 
e^{\pi i}a$, we deduce from the \eqref{prepotential} that the vector $\Pi$ transforms as $\Pi \to {\bf M}_\infty\,\Pi$, with ${\bf M}_\infty$ given by
\begin{equation}\label{m_inf}
{\bf M}_\infty=\begin{pmatrix}
			-1& 4-N_f& 0&-n_1&\cdots &0&-n_{N_f}\\
			0&-1&0&0&\cdots &0&0\\
			0&n_1&1&1&\cdots &0&0\\
			0&0&0&1&\cdots &0&0\\
			\vdots &\vdots&&&\ddots&&\\
			0&n_{N_f}&0&0&\cdots &1&1\\
			0&0&0&0&\cdots &0&1
		\end{pmatrix}.
\end{equation}
The monodromy matrix ${\bf M}_\infty$ is in $\text{SL}(2+2N_f,\BZ)$, while it acts on the couplings by a symplectic transformation, i.e. it preserves the symplectic form (\ref{sympForm}). This can be checked by requiring that any monodromy ${\bf M}_\infty$ satisfies ${\bf M}^T {\bf J} {\bf M}={\bf J} $, with
\begin{equation}
    {\bf J}=\begin{pmatrix}0&1\\-1&0\end{pmatrix}^{\oplus N_f+1}.
\end{equation} 
The action on the couplings $\tau$ \eqref{tauF}, $v_j$ and $w_{jk}$ \eqref{Defvw} is thus
\be\label{vw_inf}
{\bf M}_\infty:  \begin{cases}
\tau\to \tau+N_f-4,\\
v_j\to -v_j-n_j,\\
w_{jk}\to w_{jk}+\delta_{jk},
\end{cases}
\ee   
with $\delta_{jk}$ the Kronecker delta.

If we assume that the mass $m_j$ is large, we can also deduce the monodromies around $a=\tfrac{m_j}{\sqrt{2}}$, $j=1,\dots, N_f$ from the perturbative prepotential \eqref{prepotential}. For $a$ encircling $\tfrac{m_1}{\sqrt{2}}$ counterclockwise, $\Pi\to {\bf M}_1 \Pi$, we find for the monodromy matrix ${\bf M}_1$,
\be
{\bf M}_1=\begin{pmatrix}
			1& 1 & 0&-1&\cdots &0& 0\\
			0&1&0&0&\cdots &0&0\\
			0&-1&1& 1&\cdots &0&0\\
			0&0&0&1&\cdots &0&0\\
			\vdots &\vdots&&&\ddots&&\\
			0&0&0&0&\cdots &1&0\\
			0&0&0&0&\cdots &0&1
		\end{pmatrix},
\ee
while the ${\bf M}_j$ for other values of $j$ are given by permutations. Its action on the couplings is
\be\label{vw_Mj}
{\bf M}_j:  \begin{cases}
\tau\to \tau+1,\\
v_k\to v_k-\delta_{jk},\\
w_{kl}\to w_{kl}+\delta_{kl}\delta_{jl}.
\end{cases}
\ee 
To visualise the monodromies with large ${\rm Im}(\tau)$ for $N_f=1$, we refer to Fig. \ref{fig:nf1CutsAlt}(d). In this regime of the masses, there is one monodromy with periodicity 3 and one with periodicity 1.

Besides the monodromies ${\bf M}_\infty$ and ${\bf M}_j$, there are monodromies ${\bf M}_m$ and ${\bf M}_d$ around the points where a monopole and a dyon becomes massless, respectively. By requiring that the electro-magnetic charges of the massless particles are $(n_m,n_e)=(1,0)$ and $(1,-2)$, respectively, we can fix the upper left blocks of the monodromies. We fix the remaining entries by assuming that the masses remains invariant, $m_j\to m_j$, and that the other periods only change by a multiple of the vanishing cycle at the corresponding cusp, together with the requirement that
\be
\label{ModsIdent}
{\bf M}_\infty={\bf M}_m {\bf M}_d \prod_{j=1}^{N_f}{\bf M}_j.
\ee
For $N_f=1$ and $n_1=n$, this gives for ${\bf M}_{m}$, 
\be
{\bf M}_m=\begin{pmatrix}
			1& 0 & 0& 0 \\
			-1 &1&0&-(n+1)/2 \\
			(n+1)/2 &0&1& (n+1)^2/4\\
			0&0&0&1
		\end{pmatrix}.
\ee
This acts on the couplings as
\be\label{Mm_transformation}
{\bf M}_m:
\begin{cases}
\tau\to \frac{\tau}{-\tau+1},\\
v\to \frac{v+(n+1)\tau/2}{-\tau+1},\\
w\to w+\frac{(v+(n+1)/2)^2}{-\tau+1}.
\end{cases}
\ee
The monodromy ${\bf M}_{d}$ around the dyon singularity for $N_f=1$ is
\be
{\bf M}_d=\begin{pmatrix}
			-1& 4 & 0& -n-1 \\
			-1 &3&0&-(n+1)/2 \\
			(n+1)/2&-n-1&1& (n+1)^2/4\\
			0&0&0&1
		\end{pmatrix},
\ee
This acts on the couplings as
\be
{\bf M}_d:
\begin{cases}
\tau\to \frac{-\tau+4}{-\tau+3},\\
v\to \frac{v+(n+1)\tau-n-1}{-\tau+3},\\
w\to w+\frac{(v+(n+1)/2)^2}{-\tau+3}.
\end{cases}
\ee
We can note that all the above monodromy matrices leave the symplectic form \eqref{sympForm} invariant and are independent of the masses.

For a small mass $m$, the
fourth ``hypermultiplet'' cusp of the fundamental domain for $N_f=1$ lies naturally near the real axis, $\tau\to 1$. See for example  Fig.
\ref{fig:nf1CutsAlt}(a). Having determined ${\bf M}_\infty$,
${\bf M}_m$ and ${\bf M}_d$, we can easily determine the monodromy
${\bf \tilde M}_1$ in this regime as
\be
{\bf \tilde M}_1={\bf  M}_m^{-1} {\bf M}_\infty {\bf  M}_d^{-1}=\begin{pmatrix}
			0 & 1 & 0 & (1-n)/2 \\
			-1 & 2 &0& (1-n)/2  \\
			(n-1)/2& (1-n)/2 &1& (n-1)^2/4\\
			0&0&0&1
		\end{pmatrix}
\ee
Thus for $n=-1$, the massless particle has charge $\pm (-1,1,0,1)$. 

We get similar monodromies for $N_f=2,3$. The action on the running couplings $\tau$ are the same for all $N_f$, by construction. The transformations of $v_j$ and $w_{jk}$ also take the same form for all $N_f$ and can be summarised as
\begin{equation}
    \begin{aligned}
    &{\bf M}_m: \begin{cases}v_j&\to \frac{v_j+(n_j+1)\tau/2}{-\tau+1},\\ 
        w_{jk}& \to w_{jk}+\frac{(v_j+(n_j+1)/2)(v_k+(n_k+1)/2)}{-\tau+1},\end{cases}\\
    &{\bf M}_d: \begin{cases}
        v_j&\to \frac{v_j+(n_j+1)\tau-n_j-1}{-\tau+3},\\ 
        w_{jk}& \to w_{jk}+\frac{(v_j+(n_j+1)/2)(v_k+(n_k+1)/2)}{-\tau+3}.\end{cases}
    \end{aligned}
\end{equation}

\section{The UV theory on a four-manifold}\label{sec:UVtheory}
We review various aspects of the formulation of the UV theory on a compact smooth four-manifold.

\subsection{Aspects of four-manifolds}\label{4manifolds}
We let $X$ be a smooth, compact, oriented Riemannian four-manifold, with Euler number $\chi=\chi(X)$ and signature $\sigma=\sigma(X)=b_2^+-b_2^{-}$. The $u$-plane integral is non-vanishing only for  four-manifolds $X$ with $b_2^+\leq 1$. In this article, we consider manifolds with $b_2^+=1$. Such four-manifolds admit a linear complex structure $\CJ$ on the tangent space $TX_p$ at each point $p$ of $X$. The complex structure varies smoothly on $X$, such that $TX$ is a complex bundle. We introduce furthermore the canonical class $K_X=-c_1(TX)$ of $X$, with $c_1(TX)$  the first Chern class of $TX$. For a manifold $X$ with $(b_1,b_2^+)=(0,1)$, we have that 
\begin{equation}\label{ksigma}
	K_X^2=8+\sigma(X).
\end{equation}

The middle cohomology $H^2(X,\mathbb{Z})$ of $X$ gives rise to the uni-modular lattice $L$. More precisely, we identify $L$ with the natural embedding of $H^2(X,\mathbb{Z})$ in $H^2(X,\mathbb{Z})\otimes \mathbb{R}$, which mods out the torsion of $H^2(X,\mathbb{Z})$. A characteristic element $K\in L$ is an element which satisfies $\bfl^2+ B(K,\bfl)\in 2\mathbb Z$ for all $\bfl \in L$. The Riemann-Roch theorem demonstrates that the canonical class $K_X$ of $X$ is a characteristic element of $L$. The Wu formula furthermore shows that any characteristic vector $K$ of $L$ is a lift of $w_2(X)$.

 The quadratic form $Q$ of the lattice $L$ for a 4-manifold with $(b_1,b_2^+)=(0,1)$ can be brought to a simple standard form depending on whether $Q$ is even or odd \cite{Donaldson90}. This divides such manifolds into two classes, for which the evaluation of their $u$-plane integrals needs to be done separately \cite{Korpas:2019cwg,Aspman:2023ate}. 
 The period point $J\in H^2(X,\mathbb R)$ is defined as the unique class in the forward light cone of $H^2(X,\mathbb R)$ that satisfies $J=*J$ and $J^2=1$.

All four-manifolds without torsion and even intersection form admit a Spin structure. 
More generally, for any oriented four-manifold one can define a $\spinc$-structure. The group $\spinc(4)$ can be defined as pairs of unitary $2\times 2$ matrices with coinciding determinant, 
\begin{equation}
	\spinc(4)= \{ (u_1,u_2)\in \mathrm{U}(2)\times \mathrm{U}(2)|\det {u_1}=\det u_2\}.
\end{equation}
There exists a short exact sequence   
\begin{equation}
1 \longrightarrow	U(1)\longrightarrow \spinc(4)\longrightarrow SO(4) \longrightarrow 1.
\end{equation}
A $\spinc$-structure $\mathfrak{s}$  on a four-manifold $X$ is then a reduction of the structure group of the tangent bundle on $X$, i.e. SO(4), to the group $\spinc(4)$. The different $\spinc$-structures correspond to the inequivalent ways of choosing transition functions of the tangent bundle such that the cocycle condition is satisfied. The $\spinc$-structure defines two rank two hermitian vector bundles $W^\pm$. We let $c(\mathfrak s)$ be the first Chern class of the determinant bundles, $c(\mathfrak s)\coloneqq c_1(\det W^\pm)\in H^2(X,\BZ) $.

If $\mathfrak s$ is the canonical $\spinc$ structure associated to an almost complex structure on $X$, then $c(\mathfrak s)^2=2\chi+3\sigma$.
More generally, 
\begin{equation}\label{spinCsignature}
	c_1(\mathfrak s)^2\equiv \sigma \mod 8.
\end{equation}

\subsection{Topological twisting with background fluxes}
We discuss in this section topological twisting of theories with fundamental hypermultiplets including background fluxes. The discussion is parallel to the case of $\CN=2^*$ \cite{Manschot:2021qqe}, where the hypermultiplet is in the adjoint representation of the gauge group.

We let $(E\to X,\nabla)$ be a principal $\text{SU}(2)/\BZ_2\cong \text{SO}(3)$-bundle with connection $\nabla$. The second Stiefel-Whitney class $w_2(E)\in H^2(X,\mathbb{Z}_2)$ measures the obstruction to lift $E$ to an SU(2) bundle, which will exist locally but not globally if $w_2(E)\neq 0$. We denote a lift of $w_2(E)$ to the middle cohomology lattice $L$ by $\bar w_2(E)\in L$, and define the 't Hooft flux $\bfmu=\bar w_2(E)/2\in L/2$. The instanton number of the principal bundle is defined as $k=-\frac14\int_X p_1(E)$  and satisfies $k\in -\bfmu^2 + \mathbb{Z}$, where $p_1$ is the first Pontryagin class.

To formulate the theories with $N_f$ fundamental hypermultiplets on a compact four-manifold, we perform a topological twist. Coupling the four-dimensional $\CN=2$ $\text{SU}(2)$ theory to background fields means choosing two sets of data:
\begin{itemize}
    \item A principal $\text{SU}(2)_R$ R-symmetry bundle, with connection $\nabla_R$,
    \item and a principal bundle $\CL$ with connection for global symmetries (the flavour symmetries) \cite{Manschot:2021qqe}.
\end{itemize}
The relevant twist for the $\CN=2$ supersymmetry algebra in four dimensions is the Donaldson-Witten twist. This twist is the local identification of the SU(2)${}_+$ with the diagonal subgroup of the SU(2)${}_+\times$SU(2)${}_R$ factor of the spin lift of the local spin group $\text{Spin}(4)\cong \text{SU}(2)_+\times \text{SU}(2)_-$ \cite{Witten:1988ze}. 
Alternatively, one can view the fields as sections of a non-trivial R-symmetry bundle, isomorphic to the spin bundle  $S^+$. Application of this to the representations of the vector multiplet (\ref{repvecbos}) and (\ref{rephypferm}) gives:
\be 
\begin{aligned}
&\text{bosons:}\quad &&({\bf 2}, {\bf 2})\oplus ({\bf 1}, {\bf 1}) \oplus ({\bf 1}, {\bf 1}),  \\
&\text{fermions:}\quad &&({\bf 2},{\bf 2})\oplus ({\bf 3}, {\bf 1})\oplus ({\bf 1},{\bf 1}).
\end{aligned}
\ee 
Thus the bosons remain unchanged, a vector and a complex scalar, while the fermions reorganise to a vector, self-dual two-form and real scalar, which we denote as $\psi$, $\chi$ and $\eta$, respectively. We note that none of these fields are spinors, and can thus be considered on a non-spin four-manifold. The  original supersymmetry generators also transform in the representations for the fermions above. Thus the theory contains a scalar fermionic supercharge $\CQ=\epsilon^{\dot A \dot B}\overline{\CQ}_{\dot A\dot B}$, whose cohomology provides the operators in the topological theory \cite{Witten:1988ze}.

For the fields of a hypermultiplet, (\ref{rephypbos}) and (\ref{rephypferm}), one finds
\be 
\begin{aligned}
&\text{bosons:}\quad &&({\bf 1}, {\bf 2})\oplus ({\bf 1}, {\bf 2}),   \\
&\text{fermions:}\quad &&({\bf 2}, {\bf 1})\oplus ({\bf 1},{\bf 2})\oplus ({\bf 2}, {\bf 1})\oplus ({\bf 1}, {\bf 2}).
\end{aligned}
\ee 
Thus hypermultiplet bosons become spinors, i.e. sections of the spin bundle $S^+$, while the fermions are sections of $S^+$ and $S^-$. Thus the twisted hypermultiplets can in this case only be formulated on four-manifolds which are spin, i.e. $w_2(X)=0$ \cite{Hyun:1995mb, Moore:1997pc}. 

However, if the hypermultiplets are charged under a gauge field or flux, the product of these bundles with $S^\pm$ may be a Spin$^c$ bundle, $W^+$ or $W^-$ \cite{Hyun:1995mb,Laba05, Manschot:2021qqe}. The latter are defined for arbitrary four-manifolds. For example, an almost complex structure on $X$ determines two canonical Spin$^c$ bundles $W^\pm \simeq S^\pm\otimes K^{-1/2}_X$ with $K_X$ the canonical class determined by the almost complex structure. Since the hypermultiplets are in the fundamental, two-dimensional representation of SU(2), the topologically twisted hypermultiplets are well-defined on a non-spin four-manifold if $\bfmu=-K_X/2$ \cite{Moore:1997pc}.

Let us state this also in terms of the gauge bundle $E$. To this end, we label the two components of the fundamental, two-dimensional representation of SU(2) by $\pm$. The two components are sections of a line bundle $\CL_E^{\pm 1/2}$ with $c_1(\CL_E)=\bar w_2(E)$. Of course, the square root $\CL_E^{1/2}$ only exists if $w_2(E)\in 2L$. On the other hand, the physical requirement is that $S^+\otimes \CL_E^{1/2}$ is well defined, or $\bar w_2(X)+\bar w_2(E)\in 2L$. Therefore, the obstructions can cancel each other for a suitable choice of $w_2(E)$. Thus the topological twisted theory is not well-defined for an arbitrary choice of 't Hooft flux $\bfmu \coloneqq \frac12 \bar w_2(E)$; but rather  $\bfmu$ has to satisfy $\bfmu=\frac12 \bar w_2(X) \mod L$ \cite{Moore:1997pc}, or
\be 
\label{w2Xw2E}
\bar w_2(X)=\bar w_2(E) \mod 2L.
\ee 

To consider more general 't Hooft fluxes $\bfmu$ or equivalently $w_2(E)$, we can couple the $j$'th hypermultiplet to a background flux or line bundle $\CL_j$, with  $\CL_j$ possibly different
for each $j$. We let $\CE_j=\CL_E\otimes \CL_j$.
Then the requirement that $S^\pm\otimes \CE_j^{\pm1/2}$ is globally well-defined is that
\be
c_1(\CE_j)\in \bar w_2(X) +2L,
\ee
which can be satisfied for any $\bar w_2(E)$ for a suitable choice of $\CL_j$. Thus we can formulate the $u$-plane integral for arbitrary $\bar w_2(E)$, if we require that the background fluxes satisfy
\be \label{constraint_c1CL}
c_1(\CL_j)\in \bar w_2(X)+\bar w_2(E) +2L,
\ee 
for each $j$. This is consistent with (\ref{w2Xw2E}) for $c_1(\CL_j)=0$.

The Chern classes $c_1(\CL_j)$ can also be seen as the splitting classes of the $\spin(2N_f)$ principal bundle $\CL$. The Chern class of $\CL$ reads
\be 
c(\CL)=\sum_{l=0}^2 c_l(\CL)=\prod_{j=1}^{N_f} (1+c_1(\CL_j)).
\ee 
The scalar generators of the equivariant cohomology of $\spin(2N_f)$ are the masses $m_j$, which generate the $N_f$-dimensional Cartan subalgebra of $\spin(2N_f)$. The gauge bundle $E_k$ is also $\spin(2N_f)$ equivariant. For generic masses, the flavour group is U(1)$^{N_f}$, and is enhanced for special loci of the masses, for example to $\text{U}(N_f)$ for equal masses \cite{Seiberg:1994aj}.

The $\CQ$-fixed equations are the non-Abelian monopole equations with $N_f$ matter fields in the fundamental representation. For generic gauge group $G$ and with representation $R$, these equations read \cite{Labastida:1995zj}
\be 
\label{Qfixedeqs}
\begin{split}
&\left(F^{a}_{\dot \alpha \dot \beta}\right)^++\frac{i}{2} \sum_{j=1}^{N_f}\bar M^j_{(\dot \alpha} T^a M^j_{\dot \beta)}=0,\\
&\slashed DM^j=\sum_\mu \sigma^\mu\, D_\mu M^j=0,
\end{split}
\ee 
where $T^a$ is a generator of the Lie algebra in the representation $R$. Including the sum over matrix elements, we have
\be
M^j_{(\alpha} T^a M^j_{\beta)}=\sum_{k,l}(M^j)^k_{(\alpha} (T^a)^{kl} (M^j)^l_{\beta)}.
\ee
We denote the moduli space of solutions to \eqref{Qfixedeqs} by $\CM^{Q,N_f}_{k,\CL_j}$, and leave the dependence on the 't Hooft flux $\bfmu$ and the metric $J$ implicit. For $N_f=4$ on $X=\mathbb{CP}^2$, such moduli spaces are studied in \cite{Gorodentsev:1996cu}. 

The moduli spaces $\CM^{Q,N_f}_{k,\CL_j}$ is non-compact for vanishing masses \cite{bryan1996,Moore:1997dj, Dedushenko:2017tdw}. This is improved upon turning on masses and  localizing with respect to the $\text{U}(1)^{N_f}$ flavour symmetry,  $M^j_\alpha\to e^{i\,\varphi_j} M^j_{\alpha}$, which leave invariant the $\CQ$-fixed equations \eqref{Qfixedeqs}. 
 There are two components:
\begin{itemize}
    \item the instanton component, with $F^+=0$ and $M^j=0$, $j=1,\dots,N_f$. The moduli space for this component is denoted $\CM^{\rm i}_k$. Since the hypermultiplet fields vanish, this component is associated to the Coulomb branch.
\item the abelian or monopole component, for which a U$(1)$ subgroup of the flavour group acts as pure gauge. Here the connection is reducible, and a U(1) subgroup of the  SU(2) gauge group is preserved. For generic masses, there are $N_f$  such components, where $M^\ell$ is upper or lower triangular for some $\ell$, and $M^j=0$ for all $j\neq \ell$. The moduli space of this component is denoted $\CM^{{\rm a},j}_k$, $j=1,\dots, N_f$.  Since some of the hypermultiplet fields are non-vanishing, this component is associated to the Higgs branch \cite{Moore:1997dj,Marino:1998uy}.
\end{itemize}

The instanton component $\CM^{\rm i}_k$ is non-compact due to point-like instantons. This can be cured using the Uhlenbeck compactification or algebraic-geometric compactifications. We assume that the physical path integral chooses a specific compactification, whose details are however not manifest at the level of the low energy effective field theory other than that the compactification must be in agreement with the correlation functions. 

The topological twist for $\CN=2$ supersymmetric QCD can be further made dependent on a cocycle $\zeta_{\alpha\beta\gamma}^{\text{gauge}}$ representing the 't Hooft flux, and $\zeta_{\alpha\beta\gamma}^s$ a cocycle representing $w_2(X)$ (the cocycles are the U(1) valued functions measuring the obstruction of the cocycle condition for transition functions) \cite{Manschot:2021qqe}. Without additional line bundles, $\bar w_2(X)=\bar w_2(E)$ is equivalent to the cocycle $\zeta_{\alpha\beta\gamma}^{\text{gauge}}\zeta_{\alpha\beta\gamma}^s$ being trivialisable. We leave it for future work to explore whether the invariants depend on the choice of trivialisation.

\subsection{Correlation functions and moduli spaces}
\label{corrfunctions}
 The $\CQ$-fixed equations (\ref{Qfixedeqs}) include a Dirac equation for each hypermultiplet $j=1,\dots,N_f$ in the fundamental representation. The corresponding index bundle $W_k^j$ defines an element of the $K$-group of $\CM^{\rm i}_k$. 
  Its virtual rank $\rk(W^j_k)$ is the formal difference of two infinite dimensions. It is given by an index theorem and reads
\begin{equation}\label{rkW_k^j}
	\rk(W^j_k)=-k+\frac{1}{4}(c_1(\CL_j)^2-\sigma)\in \mathbb{Z},
\end{equation}
where $c_1(\CL_j)$ is the first Chern class of the bundle $\CL_j$.
Note that the rhs is {\it not} an integer for an arbitrary $c_1(\CL_j)\in H^2(X,\mathbb{Z})$. To verify that the rhs is integral for the $c_1(\CL_j)$'s satisfying (\ref{constraint_c1CL}), we rewrite $\rk(W^j_k)$ as
\be 
\rk(W^j_k)=-(k+\bfmu^2)-c_1(\CL_j)\cdot \bfmu+\frac{1}{4}\left((c_1(\CL_j)+2\bfmu)^2-\sigma\right).
\ee 
Then the first term on the rhs is an integer since $k\in-\frac{1}{4}w_2(E)^2+\mathbb{Z}$ for an $\text{SO}(3)$ bundle. The second term is an integer because $c_1(\CL_j)\cdot \bfmu=(\bar w_2(X)-2\bfmu)\cdot \bfmu \mod \mathbb{Z} \in \mathbb{Z}$, and the third term is an integer using (\ref{spinCsignature}) and the fact that $c_1(\CL_j)+2\bfmu$ equals the characteristic class of a $\spinc$-structure $\mathfrak s_j$ by (\ref{constraint_c1CL}),
\begin{equation}\label{k+mu=spinC}
    c_1(\CL_j)+2\bfmu=c(\mathfrak s_j),
\end{equation}
for each $j$.

The mass $m_j$ is the equivariant parameter of the U(1) flavour symmetry associated to the $j$'th hypermultiplet. 
The equivariant Chern class of $W^j_k$ reads in terms of the splitting class $x_l$,
\be 
c(W_k^j)=\prod_{l=0}^{-\rk(W^j_k)} (x_l+m_j)=m_j^{-\rk(W^j_k)} \sum_l \frac{c_l(W_k^j)}{m_j^l}.
\ee 
We abbreviate $c_l(W_k^j)$ to $c_{l,j}$, and let $c(W_k)=\prod_{j=1}^{N_f} c(W_k^j)$. 
 
The moduli space $\CM_{k,\bfmu,\CL_j}^{Q}$ for $N_f$ hypermultiplets corresponds to the vanishing locus of the obstructions for the existence of $N_f$ zero modes of the Dirac operator. As a result,  the virtual complex dimension of the moduli space $\CM_{k,\bfmu,\CL_j}^{Q}$ is that of the instanton moduli space plus the sum of (typically negative) ranks of the index bundles $W_k^j$,
$\vdim(\CM^{Q,N_f}_{k,\CL_j})=\vdim(\CM^{Q}_k)_{N_f=0}+\sum_{j=1}^{N_f}\rk(W_k^j)$ \cite{Hyun:1995hz, bryan1996, Feehan:2001jc, Manschot:2021qqe}.
This gives
\begin{equation}
\label{vdimML}
	\vdim(\CM_{k,\CL_j}^{Q,N_f})=(4-N_f)k+\frac14\left(-3\chi-(3+N_f)\sigma+\sum_{j=1}^{N_f} c_1(\CL_j)^2\right).
\end{equation}

It is argued in \cite{LoNeSha} that the inclusion of massive matter  amounts to inserting an integral of the equivariant Euler class of the Dirac index bundle over the moduli space. Therefore, the correlation functions are the generating functions for the intersection numbers of the standard Donaldson observables and the Poincar\'e duals to the Chern classes of the various vector bundles.

The correlation functions on $X$ in the theory with $N_f$ massive fundamental hypermultiplets are conjectured to be 
\begin{equation}
	\langle \CO_1\dots \CO_p\rangle=\sum_k \Lambda_{N_f}^{{\rm vdim}(\CM^{Q,N_f}_{k,\CL_j})} \, \int_{\CM_{k,\CL_j}^{Q,N_f}} c(W_k)\, \omega_1\wedge\dots\wedge\omega_p,
\end{equation}
where $\omega_i=\mu(\CO_i)$ are the Donaldson classes associated to the physical observable $\CO_i$, and $c(M)$ is the Euler class of the matter bundle \cite{Mathai:1986tc, Atiyah:1990tm, losev1998, Cordes:1994fc}. Localising to the fixed point locus in $\CM^{Q,N_f}_{k,\CL_j}$ with respect to $U(1)^{N_f}$ gives 
\begin{equation}
	\label{UVpartfunc}
	\begin{split}
	&\langle \CO_1\dots \CO_p\rangle=\sum_k  \Lambda_{N_f}^{{\rm vdim}(\CM^{Q,N_f}_{k,\CL_j})}\\  &\qquad \qquad \times \int_{\CM^{\rm i}_k \cup \CM^{\rm a}_k} \left(\prod_{j=1}^{N_f}  m_j^{-\rk( W^j_k)}\sum_l \frac{c_{l,j}}{m_j^l}\right)\, \omega_1\wedge\dots\wedge\omega_p.
	\end{split}
\end{equation}
where the integral is over the union $\CM^{\rm i}_k \cup \CM^{\rm a}_k$ of the instanton component $\CM^{\rm i}_k$ \cite[Eq. (5.13)]{LoNeSha} and the monopole component $\CM^{\rm a}_k$ \cite{Manschot:2021qqe, Vafa:1994tf}. The equation together with the dimension of the moduli spaces (\ref{vdimML}) demonstrate a selection rule for observables together with powers of $\Lambda_{N_f}$ and $m_j$.

In the decoupling limit $m_{N_f}\to \infty$, $\Lambda_{N_f}\to 0$ (\ref{decouplimit}), the only contribution for $j=N_f$ is from $l=0$, $c_{0,N_f}=1$.  The powers of $m_j$ and $\Lambda_{N_f}$ work out such that the correlation functions reduce to those of the theory with $N_f-1$ hypermultiplets \cite{Marino:1998uy}
\be \label{UV-decoupling}
\left(\frac{\Lambda_{N_f}}{\Lambda_{N_f-1}}\right)^{-\alpha} \langle \CO_1\dots \CO_p\rangle_{N_f} \to  \langle \CO_1\dots \CO_p\rangle_{N_f-1},
\ee 
We deduce from (\ref{UVpartfunc}) that
\be 
\label{alphadef}
\begin{split}
\alpha&={\rm vdim}(\CM^{Q,N_f}_k)+(4-N_f)\, \rk(W_k^{N_f})\\
&=\frac{1}{4}\Big(-3\chi-7\sigma+(5-N_f)\,c_1(\CL_{N_f})^2+\sum_{j=1}^{N_f-1}c_1(\CL_j)^2 \Big).
\end{split}
\ee 
The overall factor can be accounted for by an overall renormalization in the decoupling limit.

The correlation function has a smooth massless limit $m_j\to 0$, for which only terms with the top Chern classes contribute. These are given by
\begin{equation}
	\label{UVpartfuncmassless} 
	\langle \CO_1\dots \CO_p\rangle=\sum_k \prod_{j=1}^{N_f}   \Lambda_{N_f}^{{\rm vdim}(\CM^{Q,N_f}_{k,\CL_j})}  \int_{\CM^{\rm i}_k \cup \CM^{\rm a}_k} c_{l_j,j}\, \omega_1\wedge\dots\wedge\omega_p 
\ee 
with $l_j=-{\rm rk}(W^j_k)$ for each $c_{l,j}$. For a non-vanishing result, the degree of $\omega_1\dots \omega_p$ must equal ${\rm vdim}(\CM^{Q,N_f}_{k,\CL_j})$. 

By comparing the form degrees (or ghost numbers) of  operators  with the virtual dimension of the moduli space, one derives selection rules for correlation functions of point and surface operators. 
Since $u$ has ghost number 4, the fugacity $p$ naturally has ghost number $-4$. Similarly, we associate ghost number $-2$ to the surface $\bfx$. The mass $m_j$ finally has ghost number $2$. Thus correlation functions will evaluate to sums of monomials of the form $p^s\bfx^t\prod_{j=1}^{N_f}m_j^{r_j}$, with the selection rule
\begin{equation}
\label{selrule}
    -4s-2t+2\sum_{j=1}^{N_f}r_j=-\vdim_{\mathbb R}(\CM_{k,\CL_j}^{Q,N_f}).
\end{equation}

\section{The effective theory on a four-manifold}\label{sec:eff_theories}
We consider in this Section the low energy effective field theory on a four-manifold. We derive the semi-classical action of the theory coupled to background $\text{U}(1)$ fields. As in previous cases \cite{Witten:1995gf, Shapere:2008zf, Manschot:2021qqe,Moore:1997pc}, the final expression takes the form of a Siegel-Narain theta series multiplied by a measure factor.

\subsection{Hypermultiplets and background fields}
The effective theory coupled to $N_f$ background fluxes can be modelled as that of a theory with gauge group $\text{SU}(2)\times \text{U}(1)^{N_f}$, where the fields of the U(1) factors have been frozen in a special way \cite{Nelson:1993nf, Manschot:2021qqe}. To derive the precise form, we recall the low-energy effective Lagrangian for the $r$ multiplets $(\phi^J,\eta^J,\chi^J,\psi^J,F^J)$ of the topologically twisted $\text{U}(1)^r$ SYM theory \cite{Marino:1998bm}. Since the $u$-plane integral reduces to an integral over zero-modes \cite{Moore:1997pc}, it suffices to only include the zero-modes in the Lagrangian. For simply connected four-manifolds, there is no contribution from the one-form fields $\psi^J$. The Lagrangian is then given in terms of the prepotential $F(\{a^J\})$ and its derivatives to the vevs $\langle \phi^J \rangle=a^J$, as
\begin{equation}
\begin{split} 
	\CL&= \frac{\im}{16\pi}(\bar\tau_{JK}F^J_+\wedge F_+^K+\tau_{JK}F^J_-\wedge F^K_-)-\frac{1}{8\pi}y_{JK}D^J\wedge D^K\\ 
	&\qquad +\frac{\im\sqrt{2}}{16\pi}\bar\CF_{JKL}\eta^J\chi^K\wedge(D+F_+)^L,
\end{split}
\end{equation}
with $y_{JK}=\text{Im}(\tau_{JK})$, $\tau_{JK}=\partial_J\partial_K F(\{a^J\})$ and $\CF_{JKL}=\partial_J\partial_K\partial_L F(\{a^J\})$. It is left invariant by the BRST operator $Q$, which acts on the zero modes as
\begin{equation}
	\begin{alignedat}{2}
		[Q,A^J]&=\psi^J=0,\qquad &&[Q,\psi^J]=4\sqrt{2}da^J,\\
		[Q,a^J]&=0,\qquad &&[Q,\bar a^J]=\sqrt{2}\im\eta^J,\\
		[Q,\eta^J]&=0,\qquad &&[Q,\chi^J]=\im(F_+-D_+)^J,\\
		[Q,D^J]&=(d\psi^J)_+=0.
	\end{alignedat}
\end{equation}
Using this operator, we can write $\CL$ as the sum of a topological, holomorphic term and a $Q$-exact term,
\begin{equation}
	\CL=\frac{\im}{16\pi}\tau_{JK}F^J\wedge F^K+\{Q,W\},
\end{equation}
with
\begin{equation}
	W=-\frac{\im}{8\pi}y_{JK}\chi^J(F_++D)^K.
\end{equation}

The low-energy theory of $\text{SU}(2)$ gauge theory with $N_f$ hypermultiplets coupled to $N_f$ background fluxes can then be modelled by the above  rank $r$ description with $r=N_f+1$. We identify $F(\{a^J\})$ with $F(a,\bfm)$. We let the indices $J,K$ run from 0 to $N_f$ and identify the index 0 with the unbroken U(1) of the SU(2) gauge group and the indices $j,k,l=1,\dots, N_f$ with that of the frozen $\text{U}(1)^{N_f}$ factors. We further set $\phi^0\coloneqq \phi$ for any field $\phi$. We will proceed by using lower indices for $j,k,l$, except where the summation convention is explicitly used, to avoid confusion with powers of the fields. 

The masses of the hypermultiplets  are the vevs of the frozen scalar fields of the corresponding vector multiplets, $\tfrac{m_{j}}{\sqrt{2}}=\langle\phi_{j}\rangle=a_{j}$ \cite{Nelson:1993nf}. We set $[F_{j}]=4\pi \bfk_{j}$ with
\begin{equation}
    \bfk_{j}=c_1(\CL_j)/2\in L/2.
\end{equation}
To make the BRST variations of the fields from the frozen U(1) factors vanish, we set $\eta_j=\chi_j=0$, as well as $D^j=F_+^j$. With these identifications, the Lagrangian becomes 
\begin{equation}
	\begin{aligned}
		\CL=\,  &\frac{\im}{16\pi}\tau_{JK}F^J\wedge F^K+\frac{1}{8\pi}y_{00}F_+\wedge F_+-\frac{1}{8\pi}y_{00}D\wedge D\\
		&+\frac{\im\sqrt{2}}{16\pi}\bar\CF_{000}\eta\chi\wedge(D+F_+)+\frac{\im\sqrt{2}}{8\pi}\bar\CF_{00j}\eta\chi\wedge F_+^j\\
		&+\frac{1}{4\pi}y_{0j}(F_+-D)\wedge F_+^j.
	\end{aligned}
\end{equation}
Integrating over $D$, $\eta$ and $\chi$ in the standard way \cite{Marino:1998bm,Moore:1997pc,  Manschot:2021qqe}, we end up with 
\begin{equation}\label{totDeriv}
	\begin{aligned}
		&\int dD d\eta d\chi\, e^{-\int_X \CL} \\
		&=\frac{\partial}{\partial\bar a}\left(\im\sqrt{y_{00}}\,B\left(F+\frac{y_{0j}}{y_{00}}F^j,J\right)\right)e^{-\int_X\CL_0},
	\end{aligned}
\end{equation}
where
\begin{equation}
\begin{aligned}
	\CL_0&=\frac{\im}{16\pi}\tau_{JK}F^J\wedge F^K+\frac{1}{8\pi}y_{00}F_+\wedge F_++\frac{y_{0j}}{4\pi}F_+\wedge F_+^j+\frac{1}{8\pi}\frac{y_{0j}y_{0k}}{y_{00}}F^j_+\wedge F^k_+\\
	&= \frac{\im}{16\pi} (\bar\tau F_+\wedge F_++\tau F_-\wedge F_-)+\frac{\im}{8\pi}(v_j F_-\wedge F_-^j+\bar v_jF_+\wedge F^j_+)\\
	&\quad +\frac{\im}{16\pi} w_{jk}F^j\wedge F^k+\frac{ y}{8\pi}\text{Im}(v_j)\text{Im}(v_k)F_+^j\wedge F_+^k,
\end{aligned}
\end{equation}
and we identified $\tau\coloneqq \tau_{00}$, $y={\rm Im}(\tau)=y_{00}$, $v_j\coloneqq \tau_{0j}$ and $w_{jk}\coloneqq \tau_{jk}$. Thus the coupling $w_{jk}$ is holomorphic, but the coupling $v_j$ is {\it non-holomorphic}. This is similar to the couplings for $\CN=2^*$ \cite{Manschot:2021qqe}.

\subsection{Sum over fluxes}\label{sec:sumflux}
The path integral includes a sum over fluxes $\bfk=[F]/4\pi\in L/2$. After summing the exponentiated action \eqref{totDeriv} over the fluxes $\bfk$ and multiplying by $\frac{d\bar a }{d\bar \tau}$, we find that this takes the form
\be 
\sum_{\bfk\in L+\bfmu}\int dD d\eta d\chi\, e^{-\int_X \CL}=\left(\prod_{j,k=1}^{N_f} C_{jk}^{B(\bfk_j,\bfk_k)}\right)\,\Psi_{\bfmu}^J(\tau,\bar \tau,\bfz,\bar \bfz).
\ee
The couplings $C_{jk}$ are given in terms of $w_{jk}$ \eqref{Defvw} by 
\begin{equation}\label{cij}
	C_{jk}=e^{-\pi\im w_{jk}},
\end{equation}
for $j,k=1,\dots, N_f$. Such couplings were first put forward in \cite{Shapere:2008zf}, and were also crucial in \cite{Manschot:2021qqe}.

The term $\Psi_{\bfmu}^J$ is an example of a Siegel-Narain theta function. It reads explicitly
\begin{equation} \label{psi}
	\begin{aligned}
		\Psi_{\bfmu}^J(\tau,\bar \tau,\bfz,\bar \bfz)&=e^{-2\pi y\bfb^2_+}   \sum_{\bfk\in L+\bfmu}\partial_{\bar{\tau}}\left(4\pi\im\sqrt{y}B( \bfk+\bfb, J)\right)\\
		&\times (-1)^{B(\bfk,K)} q^{-\bfk_-^2/2}\bar q^{\bfk_+^2/2}  e^{-2\pi \im B(\bfz,\bfk_-)-2\pi \im B(\bar \bfz, \bfk_+)},\end{aligned}
\end{equation}
and discussed in more detail in Appendix \ref{app:duality}. The elliptic variable reads in terms of $v_j$ and $\bfk_j$,
\begin{equation}\label{z=v_jk_j}
\bfz=\sum_{j=1}^{N_f}v_j\bfk_j,\quad \text{and} \quad  \bfb=\frac{\text{Im}(\bfz)}{y},
\end{equation}
thus inducing a non-holomorphic dependence on $v_j$.
Furthermore, $K$ appearing in the fourth root of unity $(-1)^{B(\bfk,K)}$ is a characteristic vector of $L$. Note that $\Psi_\bfmu^J$ changes by the sign $(-1)^{B(\bfmu,K-K')}$ upon replacing $K$ by a different characteristic vector $K'$ \cite{Moore:1997pc, Cordova:2018acb, Manschot:2021qqe}.

For $N_f=0$, this phase can be understood as arising from integrating out massive fermionic modes \cite{Witten:1995gf}. It also appears naturally in decoupling the adjoint hypermultiplet in the analogous function for $\CN=2^*$ \cite{Manschot:2021qqe}. For $N_f>0$, the constant part of the couplings $v_j$ (\ref{Defvw}) effectively contribute to the phase, such that the total phase reads 
\be 
\label{phase+winding}
e^{\pi i B(\bfk,K)} \prod_{j=1}^{N_f} e^{\pi i\, n_j\,B(\bfk_j,\bfk)},
\ee 
with $n_j$ the magnetic winding numbers. For arbitrary $n_j\in \mathbb{Z}$, the phase is an eighth root of unity. It would be interesting to understand this phase from integrating out massive modes. 

We deduce from (\ref{phase+winding}) that the summand of $\Psi^J_\bfmu$ changes by a phase 
\be
e^{\pi i (n_j'-n_j) B(\bfk_j,\bfk)}
\ee
if the winding numbers $n_j$ are replaced by $n'_j$. Since $\bfk_j\in K/2-\bfmu \mod L$ (see \eqref{constraint_c1CL}) and $\bfk\in L+\bfmu$, this phase is 1 if $n_j'-n_j=0\mod 4$. We can therefore 
restrict to $n_j\in \mathbb{Z}_4$. For specific choices of $\bfmu$ and
$\bfk_j$, the $n_j$ can lie in a subgroup of $\mathbb{Z}_4$. 

The modular transformations of $\Psi^J_\bfmu$ are discussed in Appendix \ref{app:duality}, which are crucial input for single-valuedness of the $u$-plane integrand. We will demonstrate in Section \ref{sec:definitionintegrals} that the $u$-plane integrand is single-valued if we impose further constraints on the winding numbers $n_j$.

Finally, if the theory is considered on a curved background, topological couplings arise in the effective field theory \cite{Witten:1995gf}. These terms couple  to the Euler characteristic and the signature of the four-manifold $X$, respectively denoted $A$ and $B$. These take the form \cite{Witten:1995gf, Moore:1997pc},
\begin{equation}\label{ABtopological}
	A=\alpha\left(\frac{du}{da}\right)^{1/2},\qquad B=\beta\Delta_{N_f}^{1/8}.
\end{equation}
Here, $\Delta_{N_f}$ is the physical discriminant incorporating the singularities of the effective theory, while $\frac{du}{da}$ is the (reciprocal of) the periods of the SW curves as introduced in Section \ref{sec:SWtheories}. Both can be determined directly from the SW curve, as described in Section \ref{sec:order_parameters}. The prefactors $\alpha$ and $\beta$ are independent of $u$, but can be functions of other moduli such as the masses $\bfm$, the dynamical scale $\Lambda_{N_f}$ or the UV coupling $\uv$. However, it turns out that for the theories with fundamental matter they are independent of the masses and only depend on the scale \cite{Marino:1998uy, Manschot:2019pog}. They satisfy several constraints from holomorphy, RG flow,  homogeneity and dimensional analysis, and can in principle be fixed  for any Lagrangian theory from a computation in the $\Omega$-background \cite{Moore:2017cmm, Marino:1998uy,Manschot:2019pog,Closset:2021lhd}.

\subsection{Observables and contact terms}
\label{sec:contaccterm}
The observables in the topologically twisted theories are the point observable or 0-observable $u$, as well as $d$-observables supported on a $d$-dimensional submanifold of $X$. The $d$-observables are only non-vanishing if the submanifold corresponds to a non-trivial homology class. For $b_1=0$, the $d$-observables with $d$ odd therefore do not contribute.

To introduce the surface observable, let $\bfx\in H_2(X,\mathbb{Q})$. Then the surface observable reads in terms of the UV fields,
\be 
I(\bfx)=\frac{1}{4\pi^2}\int_{\bfx} {\rm Tr}\left[\psi\wedge \psi -\frac{1}{\sqrt{2}}\phi\,F \right].
\ee 
In the effective infrared theory, this operator becomes,
\be 
\tilde I(\bfx)=\frac{i}{\sqrt{2}\pi} \int_\bfx \frac{1}{32} \frac{d^2u}{da^2}\,\psi\wedge \psi-\frac{\sqrt{2}}{4}\frac{du}{da}(F_-+D).
\ee 
Generating functions of correlation functions are obtained by inserting 
\be \label{insertion_correlation_function}
e^{p\,u/\Lambda_{N_f}^2+\tilde I(\bfx)/\Lambda_{N_f}}
\ee 
in the path integral. The surface observable leads to a change in the argument of the sum over fluxes (\ref{psi}),
\be 
\bfz\to \bfz + \frac{\bfx}{2\pi\,\Lambda_{N_f}}\,\frac{du}{da},\qquad \bar \bfz \to \bar \bfz. 
\ee 
and to analytically continue $\bfb$ (\ref{z=v_jk_j}) to the complex number by setting $\bfb=(\bfz-\bar \bfz)/(2iy)$.

The inclusion of the surface observable also gives rise to a
contact term \cite{Witten:1994ev,Moore:1997pc,losev1998}, 
which in particular ensures that the $u$-plane integrand is single-valued.
For $0\leq N_f\leq 3$, the contact term is $\exp(\bfx^2\,G_{N_f})$
with  \cite{mm1998,LoNeSha,Mari_o_1999}
\begin{equation}\label{contactterm}
G_{N_f}=-\frac{1}{24\,\Lambda_{N_f}^2}E_2\left(\frac{du}{da}\right)^{2}+\frac{1}{3\,\Lambda_{N_f}^2}\left(u+\frac{\Lambda_3^2}{64}\delta_{N_f,3}\right),
\end{equation}
while for $N_f=4$  it is given by \cite{mm1998,Labastida:1998sk}
\begin{align}\label{nf4contactterm}
G_{N_f=4}&=-\frac{1}{24\,\Lambda_4^2}E_2\left(\frac{du}{da}\right)^{2}+\frac{u}{3\,\Lambda_4^2}E_2(\uv)+\frac{1}{18\,\Lambda_4^2}\left\llbracket m_{1}^{2}\right\rrbracket E_4(\tuv).
\end{align}
This expression \eqref{contactterm} is valid for the theories with $N_f$ arbitrary hypermultiplet masses. The reason for it is the following \cite{mm1998,Marino:1998bm,Marino:1998ru}: $G$ is guaranteed to be $\bar\CQ$-closed and hence locally holomorphic. First, notice that $\frac{\partial F}{\partial\tau_0}=\frac u4$, 
where $\Lambda_{N_f}^{4-N_f}\eqqcolon e^{\pi\im \tau_0}$ for the asymptotically free theories ($N_f\leq 3$) and $\tn=\tuv$ for $N_f=4$. The real part of the exponential prefactor of $\Psi_\bfmu^J$ can be added to $G$ to give a monodromy-invariant contribution $\hat G$ which multiplies the intersection $\bfx^2$. From the action of a duality transformation on $\hat G$ it can be inferred that
\begin{equation}\label{CTuniversal}
G_{N_f}=-\frac{4\im}{\pi\Lambda_{N_f}^2}\frac{\partial^2 F}{\partial\tau_0^2}.
\end{equation}
The expressions \eqref{contactterm} follow by direct computation. 
A more general scheme to fix the contact terms is proposed in \cite{LoNeSha}. Contact terms can also be derived from the corresponding Whitham hierarchies \cite{TAKASAKI_1999,Mari_o_1999}. 
In the presence of surface observables, there are additional mixed contact terms $\frac{\partial^2\CF}{\partial \tn \partial m}$ for the external fluxes $\{\bfk_j\}$ as encountered in \cite{Manschot:2021qqe} for the $\nstar$ theory.

\section{The \texorpdfstring{$u$}{u}-plane integral}
\label{sec:uplane_integral}
In this section, we set up the $u$-plane integral schematically given in \eqref{phi_mu_schematic}, and demonstrate that it is well-defined on the integration domain for any $\bfmu$ with appropriate background fluxes. The case $\bfmu=\bar w_2(X)/2$ and $\bfk_j=0$ was analyzed in \cite{Moore:1997pc}.

\label{sec:sqcdcoulombbranch}
\subsection{Definition of the integrand}
As discussed in the previous sections, the $u$-plane integral on a closed four-manifold $X$ with $(b_1, b_2^+)=(0,1)$ depends on many parameters. We summarise:
\begin{itemize}
    \item The scale $\Lambda_{N_f}$ and masses $\bfm=(m_1,\dots,m_{N_f})$ of the theory. See Section \ref{sec:SWtheories}.
        \item The magnetic winding numbers $n_j$, $j=1,\dots, N_f$. See Section \ref{sec:SWgeometry}.
    \item The four-manifold $X$, in particular its signature $\sigma=\sigma(X)$,  Euler characteristic $\chi=\chi(X)$, period point $J$ and intersection form $Q$. See Section \ref{4manifolds}.
    \item The 't Hooft flux $\bfmu$, and the  external fluxes $\{\bfk_j\}=(\bfk_1,\dots,\bfk_{N_f})$. See Section \ref{corrfunctions}. 
    \item The fugacities for the point and surface observables $p$ and $\bfx$. See Section \ref{sec:contaccterm}.
\end{itemize}

In terms of these parameters, the $u$-plane path integral reduces to the following finite dimensional integral over $\CF_{N_f}(\bfm)$,
\begin{equation}\label{generaluplaneintegral}
\boxed{\begin{aligned}
&\Phi_{\bfmu, \{\bfk_j\}}^{J}(p,\bfx, \bfm, \Lambda_{N_f})=\\
&\qquad \int_{\CF_{N_f}(\bfm)}d\tau\wedge d\bar \tau \, \nu(\tau;\{\bfk_j\})\,\Psi_\bfmu^J(\tau,\bar\tau,\bfz,\bar\bfz)\,e^{2pu/\Lambda_{N_f}^2+\bfx^2 G_{N_f}}.
\end{aligned}}
\end{equation}
We summarise the different elements on the rhs:
\begin{itemize}
\item The integration domain $\CF_{N_f}(\bfm)$ in (\ref{generaluplaneintegral}) is crucially the fundamental domain of the effective gauge coupling. As discussed in Section \ref{SecFundDom}, this domain requires new aspects compared to  integration domains for earlier discussions of $u$-plane integrals. The evaluation of integrals over $\CF_{N_f}(\bfm)$ will be discussed in more detail in Section \ref{sec:integrationFD}.
\item $\nu$ is the ``measure factor" \cite{Witten:1995gf,Moore:1997pc,Marino:1998uy,LoNeSha, Manschot:2021qqe}
\begin{equation}\label{measurefactornf}
\nu(\tau;\{\bfk_j\})= \CK_{N_f}\,\frac{da}{d\tau} \,A^\chi B^\sigma\prod_{i,j=1}^{N_f}C_{ij}^{B(\bfk_i,\bfk_j)}.
\end{equation}
It combines the topological couplings \eqref{ABtopological} and the  couplings to the background fluxes \eqref{cij} with the Jacobian  $\frac{da}{d\tau}$ of the change of variables from $a$ to $\tau$. 

$\CK_{N_f}$ is an overall normalisation factor. For $N_f=0$, it is fixed by matching to known Donaldson invariants. Since $\chi+\sigma=4$, there is an ambiguity \cite{Moore:2017cmm}
\be 
(\CK_{N_f},\alpha,\beta)\sim (\zeta^{-4}\CK_{N_f},\zeta\alpha,\zeta\beta),
\ee
with $\alpha$ and $\beta$ the $u$-independent prefactors in \eqref{ABtopological}.
\item The function $\Psi^J_\bfmu$ arises from the sum over $U(1)$ fluxes. It is a Siegel-Narain theta function \eqref{psi} and discussed in detail in Section \ref{sec:sumflux}. The elliptic parameter $\bfz$ of the Siegel-Narain theta function reads
\be 
\begin{split}\label{bfz_def}
\bfz &= \frac{\bfx}{2\pi\,\Lambda_{N_f}}\frac{du}{da}+\sum_{j=1}^{N_f} v_j\bfk_j,\\
\bar \bfz &= \sum_{j=1}^{N_f} \bar v_j\bfk_j.
\end{split}
\ee 
\item Finally, $G_{N_f}$ is the contact term, discussed in more detail in Section \ref{sec:contaccterm}.
\end{itemize}

While the path integral set up in Section \ref{sec:eff_theories} integrates the exponentiated action over the local coordinates $a$ and $\bar a$, in \eqref{generaluplaneintegral} we have changed variables to $\tau$ and $\bar\tau$. This change of variables $(a,\bar a)\to (\tau,\bar\tau)$ is valid as long as the Jacobian is nonsingular in the integration region. Since the coordinates $a$ and $\bar a$ are holomorphic and anti-holomorphic respectively, the Jacobian is diagonal and the functional determinant accordingly reads $\frac{da}{d\tau}\frac{d\bar a}{d\bar \tau}$. We thus need to show that $\frac{da}{d\tau}$ is not singular away from isolated points in $\CF_{N_f}(\bfm)$, which in \eqref{generaluplaneintegral} we remove implicitly from the integration domain.

Using $\frac{da}{du}=\frac{da}{du}\frac{du}{d\tau}$, we can study the singular points in detail. First, it is shown in \cite{Aspman:2021vhs} that the singularities of $\frac{du}{d\tau}$ are in one-to-one correspondence with the branch points. In fact, both $\frac{du}{d\tau}=0$ and $\frac{du}{d\tau}=\infty$ are realised as branch points in $\CN=2$ SQCD. In the following Section \ref{sec:integrationFD}, we remove a small circle in $\CF_{N_f}(\bfm)$ around the branch points, and show that they do not give an extra contribution. Furthermore, the solutions to $\frac{du}{da}=0$ are shown to be the Argyres-Douglas (AD) points. We exclude them from the integration region, and study their contribution also in Section \ref{sec:integrationFD}. Finally, we know that $\eta^{24}\propto \left(\frac{da}{du}\right)^{12}\Delta_{N_f}$ \cite{Aspman:2021vhs}, with $\eta$ the Dedekind eta-function as defined in \eqref{dedekindeta}. Since $\eta\neq 0$ and $\Delta_{N_f}$ does not have poles, we find that $\frac{da}{du}$ never vanishes. This agrees with the fact that $\frac{da}{du}$ is the period of a holomorphic differential and therefore is never zero. 

We conclude that the functional determinant is singular in $\mathbb H$ precisely at the branch points and AD points, however with the proper exclusion of those as done in the following Section, it is non-singular and the change of variables is well-defined. This furthermore conveniently solves the problem that there is no natural integration region in $(a, \bar a)$ space \cite{Moore:1997pc}.

\subsection{Monodromy transformations of the integrand}\label{sec:definitionintegrals}
We continue by explicitly verifying that the $u$-plane integral is single-valued around the singular points of the moduli space. We find that this puts a constraint on the magnetic winding numbers $n_j$, in addition to the constraints on the background fluxes $\bfk_j$ discussed in Section \ref{corrfunctions}.

\subsubsection*{Monodromy around infinity}
Let us determine how the $u$-plane integrand transforms under the monodromy around infinity. As a function of the effective coupling $\tau$, the measure factor \eqref{measurefactornf} is proportional to $\frac{da}{d\tau}\left(\frac{du}{da}\right)^{\frac\chi2}\Delta^{\frac\sigma8}$ times the product over the couplings $C_{ij}$. We take the monodromy at infinity to be oriented as  $u\to e^{2\pi i}u$ and $a\to e^{\pi i}a$, as in Section \ref{sec:monodromies}. Then this path also encircles all singularities $u_j$, which are the roots of the physical discriminant, $
\Delta=\prod_{j=1}^{N_f+2}(u-u_j)$. We thus have that $\Delta\to e^{2\pi i(N_f+2)}\Delta$, and hence
\begin{equation}\label{Delta_inf}
\Delta^{\frac\sigma8}\to e^{\pi i (N_f+2)\sigma/4}\Delta^{\frac\sigma8}
\end{equation}
Next, since $u\to e^{2\pi i}u$ and $a\to e^{\pi i}a$ we find $\frac{du}{da}\to e^{\pi i}\frac{du}{da}$, and therefore
\begin{equation}\label{duda_inf}
\left(\frac{du}{da}\right)^{\frac\chi2}\to e^{\pi i \chi/2}\left(\frac{du}{da}\right)^{\frac\chi2}.
\end{equation}
For $\frac{da}{d\tau}$ we have that $a\to e^{\pi i}a$, while $d\tau\to d\tau$, and thus
\begin{equation}\label{dadt_inf}
\frac{da}{d\tau}\to -\frac{da}{d\tau}.
\end{equation}
From \eqref{vw_inf} we recall that $w_{ij}\to w_{ij}+\delta_{ij}$, such that with the definition \eqref{cij} we find $C_{ij}\to e^{-\pi i \delta_{ij}}C_{ij}$. The couplings $C_{ij}$ transform in the measure factor as 
\begin{equation}\label{cij_inf}
\prod_{i,j=1}^{N_f}C_{ij}^{B(\bfk_i,\bfk_j)}\to e^{-\pi i\sum_{j}\bfk_j^2}\prod_{i,j=1}^{N_f}C_{ij}^{B(\bfk_i,\bfk_j)}.
\end{equation}
Combining \eqref{Delta_inf}, \eqref{duda_inf}, \eqref{dadt_inf},  \eqref{cij_inf}, and using $\chi=4-\sigma$, we obtain
\begin{equation}\label{nu_inf}
\nu\to -e^{\pi iN_f\sigma /4}e^{-\pi i\sum_{j}\bfk_j^2}\, \nu.
\end{equation}
This phase for $\bfk_j=0$ can be checked directly by taking $q$-expansions from the SW curves, for generic masses.

From \eqref{vw_inf} we recall  that under the monodromy around infinity  $v_j \to -v_j-n_j$, and thus
\begin{equation}
 \bfz\to -\bfz-\sum_{j=1}^{N_f}n_j\bfk_j.
\end{equation}
For the sum over fluxes we can now deduce using (\ref{tshift}) that
\begin{equation}\label{psiMinf1}
\begin{split}
&\Psi_\bfmu^J\Big(\tau+N_f-4,-\bfz-\sum_{j=1}^{N_f}n_j\bfk_j\Big)\\
&\quad =e^{\pi\im (N_f-4)(\bfmu^2-\bfmu\cdot K)}\,\Psi^J_\bfmu\Big(\tau,-\bfz-\sum_{j=1}^{N_f}n_j\bfk_j+(N_f-4)(\bfmu-\tfrac{K}{2})\Big),
\end{split}
\end{equation}
where we suppressed the dependence on the anti-holomorphic parts. Recall from \eqref{constraint_c1CL}  that  
\begin{equation}
    c_1(\CL_j)\equiv K-2\bfmu\mod 2L.
\end{equation}
and as such we can express $\bfk_j=c_1(\CL_j)/2$ as
\be
\label{bfkjell}
\bfk_j=\frac{K}{2}-\bfmu+\bfell_j,\qquad \bfell_j\in L. 
\ee
We have then 
\be
\label{bfk2}
\bfk^2_j=\frac{\sigma}{4}-K\cdot \bfmu+\bfmu^2-2\bfmu \cdot \bfell_j \mod 2\mathbb{Z},
\ee
 where we used that $K$ is a characteristic vector of $L$, and $K^2=\sigma \mod 8$.
Using (\ref{tshift}) and substitution of (\ref{bfkjell}) in this expression, \eqref{psiMinf1} equals
\begin{equation}
	e^{\pi\im(N_f-4)(\bfmu^2-\bfmu\cdot K)}\,\Psi^J_\bfmu\Big(\tau,-\bfz -\sum_{j=1}^{N_f}n_j\bfell_j+(N_f-4+\sum_j n_j)(\bfmu-\tfrac{K}{2})\Big).
\end{equation}
Our aim is to write this as a phase times $\Psi^J_\bfmu(\tau,\bfz)$. The constraints on the winding numbers should be independent of $\bfmu$ and $\bfk_j$, since the prepotential is. From \eqref{zshift}, we therefore get the first constraint 
\be\label{njcondition}
\sum_j n_j=N_f\mod 2.
\ee
Using identity \eqref{zshift}, $2\bfmu^2-K\cdot \bfmu\in \mathbb{Z}$ and $4(\bfmu-\tfrac{K}{2})\in 2L$, this simplifies to
\be
\begin{split}
&e^{\pi\im (N_f-4)(\bfmu^2-\bfmu\cdot K)+2\pi i \bfmu\cdot \sum_j n_j\bfell_j-2\pi i (N_f+\sum_j n_j)(\bfmu^2-\bfmu\cdot K/2)}\Psi^J_\bfmu(\tau,-\bfz)\\
&=-e^{\pi\im N_f(\bfmu^2-\bfmu\cdot K)+2\pi i \bfmu\cdot \sum_j n_j \bfell_j-2\pi i (N_f+\sum_j n_j)(\bfmu^2-\bfmu\cdot K/2)}\,\Psi^J_\bfmu(\tau,\bfz)\\
&=-e^{-\pi\im N_f\bfmu^2+2\pi i \bfmu\cdot \sum_j n_j \bfell_j-2\pi i \sum_j n_j(\bfmu^2-\bfmu\cdot K/2)}\,\Psi^J_\bfmu(\tau,\bfz).
\end{split}
\ee
Finally using (\ref{bfk2}), we can express the phase in terms of $\bfk_j$,
\begin{equation} \label{psi_inf}
{\bf M}_\infty:\qquad  \Psi^J_\bfmu(\tau,\bfz)\to -e^{-\pi i N_f \bfmu^2 - \pi i \sum_j n_j (\bfk_j^2+\bfmu^2-\sigma/4) }\,\Psi_\bfmu^J(\tau,\bfz).
\end{equation}
By multiplying \eqref{nu_inf} with \eqref{psi_inf}, we find 
\begin{equation}\label{nuPsi_inf1}
    \nu(\tau;\{\bfk_j\})\, \Psi_{\bfmu}^J(\tau,\bfz) \to e^{-\pi i \sum_j(n_j+1)\bfk_j^2+\frac{\pi i}{4}\sum_j (\sigma-4\bfmu^2)(n_j+1)}\, \nu(\tau;\{\bfk_j\})\,  \Psi_{\bfmu}^J(\tau,\bfz). 
\end{equation}
Combining \eqref{k+mu=spinC} with \eqref{spinCsignature}, we have that $    4(\bfk_j+\bfmu)^2 \equiv \sigma\mod 8$ 
for every $j=1,\dots, N_f$. We insert this into the second exponential of \eqref{nuPsi_inf1}, such that
\begin{equation}
  {\bf M}_\infty:\qquad    \nu(\tau;\{\bfk_j\})\, \Psi_{\bfmu}^J(\tau,\bfz) \to e^{2\pi i \bfmu \sum_j(n_j+1)\bfk_j} \,\nu(\tau;\{\bfk_j\}) \,\Psi_{\bfmu}^J(\tau,\bfz),
\end{equation}
and the $u$-plane integrand is invariant under $T^{N_f-4}$ if and only if $\bfmu \sum_j(n_j+1)\bfk_j\in \mathbb Z$. Using \eqref{bfkjell} and the fact that $K$ is a characteristic vector of $L$, we find
\begin{equation}\label{constraint2}
    n_j=1 \mod 2
\end{equation}
for all $j=1, \dots, N_f$, which implies the above constraint (\ref{njcondition}).

\subsubsection*{Monodromy ${\bf M}_j$}
Let us determine how the integrand transforms under the monodromy $M_j$ around the mass singularity $m_j/\sqrt{2}$. Since the mass singularity   corresponds to a singularity $u_j$ on the $u$-plane, we have that $(u-u_j)\to e^{2\pi i}(u-u_j)$. This implies that $\Delta=(u-u_j)\prod_{i\neq j}^{2+N_f}(u-u_i)\to e^{2\pi i}\Delta$, such that $\Delta^{\frac\sigma8}\to e^{\pi i \sigma/4}\Delta^{\frac \sigma8}$. The transformation of $\frac{da}{du}$ can be determined from \eqref{dadu_def}: While $u\to u_j$, both $g_2$ and $g_3$ remain finite and nonzero (otherwise $u_j$ would be an AD point). This implies that $\frac{g_3}{g_2}$ contains no factors of $(u-u_j)$, and thus $\frac{du}{da}\to \frac{du}{da}$. Similarly, we have that $\frac{da}{d\tau}\to \frac{da}{d\tau}$. From \eqref{vw_Mj} we finally have that $w_{ik}\to w_{ik}+\delta_{ij}\delta_{ik}$. We combine
\begin{equation}\label{nu_M_j}
{\bf M}_j: \quad \nu\to e^{\pi i\sigma/4}e^{-\pi i\bfk_j^2}\nu.
\end{equation}

For the monodromy around the mass singularity $m_j/\sqrt{2}$, we find for $\Psi_\bfmu^J$ with (\ref{tshift})
\be
\begin{split}
&\Psi_{\bfmu}^{J}(\tau+1,\bfz-\bfk_j)=e^{\pi i (\bfmu^2-\bfmu\cdot K)}\Psi_{\bfmu}^{J}(\tau,\bfz-\bfell_j)\\
&=e^{\pi i (\bfmu^2-\bfmu\cdot K)+2\pi i B(\bfmu,\bfell_j)}\,\Psi_{\bfmu}^{J}(\tau,\bfz)\\
&=e^{-\pi i \sigma/4+\pi i \bfk_j^2}\,\Psi_{\bfmu}^{J}(\tau,\bfz).
\end{split}
\ee
The phases thus cancel precisely,
\begin{equation}
   {\bf M}_j: \quad  \nu(\tau+1)\,\Psi_{\bfmu}^{J}(\tau+1,\bfz-\bfk_j)=\nu(\tau)\,\Psi_{\bfmu}^{J}(\tau,\bfz),
\end{equation}
without any constraints.

\subsubsection*{Monodromy ${\bf M}_m$}
For the monopole singularity in $N_f=1$ we find  that
\begin{equation}
	\begin{aligned}
&		\Psi_{\bfmu}^J(\tfrac{\tau}{-\tau+1},\tfrac{v\bfk_1+(n+1)/2\,\tau\bfk_1}{-\tau+1})=e^{\pi i  (n+1)\frac{v}{-\tau+1}\bfk_1^2+\pi i \frac{(n+1)^2}{4}\frac{\tau}{-\tau+1}\bfk_1^2}\,(-1)^{(n+1)\bfk_1\cdot K/2}\\
&\qquad \qquad \times \Psi_{\bfmu+(n+1)\bfk_1/2}^J(\tfrac{\tau}{-\tau+1},\tfrac{v\bfk_1}{-\tau+1}),
	\end{aligned}
\end{equation}
where we have used \eqref{zshifttau}.
Then using \eqref{psists}, we arrive at 
\begin{equation}
	\begin{aligned}
&\Psi_{\bfmu}^J(\tfrac{\tau}{-\tau+1},\tfrac{v\bfk_1+(n+1)/2\,\tau\bfk_1}{-\tau+1})
=(-\tau+1)^{b_2/2}(-\bar\tau+1)^2\,e^{\pi i (n+1)\bfk_1\cdot
  K/2-(n+1)^2\bfk_1^2/4}\,e^{-\pi i \sigma/4}\\
&\times \exp\!\left[\pi i \frac{(v+(n+1)/2)^2}{-\tau+1}\bfk_1^2
\right]\Psi_{\bfmu+(n+1)\bfk_1/2}^J(\tau,\bfz).
	\end{aligned}
\end{equation}
Since $\Psi_\bfmu^J$ is required to transform to itself up to an
overall factor, we must demand that $(n+1)\bfk_1/2 \in L$. Therefore for
$\bfk_1\in L/2$, we find the requirement that $n=-1 \in
\mathbb{Z}_4$. This simplifies the transformations considerably, and we find
\begin{equation}\label{psi_Mm}
    \Psi_{\bfmu}^J(\tau,\bfz)\to (-\tau+1)^{b_2/2}(-\bar\tau+1)^2e^{-\pi i \sigma/4} e^{\pi i\bfk_1^2\frac{v^2}{-\tau+1}} \Psi_{\bfmu}^J(\tau,\bfz).
\end{equation}
The $\bfk_j$-independent part of the measure factor transforms precisely as under ${\bf M}_j$ (see \eqref{nu_M_j}),  as the same argument holds. However, due to the transformation $\tau\to\frac{\tau}{-\tau+1}$, the measure also picks up its modular weight $\frac\sigma2+1$.
From \eqref{Mm_transformation} we furthermore find the transformation of $C_{11}$, such that 
\begin{equation}\label{nu_Mm}
    \nu(\tau,\bfk_1)\to e^{\pi i \sigma/4}e^{-\pi i\bfk_1^2\frac{v^2}{-\tau+1}}(-\tau+1)^{\frac\sigma2+1}\nu(\tau,\bfk_1),
\end{equation}
where we have already used $n=-1 \in
\mathbb{Z}_4$. If we multiply  \eqref{psi_Mm} and \eqref{nu_Mm} with $d\tau\wedge d\bar\tau$ (which has modular weight $(-2,-2)$), then
\begin{equation}
  {\bf M}_m:\quad   d\tau\wedge d\bar\tau \nu(\tau,\bfk_1)\,\Psi_{\bfmu}^J(\tau,\bfz)\to d\tau\wedge d\bar\tau \nu(\tau,\bfk_1)\,\Psi_{\bfmu}^J(\tau,\bfz),
\end{equation}
where we have used $\sigma+b_2=2$. Thus, the $u$-plane integrand is also invariant under ${\bf M}_m$.

For $N_f>1$ we find the same condition, namely that $n_j=-1 \mod 4$ for all $j$.

\subsubsection*{Monodromy ${\bf M}_d$}
Given the relation (\ref{ModsIdent}), it is not necessary to
explicitly check single-valuedness of the integrand under this monodromy, as it is a product of the above monodromies. 
\vspace{.3cm}\\
To conclude this section, let us stress the constraints for the winding number $n_j$, such that the $u$-plane integral is invariant under all monodromies in $N_f\leq 3$. To this end, we need to satisfy the constraints $n_j=1\mod 2$ (\ref{constraint2}) from ${\bf M}_\infty$, and $n_j=-1\mod 4$ for ${\bf M}_m$. Since the latter is the stronger constraint, we require
\begin{equation}\label{u_plane_contraint}
    n_j=-1\mod 4,
\end{equation}
for all $j=1,\dots,N_f$.

\section{Integration over fundamental domains}\label{sec:integrationFD}
As discussed in Sections \ref{sec:SWtheories} and \ref{sec:uplane_integral}, $u$-plane integrals for massive $\CN=2$ theories with fundamental hypermultiplets include new aspects. This section discusses how to evaluate such integrals \eqref{generaluplaneintegral}. More concretely, we aim to define and evaluate integrals of the form
\be\label{CI_f}
\CI_f=\int_{\CF(\bfm)} d\tau\wedge d\bar \tau\,y^{-s}\,f(\tau,\bar \tau),
\ee
with $s\leq 1$. The domain $\CF(\bfm)$ is the fundamental domain for the effective coupling
constant as discussed in Section \ref{SecFundDom}, and $f$ a
non-holomorphic function of weight $(2-s,2-s)$ arising from the topologically twisted
Yang-Mills theory. For $\CF(\bfm)$ a fundamental domain of a congruence subgroup, such integrals \eqref{CI_f} have been studied in the context of theta lifts of weakly holomorphic modular forms and harmonic Maass forms \cite{Borcherds:1996uda,bringmann2017regularized,bruinier2004two} as well as one-loop amplitudes in string theory  \cite{Lerche:1988np,Dixon:1990pc,Harvey:1995fq}. 

We assume that the integrand $y^{-s}\,f(\tau,\bar \tau)$ can be expressed as 
\be
\partial_{\bar \tau} \widehat h(\tau,\bar \tau)=y^{-s}\,f(\tau,\bar \tau),
\ee
for a suitable function $\widehat h(\tau,\bar \tau)$ using mock modular forms. This was indeed the case in  \cite{Korpas:2017qdo, Korpas:2019ava, Manschot:2021qqe}, and will be demonstrated for massive $\CN=2$ theories with fundamental hypermultiplets. 
The integral $\CI_f$ then reads 
\be
\label{CIfStokes}
\CI_f=-\int_{\partial \CF(\bfm)} d\tau\,\widehat h(\tau,\bar \tau),
\ee
with $\partial \CF(\bfm)$ the boundary of $\CF(\bfm)$. We will carry this out evaluation in Part II \cite{Aspman:2023ate}.

There are a number of aspects to be addressed in order to evaluate
integrals over $\CF(\bfm)$: 
\begin{enumerate}
\item Identifications of boundary components of $\CF(\bfm)$ due to monodromies on the $u$-plane.
\item Contributions from the cusps, that is $\tau\to i\infty$ or
  $\tau\to \gamma(i\infty)\in\mathbb{Q}$ for an element $\gamma\in \psl$.
\item Contributions from a singular point in the interior of $\CF(\bfm)$.
\item Contributions from an elliptic point $p\in \mathbb{H}$ of
  $\psl$.
  \item Branch points and branch cuts.
\end{enumerate}
We will discuss these aspects 1.--5. in the following.
\vspace{.3cm}\\
1. {\it Identifications}\\ 
The modular transformation induced by monodromies identify components of the boundary of the fundamental
domain $\partial\CF(\bfm)$ pairwise. Their contributions to the integral
(\ref{CIfStokes}) vanish, which is, for example, familiar from deriving 
valence formulas for modular forms \cite[Fig. 2]{Bruinier08}. See Fig. \ref{fig:nf1CutsAlt} for an example. 
\vspace{.3cm}\\
2. {\it Cusps}\\
Contributions near the cusps require a regularisation
  \cite{Moore:1997pc,  Korpas:2019ava}. Such regularisations have been
  developed in the context of string amplitudes \cite{Lerche:1988np,
    Dixon:1990pc, Harvey:1995fq} and analytic number
  theory \cite{Petersson1950, Bruinier08, 1603.03056}.

  Let us first consider the cusp
  $\tau\to i\infty$. To regularise the divergence, one introduces 
  a cut-off $\text{Im}\, \tau= Y\gg 1$, and takes the limit $Y \to \infty$ after evaluation. 
 We require that $f$ near $i\infty$ has a Fourier expansion of the
 form\footnote{Also if $f$ does not satisfy this
  requirement, the integral can be regularised as explained in
  \cite{Korpas:2019ava, 1603.03056}. We do not need this
  regularisation for the correlators in this paper.}
\be 
f(\tau,\bar \tau)=\sum_{m\gg -\infty,n\geq 0} c(m,n)\,q^m\,\bar q^n.
\ee  
Then the function $\widehat h$ has the form,
\be
\widehat h(\tau,\bar \tau)=h(\tau)+2^s\int^{i\infty}_{-\bar \tau} \frac{f(\tau,-v)}{(-i(v+\tau))^s}dv,
\ee
where $h(\tau)$ is a weakly holomorphic $q$-series, with expansion
\be
\label{htau}
h(\tau)=\sum_{m\gg-\infty} d(m)\,q^m.
\ee
The cusp $\tau\to i\infty$ then contributes
\be
\left[\CI_f \right]_\infty= w_\infty\,d(0),
\ee
with $d(0)$ the constant term of $h(\tau)$ (\ref{htau}), and
$w_\infty$ the width of the cusp $\CF(\bfm)$ at $i\infty$. For $N_f\leq 3$, $w_\infty$ is $4-N_f$ \cite{Aspman:2021vhs}.

The other cusps can be treated in a similar fashion using modular
transformations. We label the $n_c$ cusps in $\CF(\bfm)$ by
$j=1,\dots,n_c$. If the cusp is on the horizontal axis at
$-\frac{d_j}{c_j}\in \mathbb{Q}$ with relative prime $(c_j,d_j)\in
\mathbb{Z}^2$, we can map the cusp to $i\infty$ by a modular
transformation
\be
\gamma_j=\begin{pmatrix} a_j & b_j \\ c_j & d_j \end{pmatrix}.
\ee 
We let $\tau_j=\gamma_j \tau$. Then the
holomorphic part $h_j(\tau_j)$ of $(c_j\tau+d_j)^{-2}\,\widehat
h(\gamma\tau_j,\gamma \bar \tau_j)$ can be expanded for $\tau$ near
$-\frac{d_j}{c_j}$ as
\be
h_j(\tau_j)=\sum d_j(n)\,q_j^n,\qquad q_j=e^{2\pi i \tau_j}.
\ee
As a result, the cusp $j$ contributes
\be
\left[\CI_f \right]_j= w_j\,d_j(0).
\ee
\vspace{.3cm}\\
3. {\it Singular points in the interior of  $\CF(\bfm)$}\\
The integrand can be singular at a point $\tau_{\rm s}$ in the interior of
  $\CF(\bfm)$. Such singularities appear typically for deformations of
  superconformal theories, such as the $\CN=2^*$ theory and the
  $N_f=4$ theory, where the UV
  coupling $\tuv$ gives rise to such a singularity 
  \cite{Manschot:2021qqe, Aspman:2021evt}. See Fig. \ref{fig:nf4mm00} for an example.  We
  require that the expansion of $f$ near such a singularity reads, 
  \be
  \label{fexpint}
f(\tau,\bar \tau)=\sum_{m\gg -\infty,n\geq 0} c_s(m,n)\,(\tau-\tau_{\rm s})^m\,(\bar \tau-\bar \tau_{\rm s})^n.
\ee
Then, the anti-derivative $\widehat h(\tau,\bar \tau)$ has similar
expansion,
\be
  \label{hexpint}
\widehat h(\tau,\bar \tau)=\sum_{m\gg -\infty,n\geq 0} d_s(m,n)\,(\tau-\tau_{\rm s})^m\,(\bar \tau-\bar \tau_{\rm s})^n.
\ee
The contour integral for a small contour around $\tau_{\rm s}$,
\be 
C_\varepsilon(\tau_{\rm s})= \left\{ \tau=\tau_{\rm s}+\varepsilon\, e^{\im\varphi},
\varphi\in [0,2\pi)\right\},
\ee 
is bounded for such a function. Moreover, in the limit $\varepsilon \to
0$, the contour integral is finite. We define the ``residue'' of a non-holomorphic function
$g(\tau,\bar \tau)$
\begin{equation}\label{nresdef} 
\begin{split}
  \nres{\tau=\tau_{\rm s}}\left[g(\tau,\bar\tau)\right]&=\frac{1}{2\pi
  i}\lim_{\varepsilon\to0}\oint_{C_\varepsilon(\tau_{\rm
    s})}g(\tau,\bar\tau)\,d\tau.
\end{split} 
\end{equation}
For the expansion (\ref{fexpint}) this evaluates to
\be
\left[ \CI_f\right]_s=2\pi i\,  \nres{\tau=\tau_{\rm s}}\left[\widehat h(\tau,\bar\tau)\right]=d_s(-1,0),
\ee
with $d_s(-1,0)$ the coefficient in the expansion (\ref{hexpint}). 
\vspace{.3cm}\\
4. {\it Elliptic points}\\
  For $\CN=2$ QCD,  AD points  are the
  elliptic points of the duality group, and lie on the boundary 
  of $\CF(\bfm)$. See Fig. \ref{fig:nf2AD} for an example. The elliptic points are $\alpha=e^{\pi i/3}$ and $i$,
  and their images under $\psl$. Contour integrals around such points can be regularised using a
  cut-off $\varepsilon$. We assume that the anti-derivative $\widehat h$ has 
  the following expansion near an elliptic point $\tau_e$,
  \be
  \label{hexpe}
\widehat h(\tau,\bar \tau)=\sum_{m\gg -\infty,n\geq 0} d_{\rm e}(m,n)\,(\tau-\tau_{\rm e})^m\,(\bar \tau-\bar \tau_{\rm e})^n.
\ee

As a result, the boundary arc around $\tau_{\rm AD}$ in $\mathbb{H}$ is a
fraction of $2\pi$, which needs to be properly accounted for.
These neighbourhoods have an angle $\frac{2\pi}{k_e}$, with $k_e=2$ for
$\tau_e=\im$, and $k_e=6$ for $\tau_e=\alpha$
\cite{Diamond}. Furthermore, it is 
important how many images of $\CF$ in $\CF(\bfm)$ coincide at the elliptic point.
We denote this number by $n_e$. For $\CN=2$ SQCD, we found examples with
$n_e=2$ and $4$ for $\tau_e\sim\alpha$, while for $\tau_e\sim \im$,
$n_e=1$ \cite{Aspman:2021vhs}. The contribution
from an elliptic point is then, 
\begin{equation}\label{regulariseduplane}
\left[\CI_f\right]_e=2\pi i \frac{n_{\text{e}}}{k_{\rm e}}\, \nres{\tau=\tau_{\rm
    e}}\left[\widehat h(\tau,\bar \tau)\right]=\frac{n_{\text{e}}}{k_{\rm e}}\,d_e(-1,0),
\end{equation}
\vspace{.3cm}\\
5. { \it  Branch points and cuts}\\
Branch points and cuts are a new aspect compared to previous
analyses (see for instance Fig. \ref{fig:nf1CutsAlt} and \ref{fig:nf2domain}). We will demonstrate that their contribution vanishes for the
integrands of interest. 

We assume that the integrand $f$ satisfies
\be\label{branch_point_singularity}
\widehat h(\tau,\bar \tau)= (\tau-\tau_{\bp})^{n}\,g(\tau,\bar \tau),
\ee
with $n\in \mathbb{Z}/2$ and $n\geq -1/2$, $g(\tau,\bar \tau)$ being a real
analytic function near $\tau_\bp$. This assumption is satisfied for the twisted Yang-Mills theories \cite{Aspman:2023ate}. 
To treat this type of singularity, we remove a $\delta$ neighbourhood
and analyse the $\delta\to 0$ limit.
Let $C_\delta$ be the contour 
\be
C_\delta =\{ \tau_\bp+\delta\, e^{i\theta}\,\vert \,  \theta\in (0,2\pi)\}
\ee
around $\tau_\bp$ with radius $\delta>0$.
Therefore, on the contour $|y^{-s} f|$ is bounded by
\be
|\widehat h|\leq \delta^n\,K
\ee
for some $K>0$. The integral around the branch point therefore vanishes in the limit,
\be
\begin{split}
\CI^{\rm bp}_f&=\lim_{\delta\to 0}\,\int_{C_\delta} \widehat h\,|d\tau|\leq \lim_{\delta\to 0}\,\int_0^{2\pi} \delta^n\,K\,\delta\,d\theta\\
&=\lim_{\delta\to 0}\,2\pi K \delta^{n+1}=0.
\end{split}
\ee
 The branch points  necessarily give  rise to branch cuts. For the
purpose of integration, we remove a neighbourhood with distance $r$
from the cut, and take the limit $r\to 0$ after determining the
integral. Since the value of the integrand is finite near the branch
cut, the contribution to the integral vanishes. 
\vspace{.3cm}\\
{\it Summary}\\
Combining all the contributions discussed above, we find
\be\label{integrationresult}
\CI_f=\sum_{j=1}^n w_j\,d_j(0)+\sum_{s} d_{s}(-1,0)+\sum_e\, \frac{n_{\text{e}}}{k_{\rm e}}\,d_e(-1,0).
\ee
This formula generalises 
\cite{Moore:1997pc} for the pure $N_f=0$ theory on a smooth four-manifold $X$ that admits a metric of positive scalar curvature, \cite[Equation (5.10)]{Korpas:2019cwg} for the pure theory on  generic $X$,  
\cite[Equation (4.88)]{Manschot:2021qqe} for the $\nstar$ theory on $X$,  and \cite{Malmendier:2008db} for the massless $N_f=2$ and $N_f=3$ theories on $X=\mathbb{CP}^2$.

\acknowledgments
We are happy to thank Cyril Closset, Horia Magureanu, Greg Moore and Samson Shatashvili for correspondence and discussions. JM thanks Greg Moore for collaboration on related projects. JA is supported by the Government of Ireland Postgraduate Scholarship Programme GOIPG/2020/910 of the Irish Research Council. EF is supported by the TCD Provost’s PhD Project Award. JM is supported by the Laureate Award 15175 “Modularity in Quantum Field Theory and Gravity” of the Irish Research Council.

\appendix
\section{Modular forms}\label{app:modularforms}
In this Appendix, we collect some properties of modular forms for subgroups of $\psl$. For further reading, see \cite{Bruinier08,ono2004,Zagier92,Diamond,schultz2015,koblitz1993}.

\subsection{Theta functions and Eisenstein series}\label{sec:jacobitheta}
We make use of modular forms for the congruence subgroups $\Gamma_0(n)$ and $\Gamma^0(n)$  of $\psl$. These subgroups are defined as 
\be\begin{aligned}
\Gamma_0(n) = \left\{\begin{pmatrix}a&b\\c&d\end{pmatrix}\in \slz\big| \, c\equiv0 \; \mod n\right\},\\
\Gamma^0(n) = \left\{\begin{pmatrix}a&b\\c&d\end{pmatrix}\in \slz\big| \, b\equiv0 \; \mod n\right\},
\end{aligned}\ee
and are related by conjugation with the matrix $\text{diag}(n,1)$. We furthermore define the \emph{principal congruence subgroup} $\Gamma(n)$ as the subgroup of $\slz\ni A$ with $A\equiv\mathbbm 1\mod n$. A subgroup $\Gamma$ of $\slz$ is called a congruence subgroup if it contains $\Gamma(n)$ for some $n\in \mathbb N$.

The above introduced congruence subgroups host a number of interesting modular forms. 
The Jacobi theta functions $\vartheta_j:\mathbb{H}\to \mathbb{C}$,
$j=2,3,4$, are defined as
\be
\label{Jacobitheta}
\begin{split}
\vartheta_2(\tau)= \sum_{r\in
  \mathbb{Z}+\frac12}q^{r^2/2},\quad 
\vartheta_3(\tau)= \sum_{n\in
  \mathbb{Z}}q^{n^2/2},\quad
\vartheta_4(\tau)= \sum_{n\in 
  \mathbb{Z}} (-1)^nq^{n^2/2},
\end{split}
\ee
with $q=e^{2\pi i\tau}$. These functions transform under $T,S\in \slz$ as
\be
\begin{split}
S:\quad& \begin{array}{l}
  \vartheta_2(-1/\tau)=\sqrt{-i\tau}\vartheta_4(\tau), \\
  \vartheta_3(-1/\tau)=\sqrt{-i\tau}\vartheta_3(\tau), \\ \vartheta_4(-1/\tau)=\sqrt{-i\tau}\vartheta_2(\tau),\end{array}\\
T:\quad& \begin{array}{l}\vartheta_2(\tau+1)=e^{\frac{\pi
      i}{4}}\vartheta_2(\tau), \\
  \vartheta_3(\tau+1)=\vartheta_4(\tau), \\
  \vartheta_4(\tau+1)=\vartheta_3(\tau). \end{array} \label{jttransformations}
\end{split}
\ee
They furthermore satisfy the Jacobi abstruse identity
\begin{equation}\label{jacobiabstruseidentity}
\vartheta_2^4+ \vartheta_4^4= \vartheta_3^4.
\end{equation}
The Dedekind eta function $\eta: \mathbb H\to \mathbb C$ is defined as the infinite product
\begin{equation}\label{dedekindeta}
\eta(\tau)=q^{\frac{1}{24}}\prod_{n=1}^{\infty}(1-q^n), \quad q=e^{2\pi i\tau}.
\end{equation}
It transforms under the generators of $\slz$ as
\be\begin{aligned}\label{etatransformation}
S: \quad& \eta(-1/\tau)=\sqrt{-i\tau }\, \eta(\tau),\\
T: \quad& \eta(\tau+1)=e^{\frac{\pi i}{12}}\, \eta(\tau),
\end{aligned}\ee
and relates to the Jacobi theta series as $
\eta^{3}=\frac{1}{2}\jt_2\jt_3\jt_4$.

\subsubsection*{Eisenstein series}\label{sec:eisenstein}
We let $\tau\in \mathbb{H}$ and define $q=e^{2\pi i \tau}$. Then the Eisenstein series $E_k:\mathbb{H}\to \mathbb{C}$ for even $k\geq 2$ are defined as the $q$-series 
\be
\label{Ek}
E_{k}(\tau)=1-\frac{2k}{B_k}\sum_{n=1}^\infty \sigma_{k-1}(n)\,q^n,
\ee
where $B_k$ are the Bernoulli numbers and $\sigma_k(n)=\sum_{d|n} d^k$ is the divisor sum. For $k\geq 4$ even, $E_{k}$ is a modular form  of weight $k$ for
$\operatorname{SL}(2,\mathbb{Z})$.  Any modular form for $\slz$ can be related to the Jacobi theta functions \eqref{Jacobitheta} by 
\begin{equation}\label{e4e6jacobi}
E_4=\frac12(\jt_2^8+\jt_3^8+\jt_4^8),\qquad E_6=\frac12(\jt_2^4+\jt_3^4)(\jt_3^4+\jt_4^4)(\jt_4^4-\jt_2^4).
\end{equation}
With our normalisation \eqref{Ek}, the $j$-invariant can be written as 
\begin{equation}\label{je4e6}
j=1728\frac{E_4^3}{E_4^3-E_6^2}=256\frac{(\jt_3^8-\jt_3^4\jt_4^4+\jt_4^8)^3}{\jt_2^8\jt_3^8\jt_4^8}.
\end{equation}

\subsection{Siegel-Narain theta function}\label{app:duality}
Let $L$ be an $n$-dimensional uni-modular lattice with signature $(1,n-1)$. For the application to the $u$-plane integral, $n=b_2(X)$.
Let $K$ be a characteristic vector of $L$. Its defining property is $\bfl^2=\bfl\cdot K\mod 2$ for every $\bfl\in L$. 
Furthermore, we have that $\bfmu\in L/2$.

We consider the Siegel-Narain theta function $\Psi_{\bfmu}^J:\mathbb H\times \mathbb C\to\mathbb C$ defined in the main text in  (\ref{psi}). We repeat it here for convenience,
\begin{equation} \label{psiA}
	\begin{aligned}
		\Psi_{\bfmu}^J(\tau,\bar \tau,\bfz,\bar \bfz)&=e^{-2\pi y\bfb^2_+}   \sum_{\bfk\in L+\bfmu}\partial_{\bar{\tau}}\left(4\pi\im\sqrt{y}B( \bfk+\bfb,J)\right)\\
		&\times (-1)^{B(\bfk,K)} q^{-\bfk_-^2/2}\bar q^{\bfk_+^2/2}  e^{-2\pi \im B(\bfz,\bfk_-)-2\pi \im B(\bar \bfz, \bfk_+)},
		\end{aligned}
\end{equation}
where $J$ is a normalized positive vector in $L\otimes \mathbb{R}$, $\bfk_+=B(\bfk,J)\,J$, $\bfk_-=\bfk-\bfk_+$ and $\bfb=\text{Im}(\bfz)/y$.
The transformations under the generators $S$ and $T$ of $\psl$ are most easily determined if we shift $\bfmu\to \bfmu+K/2$. One finds \cite{Korpas:2017qdo, Korpas:2019cwg} 
\be
\label{Psi_trafos}
\begin{split}
 &S:\qquad \Psi^J_{\bfmu+K/2}(-1/\tau,-1/\bar \tau,\bfz/\tau,\bar \bfz/\bar \tau)= -\im(-\im\tau)^{n/2}(\im\bar \tau)^2\\
 &\hspace{3cm}\times\,e^{-\pi
  \im \bfz^2/\tau+\pi \im K^2/2}\,(-1)^{B(\bfmu,K)} \,\Psi_{K/2}^J(\tau,\bar \tau,\bfz-\bfmu,\bar\bfz-\bfmu), \\
 &T:\qquad \Psi^J_{\bfmu+K/2}(\tau+1,\bar \tau+1,\bfz,\bar \bfz)=\\
&\hspace{3cm} e^{\pi \im(\bfmu^2-K^2/4)}\,\Psi^J_{\bfmu+K/2}(\tau,\bar \tau,\bfz+\bfmu,\bar \bfz+\bfmu).
\end{split}
\ee 

Using these transformations, one finds for the periodicity in $\tau$,
\begin{equation}\label{tshift}
\begin{split}
\Psi_{\bfmu}^J(\tau+1,\bar \tau+1,\bfz,\bar \bfz)&=e^{\pi\im(\bfmu^2-B(\bfmu,K))}\Psi_{\bfmu}^J(\tau,\bar \tau,\bfz+\bfmu-K/2,\bar \bfz+\bfmu-K/2) \\
\end{split}
\end{equation}
and for $S^{-1}T^{-k}S=\left(\begin{smallmatrix}1&0\\ k&1\end{smallmatrix}\right)$,
\begin{equation}\label{psists}
\Psi^J_\bfmu\!\left( \tfrac{\tau}{k\tau+1},\tfrac{\bar \tau}{k\bar \tau+1}, \tfrac{\bfz}{k\tau+1},\tfrac{\bar \bfz}{k\bar \tau+1} \right)=(k\tau+1)^{\frac{n}{2}}(k\bar \tau+1)^2e^{-\frac{\pi \im k\bfz^2}{k\tau+1}}e^{\frac{\pi \im}{4}kK^2}\Psi^J_{\bfmu}(\tau,\bar \tau,\bfz,\bar \bfz).
\end{equation}

We furthermore list the following transformations for $\bfz$: 
\begin{itemize}
\item For the reflection $\bfz\to -\bfz$,
\begin{equation}\label{ellparamminus}
\Psi^J_\bfmu(\tau,\bar \tau,-\bfz,-\bar \bfz)=-e^{2\pi \im B(\bfmu,K)}\,\Psi^J_\bfmu(\tau,\bar \tau,\bfz,\bar \bfz).
\end{equation}
\item For shifting $\bfz\to \bfz+\bfnu$ with $\bfnu\in L$,
\be\label{zshift}
\Psi^J_\bfmu(\tau,\bar \tau,\bfz+\bfnu,\bar \bfz+\bfnu)=e^{-2\pi \im B(\bfnu,\bfmu)}\, \Psi^J_{\bfmu}(\tau,\bar \tau,\bfz,\bar \bfz).
\ee
\item For shifting $\bfz\to \bfz+\bfnu\tau$ with $\bfnu\in L \otimes
\mathbb{R}$,
\be\label{zshifttau}
\Psi^J_\bfmu(\tau,\bfz+\bfnu\tau )=e^{2\pi \im B(\bfz,\bfnu)} q^{\bfnu^2/2} (-1)^{-B(\bfnu,K)}\, \Psi^J_{\bfmu+\bfnu}(\tau,\bar \tau,\bfz,\bar \bfz).
\ee
We can restrict to $\bfnu\in L/2$, if the characteristic ${\bfmu+\bfnu}$ is required to be in $L/2$.
\end{itemize}

\section{Class \texorpdfstring{$\mathcal{S}$}{S} representation}\label{app:class_S}
A different parametrisation of the SW curves \eqref{eq:curves} is the class $\CS$ representation. This representation gives the SW differential in a canonical form
\begin{equation}\label{classScurve}
	\lambda^2=p_{N_f}(z,u,\Lambda_{N_f},\bfm)dz^2,
\end{equation}
where the Laurent polynomials $p_{N_f}$ read \cite[Section 10]{Gaiotto:2009hg}
\begin{equation}\begin{aligned}
		p_0&=\frac{\Lambda_0^2}{z^3}+\frac{2u}{z^2}+\frac{\Lambda_0^2}{z},\\
		p_1&=\frac{\Lambda_1^2}{z^3}+\frac{3u}{z^2}+\frac{2\Lambda_1m}{z}+\Lambda_1^2,\\
		p_2&=\frac{\Lambda_2^2}{z^4}+\frac{2\Lambda_2 m_1}{z^3}+\frac{4u}{z^2}+\frac{2\Lambda_2m_2}{z}+\Lambda_2^2.
\end{aligned}\end{equation}
The corresponding elliptic curves can  be found as $x^2=z^4\,p_{N_f}$, which is quartic for $N_f=0,1$ and cubic for $N_f=2$. By comparing invariants of the SW curves \eqref{eq:curves} with those of the class $\CS$ curve \eqref{classScurve}, one finds the dictionary
\begin{equation}\begin{aligned}
		N_f=0: \quad& u_{\CS}=u_{\text{SW}}, \\
		N_f=1: \quad& u_{\CS}=\frac43 u_{\text{SW}}, \quad m_{\CS}=2m_{\text{SW}}, \\
		N_f=2: \quad& u_{\CS}=4 u_{\text{SW}}, \quad m_{\CS,i}=4m_{\text{SW},i}.
\end{aligned}\end{equation}
These relations merely amount to a rescaling of the parametrisation of the Coulomb branch, and in particular leave its geometry invariant. For this reason, we proceed above with using the SW curves \eqref{eq:curves}.

\section{Winding numbers}\label{sec:winding}
In this Appendix, we discuss the winding numbers appearing in \eqref{prepotential} in some more detail. As mentioned in Section \ref{sec:SWgeometry}, generally the theory admits $N_f$ electric winding numbers and $N_f$ magnetic winding numbers. While in the main text we set the electric winding numbers to zero, in this Appendix we keep the possibility open for them to be nonzero. 

The $\CN=2$ supersymmetry algebra requires that the central charge $Z$ is a linear combination of conserved charges. The U(1) conserved charges $S_i$ of the hypermultiplets must appear in $Z$ as follows \cite{Seiberg:1994aj},
\begin{equation}\label{centralcharge}
Z= n_m a_D+n_e a +\sum_{j=1}^{N_f} S_j \frac{m_j}{\sqrt2}.
\end{equation}
We note that  $Z$ is an inner product of periods $(a_D,a, \frac{1}{\sqrt2}\bfm)$ and conserved charges $(n_m, n_e,\boldsymbol S)$. 
The periods are given by contour integrals 
\begin{equation}\label{periods_lambda}
a=\int_{\gamma_1}\lambda, \qquad a_D=\int_{\gamma_2}\lambda,
\end{equation}
where $\gamma_1,\gamma_2$ generate $H_1(E,\mathbb Z)$, and $\lambda$ is a meromorphic 1-form, the SW differential. 
In massive SQCD, $\lambda$ has poles with nonzero residues. If the contours $\gamma_i$ are deformed across poles of $\lambda$, then 
$a$ and $a_D$ pick up  contributions from the residues. By invariance of \eqref{centralcharge}, these jumps are  linear combinations of the  masses, 
 \begin{equation}
\sum_{j=1}^{N_f} \Delta S_j \frac{m_j}{\sqrt2},
\end{equation}
i.e. the global charges of the hypermultiplets are shifted as $S_j\to S_j+\Delta S_j$. From the residue theorem, the  shift in the periods gives 
\begin{equation}
2\pi i \sum_{i=1}^{N_f}n_i\text{Res}(\lambda,x_i),
\end{equation}
where $x_i$ are the $N_f$ poles of $\lambda$, and $n_i\in\mathbb Z$ are the winding numbers of the contour deformation. Since both $a_D$ and $a$ are given by contour integrals of the same differential, a deformation of either $\gamma_1$ or $\gamma_2$ or both results in such a shift. \\

\emph{Example.} 
In order to see how $n_i$ and $\Delta S_j$ are related, let us study $N_f=2$ as an example. In $N_f=2$, the $S_j$ are integers for fundamental particles and half-integers for monopoles \cite{Seiberg:1994aj}.
The two poles of $\lambda$ are $x^\pm=\pm \frac18\Lambda_2^2$, with  residue \cite[(15.4)]{Seiberg:1994aj}
\begin{equation}\label{residuenf=2}
2\pi i\text{Res}(\lambda,x^\pm)=\pm \frac{m_1+m_2}{2\sqrt2}. 
\end{equation}
Then, 
 \begin{equation}
\sum_{j=1}^{2} \Delta S_j \frac{m_j}{\sqrt2}=2\pi i \sum_{i=1}^{2}n_i\text{Res}(\lambda,x_i)=\sum_{j=1}^2\frac{n_1-n_2}{2}\frac{m_j}{\sqrt2},
\end{equation}
such that $\Delta S_j=\frac{n_1-n_2}{2}\in\mathbb Z/2$ for both $j=1,2$. Constraints on the $n_i$ entail constraints on the change $\Delta S_j$. For instance, if both $n_1$ and $n_2$ are odd, or if both are even, then  $\Delta S_1$ and $\Delta S_2$ are integers. In the main text of this paper, we find such a constraint.

For generic $N_f\leq 3$, by construction of the SW curves the residues of $\lambda$ are linear combinations of the masses \begin{equation}
2\pi i\text{Res}(\lambda, x_i)=\sum_{j=1}^{N_f}l_{i j} \frac{m_j}{\sqrt2},
\end{equation}
with $i=1,\dots, N_f$, and $l_{ij}\in\mathbb Z/2$ \cite[(17.1)]{Seiberg:1994aj}.
Then we can compute
\begin{equation}
\sum_{j=1}^{N_f} \Delta S_j \frac{m_j}{\sqrt2}=2\pi i \sum_{i=1}^{N_f}n_i\text{Res}(\lambda,x_i)=\sum_{i=1}^{N_f}\sum_{j=1}^{N_f} n_i l_{i j} \frac{m_j}{\sqrt2}=\sum_{j=1}^{N_f} \frac{m_j}{\sqrt 2} \sum_{i=1}^{N_f}l_{i j}n_i.
\end{equation}
We  find that the change in the abelian global charges $\Delta S_j$ is given by a half-integral linear combination of winding numbers,
\begin{equation}
\Delta S_j=\sum_{i=1}^{N_f} l_{i j}n_i.
\end{equation}

In order to relate the winding numbers $n_i$ to those for the periods $a_i$ and $a_{D,i}$ as found in \cite{Ohta_1997}, we can expand 
\begin{equation}\label{aD_m_inf}
    {\bf M}_\infty^{\text{Ohta}}: \quad a_D\to -a_D+ (4-N_f)a-\frac{1}{\sqrt2}\left(\frac{n^{a_D}_{N_f}}{n^a_{N_f}}-\frac{4-N_f}{2}\right)\sum_{j=1}^{N_f}n_j^am_j,
\end{equation}
with $n_j^a$ being the coefficients of $-\frac{1}{2\sqrt2}m_j$ in the constant term of $a$, and similarly $n_j^{a_D}$ those for $a_D$ \cite[(5.2)]{Ohta_1997}\footnote{In \cite{Ohta_1997}, the $n_j^a$ are called $n_j$ and $n_j^{a_D}$ are called $n_j'$.}. We are aiming to compare this to  \eqref{m_inf},
\begin{equation}
    {\bf M}_\infty: \quad a_D\to -a_D+ (4-N_f)a-\frac{1}{\sqrt2}\sum_{j=1}^{N_f}n_jm_j.
\end{equation}
Since in \eqref{aD_m_inf} we divide and multiply by $n_j^a$, we cannot directly relate the two expressions by setting $n_j^a=0$. However, if we relate first 
\begin{equation}\label{winding_relations}
    n_j^a = \frac{n^a_{N_f}}{n^{a_D}_{N_f}} n_j^{a_D}, \qquad j=1,\dots N_f-1,
\end{equation}
then we find
\begin{equation}
    \begin{aligned}
    {\bf M}_\infty^{\text{Ohta}}: \quad a_D\to &-a_D+ (4-N_f)a-\frac{1}{\sqrt2}\sum_{j=1}^{N_f}\left(n_j^{a_D}-\frac{4-N_f}{2}n_j^a\right)m_j.
\end{aligned}
\end{equation}
In this case, it is well-defined to set $n_j^a=0$, such that the action of ${\bf M}_\infty^{\text{Ohta}}$ and $ {\bf M}_\infty$ coincide for $n_j=n_j^{a_D}$, as anticipated.

The $N_f-1$ conditions \eqref{winding_relations}
reproduce the constraints on the winding numbers in \cite[(5.3)]{Ohta_1997}: For $N_f=1$, the condition is empty. For $N_f=2$, it agrees precisely with Ohta. For $N_f=3$ finally, the $N_f-1=2$ equations \eqref{winding_relations} are equivalent to those found in \cite{Ohta_1997}. It seems however that our result $n_j^a =0, n_j^{a_D}=-1$ \eqref{u_plane_contraint} is merely one particular of the infinitely many  solutions to the geometric constraint \eqref{winding_relations}.\footnote{One other simple solution is $n_j^a=n$ and $n_j^{a_D}=\tilde n$, with arbitrary $n,\tilde n\in\mathbb Z$.} Thus it appears that the formulation of the theory on a compact four-manifold introduces further constraints. It would be interesting to understand whether it is possible to introduce non-vanishing electric winding numbers in the $u$-plane integral, and whether this leads to different correlation functions.

\providecommand{\href}[2]{#2}\begingroup\raggedright\endgroup

\begin{thebibliography}{100}
	
	\bibitem{Witten:1988ze}
	E.~Witten, \emph{{Topological Quantum Field Theory}},
	\href{http://dx.doi.org/10.1007/BF01223371}{\emph{Commun. Math. Phys.} {\bf
			117} (1988) 353}.
	
	\bibitem{Witten:1994ev}
	E.~Witten, \emph{{Supersymmetric Yang-Mills theory on a four manifold}},
	\href{http://dx.doi.org/10.1063/1.530745}{\emph{J. Math. Phys.} {\bf 35}
		(1994) 5101--5135}, [\href{https://arxiv.org/abs/hep-th/9403195}{{\tt
			hep-th/9403195}}].
	
	\bibitem{Witten:1994cg}
	E.~Witten, \emph{{Monopoles and four manifolds}},
	\href{http://dx.doi.org/10.4310/MRL.1994.v1.n6.a13}{\emph{Math. Res. Lett.}
		{\bf 1} (1994) 769--796}, [\href{https://arxiv.org/abs/hep-th/9411102}{{\tt
			hep-th/9411102}}].
	
	\bibitem{YAMRON1988325}
	J.~P. Yamron, \emph{Topological actions from twisted supersymmetric theories},
	\href{http://dx.doi.org/https://doi.org/10.1016/0370-2693(88)91769-8}{\emph{Physics
			Letters B} {\bf 213} (1988) 325 -- 330}.
	
	\bibitem{anselmi1993}
	D.~Anselmi and P.~Fr{\'e}, \emph{Topological twist in four dimensions,
		r-duality and hyperinstantons},
	\href{http://dx.doi.org/10.1016/0550-3213(93)90481-4}{\emph{Nuclear Physics
			B} {\bf 404} (08, 1993) 288--320}.
	
	\bibitem{anselmi1994}
	D.~Anselmi and P.~Fr{\'e}, \emph{Topological sigma-models in four dimensions
		and triholomorphic maps},
	\href{http://dx.doi.org/10.1016/0550-3213(94)90585-1}{\emph{Nuclear Physics
			B} {\bf 416} (03, 1994) 255--300}.
	
	\bibitem{ANSELMI1995247}
	D.~Anselmi and P.~Fr{\'e}, \emph{Gauged hyperinstantons and monopole
		equations},
	\href{http://dx.doi.org/https://doi.org/10.1016/0370-2693(95)00033-H}{\emph{Physics
			Letters B} {\bf 347} (1995) 247 -- 254}.
	
	\bibitem{ALVAREZ1993251}
	M.~Alvarez and J.~Labastida, \emph{Breaking of topological symmetry},
	\href{http://dx.doi.org/https://doi.org/10.1016/0370-2693(93)91609-Q}{\emph{Physics
			Letters B} {\bf 315} (1993) 251 -- 257}.
	
	\bibitem{Alvarez:1994ii}
	M.~Alvarez and J.~Labastida, \emph{{Topological matter in four-dimensions}},
	\href{http://dx.doi.org/10.1016/0550-3213(94)00512-D}{\emph{Nucl. Phys. B}
		{\bf 437} (1995) 356--390}, [\href{https://arxiv.org/abs/hep-th/9404115}{{\tt
			hep-th/9404115}}].
	
	\bibitem{Seiberg:1994rs}
	N.~Seiberg and E.~Witten, \emph{{Electric - magnetic duality, monopole
			condensation, and confinement in N=2 supersymmetric Yang-Mills theory}},
	\href{http://dx.doi.org/10.1016/0550-3213(94)90124-4,
		10.1016/0550-3213(94)00449-8}{\emph{Nucl. Phys.} {\bf B426} (1994) 19--52},
	[\href{https://arxiv.org/abs/hep-th/9407087}{{\tt hep-th/9407087}}].
	
	\bibitem{Seiberg:1994aj}
	N.~Seiberg and E.~Witten, \emph{{Monopoles, duality and chiral symmetry
			breaking in N=2 supersymmetric QCD}},
	\href{http://dx.doi.org/10.1016/0550-3213(94)90214-3}{\emph{Nucl. Phys.} {\bf
			B431} (1994) 484--550}, [\href{https://arxiv.org/abs/hep-th/9408099}{{\tt
			hep-th/9408099}}].
	
	\bibitem{Moore:1997pc}
	G.~W. Moore and E.~Witten, \emph{{Integration over the $u$-plane in Donaldson
			theory}}, {\emph{Adv. Theor. Math. Phys.} {\bf 1} (1997) 298--387},
	[\href{https://arxiv.org/abs/hep-th/9709193}{{\tt hep-th/9709193}}].
	
	\bibitem{LoNeSha}
	A.~Losev, N.~Nekrasov and S.~L. Shatashvili, \emph{{Issues in topological gauge
			theory}}, \href{http://dx.doi.org/10.1016/S0550-3213(98)00628-2}{\emph{Nucl.
			Phys.} {\bf B534} (1998) 549--611},
	[\href{https://arxiv.org/abs/hep-th/9711108}{{\tt hep-th/9711108}}].
	
	\bibitem{labastida1995}
	J.~Labastida and M.~Marino, \emph{A topological lagrangian for monopoles on
		four-manifolds},
	\href{http://dx.doi.org/10.1016/0370-2693(95)00411-D}{\emph{Physics Letters
			B} {\bf 351} (03, 1995) 146--152}.
	
	\bibitem{Labastida:1995zj}
	J.~M.~F. Labastida and M.~Marino, \emph{{NonAbelian monopoles on four
			manifolds}},
	\href{http://dx.doi.org/10.1016/0550-3213(95)00300-H}{\emph{Nucl. Phys. B}
		{\bf 448} (1995) 373--398}, [\href{https://arxiv.org/abs/hep-th/9504010}{{\tt
			hep-th/9504010}}].
	
	\bibitem{Hyun:1995mb}
	S.~Hyun, J.~Park and J.-S. Park, \emph{{Spin-c Topological QCD}},
	\href{http://dx.doi.org/10.1016/0550-3213(95)00404-G}{\emph{Nucl. Phys.} {\bf
			B453} (1995) 199--224}, [\href{https://arxiv.org/abs/hep-th/9503201}{{\tt
			hep-th/9503201}}].
	
	\bibitem{Hyun:1995hz}
	S.~Hyun, J.~Park and J.-S. Park, \emph{{N=2 supersymmetric QCD and four
			manifolds: 1. The Donaldson and Seiberg-Witten invariants}},
	\href{https://arxiv.org/abs/hep-th/9508162}{{\tt hep-th/9508162}}.
	
	\bibitem{laba1998}
	J.~Labastida and C.~Lozano, \emph{Mass perturbations in twisted n = 4
		supersymmetric gauge theories},
	\href{http://dx.doi.org/10.1016/S0550-3213(98)00135-7}{\emph{Nuclear Physics
			B} {\bf 518} (05, 1998) 37--58}.
	
	\bibitem{Dijkgraaf:1997ce}
	R.~Dijkgraaf, J.-S. Park and B.~J. Schroers, \emph{{N=4 supersymmetric
			Yang-Mills theory on a Kahler surface}},
	\href{https://arxiv.org/abs/hep-th/9801066}{{\tt hep-th/9801066}}.
	
	\bibitem{Kanno:1998qj}
	H.~Kanno and S.-K. Yang, \emph{{Donaldson-Witten Functions of Massless N=2
			Supersymmetric QCD}},
	\href{http://dx.doi.org/10.1016/S0550-3213(98)00560-4}{\emph{Nucl. Phys. B}
		{\bf 535} (1998) 512--530}, [\href{https://arxiv.org/abs/hep-th/9806015}{{\tt
			hep-th/9806015}}].
	
	\bibitem{Marino:1998uy}
	M.~Marino, G.~W. Moore and G.~Peradze, \emph{{Superconformal invariance and the
			geography of four manifolds}},
	\href{http://dx.doi.org/10.1007/s002200050694}{\emph{Commun. Math. Phys.}
		{\bf 205} (1999) 691--735}, [\href{https://arxiv.org/abs/hep-th/9812055}{{\tt
			hep-th/9812055}}].
	
	\bibitem{mm1998}
	M.~Marino and G.~Moore, \emph{Integrating over the coulomb branch in n = 2
		gauge theory},
	\href{http://dx.doi.org/https://doi.org/10.1016/S0920-5632(98)00168-6}{\emph{Nuclear
			Physics B - Proceedings Supplements} {\bf 68} (1998) 336 -- 347}.
	
	\bibitem{Moore:2017cmm}
	G.~W. Moore and I.~Nidaiev, \emph{{The Partition Function Of Argyres-Douglas
			Theory On A Four-Manifold}},  \href{https://arxiv.org/abs/1711.09257}{{\tt
			1711.09257}}.
	
	\bibitem{Dedushenko:2017tdw}
	M.~Dedushenko, S.~Gukov and P.~Putrov, \emph{{Vertex algebras and 4-manifold
			invariants}},  in \emph{{Nigel Hitchin's 70th Birthday Conference}}, vol.~1,
	pp.~249--318, 5, 2017.
	\newblock \href{https://arxiv.org/abs/1705.01645}{{\tt 1705.01645}}.
	
	\bibitem{Manschot:2021qqe}
	J.~Manschot and G.~W. Moore, \emph{{Topological correlators of $SU(2)$,
			$\mathcal{N}=2^*$ SYM on four-manifolds}},
	\href{https://arxiv.org/abs/2104.06492}{{\tt 2104.06492}}.
	
	\bibitem{Edelstein:2000wg}
	J.~D. Edelstein, M.~Gomez-Reino and M.~Marino, \emph{{Remarks on twisted
			theories with matter}},
	\href{http://dx.doi.org/10.1088/1126-6708/2001/01/004}{\emph{JHEP} {\bf 01}
		(2001) 004}, [\href{https://arxiv.org/abs/hep-th/0011227}{{\tt
			hep-th/0011227}}].
	
	\bibitem{gttsche2010}
	L.~G{\"o}ttsche, H.~Nakajima and K.~Yoshioka, \emph{{Donaldson = Seiberg-Witten
			from Mochizuki's formula and instanton counting}},
	\href{https://arxiv.org/abs/1001.5024}{{\tt 1001.5024}}.
	
	\bibitem{Feehan:1997gj}
	P.~M.~N. Feehan and T.~G. Leness, \emph{{$\rm PU(2)$ monopoles. I. Regularity,
			Uhlenbeck compactness, and transversality}}, {\emph{J. Diff. Geom.} {\bf 49}
		(1998) 265--410}, [\href{https://arxiv.org/abs/dg-ga/9710032}{{\tt
			dg-ga/9710032}}].
	
	\bibitem{Feehan:1997rt}
	P.~M.~N. Feehan and T.~G. Leness, \emph{{PU(2) Monopoles and Relations between
			Four-Manifold Invariants}},
	\href{http://dx.doi.org/10.1016/S0166-8641(97)00201-0}{\emph{Topology and its
			Applications} {\bf 88} (1998) 111--145},
	[\href{https://arxiv.org/abs/dg-ga/9709022}{{\tt dg-ga/9709022}}].
	
	\bibitem{Feehan:1997vp}
	P.~M.~N. Feehan and T.~G. Leness, \emph{{PU(2) monopoles. 2. Top level
			Seiberg-Witten moduli spaces and Witten's conjecture in low degrees}},
	\href{https://arxiv.org/abs/dg-ga/9712005}{{\tt dg-ga/9712005}}.
	
	\bibitem{Feehan:2001jc}
	P.~M.~N. Feehan and T.~G. Leness, \emph{{SO(3) monopoles, level one
			Seiberg-Witten moduli spaces, and Witten's conjecture in low degrees}},
	\href{http://dx.doi.org/10.1016/S0166-8641(01)00233-4}{\emph{Topology Appl.}
		{\bf 124} (2002) 221--326}, [\href{https://arxiv.org/abs/math/0106238}{{\tt
			math/0106238}}].
	
	\bibitem{Gorodentsev:1996cu}
	A.~L. Gorodentsev and M.~I. Leenson, \emph{{How to calculate the correlation
			function in twisted SYM N=2, $N_f=4$ QFT on projective plane}},
	\href{https://arxiv.org/abs/alg-geom/9604011}{{\tt alg-geom/9604011}}.
	
	\bibitem{Nakajima:2005fg}
	H.~Nakajima and K.~Yoshioka, \emph{{Instanton counting on blowup. II.
			K-theoretic partition function}}, {\emph{Transformation groups} {\bf 10}
		(2005) 489--519}.
	
	\bibitem{nakajima2011perverse}
	H.~Nakajima and K.~Yoshioka, \emph{Perverse coherent sheaves on blowup, iii:
		Blow-up formula from wall-crossing}, {\emph{Kyoto Journal of Mathematics}
		{\bf 51} (2011) 263--335}.
	
	\bibitem{marcolli_2011}
	M.~Marcolli, \emph{Seiberg-Witten gauge theory}.
	\newblock Hindustan Book Agency, 2011.
	
	\bibitem{bryan1996}
	J.~Bryan and R.~Wentworth, \emph{{The multi-monopole equations for K{\"a}hler
			surfaces}}, {\emph{Turkish J. Math.} {\bf 20} (1996) 119--128}.
	
	\bibitem{pidstrigach1995localisation}
	V.~Pidstrigach and A.~Tyurin, \emph{Localisation of the donaldson's invariants
		along seiberg-witten classes}, {\emph{arXiv preprint dg-ga/9507004} (1995) }.
	
	\bibitem{okonek1996recent}
	C.~Okonek and A.~Teleman, \emph{Recent developments in seiberg-witten theory
		and complex geometry}, {\emph{arXiv preprint alg-geom/9612015} (1996) }.
	
	\bibitem{Laba05}
	J.~Labastida and M.~Marino, \emph{Topological quantum field theory and four
		manifolds}.
	\newblock Springer Netherlands, 2005.
	
	\bibitem{Marino:2000hx}
	M.~Marino, \emph{{Topological quantum field theory and four manifolds}},  in
	\emph{{3rd European Congress of Mathematics: Shaping the 21st Century}}, 8,
	2000.
	\newblock \href{https://arxiv.org/abs/hep-th/0008100}{{\tt hep-th/0008100}}.
	
	\bibitem{Nakajima:2003uh}
	H.~Nakajima and K.~Yoshioka, \emph{{Lectures on instanton counting}},  in
	\emph{{CRM Workshop on Algebraic Structures and Moduli Spaces Montreal,
			Canada, July 14-20, 2003}}, 2003.
	\newblock \href{https://arxiv.org/abs/math/0311058}{{\tt math/0311058}}.
	
	\bibitem{moore2017}
	G.~Moore, ``Lectures on the physical approach to donaldson and seiberg-witten
	invariants of four-manifolds.'' URL:
	\url{https://www.physics.rutgers.edu/~gmoore/SCGP-FourManifoldsNotes-2017.pdf},
	2017.
	
	\bibitem{Taubes1994}
	C.~H. Taubes, \emph{{The Seiberg-Witten invariants and symplectic forms}},
	\href{http://dx.doi.org/10.4310/mrl.1994.v1.n6.a15}{\emph{Mathematical
			Research Letters} {\bf 1} (1994) 809--822}.
	
	\bibitem{Bershadsky:1995vm}
	M.~Bershadsky, A.~Johansen, V.~Sadov and C.~Vafa, \emph{{Topological reduction
			of 4D SYM to 2D $\sigma$-models}},
	\href{http://dx.doi.org/10.1016/0550-3213(95)00242-K}{\emph{Nucl. Phys. B}
		{\bf 448} (1995) 166--186}, [\href{https://arxiv.org/abs/hep-th/9501096}{{\tt
			hep-th/9501096}}].
	
	\bibitem{Gottsche:2006}
	L.~Göttsche, H.~Nakajima and K.~Yoshioka, \emph{{K-theoretic Donaldson
			Invariants Via Instanton Counting}},
	\href{http://dx.doi.org/10.4310/PAMQ.2009.v5.n3.a5}{\emph{Pure and Applied
			Mathematics Quarterly} {\bf 5} (12, 2006) }.
	
	\bibitem{Harvey_1995}
	J.~A. Harvey, G.~W. Moore and A.~Strominger, \emph{{Reducing S-duality to
			T-duality}}, \href{http://dx.doi.org/10.1103/PhysRevD.52.7161}{\emph{Phys.
			Rev. D} {\bf 52} (1995) 7161--7167},
	[\href{https://arxiv.org/abs/hep-th/9501022}{{\tt hep-th/9501022}}].
	
	\bibitem{donaldson1995floer}
	S.~K. Donaldson, \emph{Floer homology and algebraic geometry}, pp.~119--138.
	\newblock London Mathematical Society Lecture Note Series.
	\newblock Cambridge University Press, 1995.
	
	\bibitem{kim2023}
	H.~Kim, J.~Manschot, G.~W. Moore, R.~Tao and X.~Zhang, \emph{{Path Integral
			Derivations of K-Theoretic Donaldson Invariants}},
	\href{https://arxiv.org/abs/To appear}{{\tt To appear}}.
	
	\bibitem{Gadde:2013sca}
	A.~Gadde, S.~Gukov and P.~Putrov, \emph{{Fivebranes and 4-manifolds}},
	\href{http://dx.doi.org/10.1007/978-3-319-43648-7_7}{\emph{Prog. Math.} {\bf
			319} (2016) 155--245}, [\href{https://arxiv.org/abs/1306.4320}{{\tt
			1306.4320}}].
	
	\bibitem{Malmendier:2008db}
	A.~Malmendier and K.~Ono, \emph{{SO(3)-Donaldson invariants of $\mathbb{P}^2$
			and Mock Theta Functions}},
	\href{http://dx.doi.org/10.2140/gt.2012.16.1767}{\emph{Geom. Topol.} {\bf 16}
		(2012) 1767--1833}, [\href{https://arxiv.org/abs/0808.1442}{{\tt
			0808.1442}}].
	
	\bibitem{Griffin:2012kw}
	M.~Griffin, A.~Malmendier and K.~Ono, \emph{{SU(2)-Donaldson invariants of the
			complex projective plane}},
	\href{http://dx.doi.org/10.1515/forum-2013-6013}{\emph{Forum Math.} {\bf 27}
		(2015) 2003--2023}, [\href{https://arxiv.org/abs/1209.2743}{{\tt
			1209.2743}}].
	
	\bibitem{Malmendier:2012zz}
	A.~{Malmendier} and K.~{Ono}, \emph{{Moonshine and Donaldson invariants of
			$\mathbb {CP}^2$}},  \href{https://arxiv.org/abs/1207.5139}{{\tt 1207.5139}}.
	
	\bibitem{Malmendier:2010ss}
	A.~Malmendier, \emph{{Donaldson invariants of $\mathbb{P}^1 \times
			\mathbb{P}^1$ and Mock Theta Functions}},
	\href{http://dx.doi.org/10.4310/CNTP.2011.v5.n1.a5}{\emph{Commun. Num. Theor.
			Phys.} {\bf 5} (2011) 203--229}, [\href{https://arxiv.org/abs/1008.0175}{{\tt
			1008.0175}}].
	
	\bibitem{Korpas:2017qdo}
	G.~Korpas and J.~Manschot, \emph{{Donaldson-Witten theory and indefinite theta
			functions}}, \href{http://dx.doi.org/10.1007/JHEP11(2017)083}{\emph{JHEP}
		{\bf 11} (2017) 083}, [\href{https://arxiv.org/abs/1707.06235}{{\tt
			1707.06235}}].
	
	\bibitem{Korpas:2019ava}
	G.~Korpas, J.~Manschot, G.~Moore and I.~Nidaiev, \emph{{Renormalization and
			BRST Symmetry in Donaldson--Witten Theory}},
	\href{http://dx.doi.org/10.1007/s00023-019-00835-x}{\emph{Annales Henri
			Poincare} {\bf 20} (2019) 3229--3264},
	[\href{https://arxiv.org/abs/1901.03540}{{\tt 1901.03540}}].
	
	\bibitem{Korpas:2019cwg}
	G.~Korpas, J.~Manschot, G.~W. Moore and I.~Nidaiev, \emph{Mocking the u-plane
		integral}, \href{http://dx.doi.org/10.1007/s40687-021-00280-5}{\emph{Research
			in the Mathematical Sciences} {\bf 8} (2021) 43}.
	
	\bibitem{Aspman:2021kfp}
	J.~Aspman, E.~Furrer, G.~Korpas, Z.-C. Ong and M.-C. Tan, \emph{{The $u$-plane
			integral, mock modularity and enumerative geometry}},
	\href{http://dx.doi.org/10.1007/s11005-022-01520-7}{\emph{Lett. Math. Phys.}
		{\bf 112} (2022) 30}, [\href{https://arxiv.org/abs/2109.04302}{{\tt
			2109.04302}}].
	
	\bibitem{Korpas:2022tij}
	G.~Korpas, \emph{{Mock modularity and surface defects in topological $\mathcal
			{N}$ =2 super Yang-Mills theory}},
	\href{http://dx.doi.org/10.1103/PhysRevD.105.026025}{\emph{Phys. Rev. D} {\bf
			105} (2022) 026025}.
	
	\bibitem{Aspman:2021vhs}
	J.~Aspman, E.~Furrer and J.~Manschot, \emph{{Cutting and gluing with running
			couplings in $\mathcal{N}=2$ QCD}},
	\href{http://dx.doi.org/10.1103/PhysRevD.105.025021}{\emph{Phys. Rev. D} {\bf
			105} (2022) 025021}, [\href{https://arxiv.org/abs/2107.04600}{{\tt
			2107.04600}}].
	
	\bibitem{Aspman:2020lmf}
	J.~Aspman, E.~Furrer and J.~Manschot, \emph{{Elliptic Loci of SU(3) Vacua}},
	\href{http://dx.doi.org/10.1007/s00023-021-01040-5}{\emph{Annales Henri
			Poincare} {\bf 22} (2021) 2775--2830},
	[\href{https://arxiv.org/abs/2010.06598}{{\tt 2010.06598}}].
	
	\bibitem{Aspman:2021evt}
	J.~Aspman, E.~Furrer and J.~Manschot, \emph{{Four flavours, triality, and
			bimodular forms}},
	\href{http://dx.doi.org/10.1103/PhysRevD.105.025017}{\emph{Phys. Rev. D} {\bf
			105} (2022) 025017}, [\href{https://arxiv.org/abs/2110.11969}{{\tt
			2110.11969}}].
	
	\bibitem{Shapere:2008zf}
	A.~D. Shapere and Y.~Tachikawa, \emph{{Central charges of N=2 superconformal
			field theories in four dimensions}},
	\href{http://dx.doi.org/10.1088/1126-6708/2008/09/109}{\emph{JHEP} {\bf 09}
		(2008) 109}, [\href{https://arxiv.org/abs/0804.1957}{{\tt 0804.1957}}].
	
	\bibitem{Ohta_1997}
	Y.~Ohta, \emph{{Prepotentials of $N=2$ $SU(2)$ Yang--Mills Theories Coupled
			with Massive Matter multiplets}},
	\href{http://dx.doi.org/10.1063/1.531858}{\emph{Journal of Mathematical
			Physics} {\bf 38} (Feb, 1997) 682--696}.
	
	\bibitem{Argyres:1995jj}
	P.~C. Argyres and M.~R. Douglas, \emph{{New phenomena in $SU(3)$ supersymmetric
			gauge theory}},
	\href{http://dx.doi.org/10.1016/0550-3213(95)00281-V}{\emph{Nucl. Phys.} {\bf
			B448} (1995) 93--126}, [\href{https://arxiv.org/abs/hep-th/9505062}{{\tt
			hep-th/9505062}}].
	
	\bibitem{Argyres:1995xn}
	P.~C. Argyres, M.~R. Plesser, N.~Seiberg and E.~Witten, \emph{{New $\mathcal
			N=2$ superconformal field theories in four-dimensions}},
	\href{http://dx.doi.org/10.1016/0550-3213(95)00671-0}{\emph{Nucl. Phys.} {\bf
			B461} (1996) 71--84}, [\href{https://arxiv.org/abs/hep-th/9511154}{{\tt
			hep-th/9511154}}].
	
	\bibitem{Klemm:1997gg}
	A.~Klemm, \emph{{On the geometry behind N=2 supersymmetric effective actions in
			four-dimensions}},  in \emph{{33rd Karpacz Winter School of Theoretical
			Physics: Duality - Strings and Fields}}, 5, 1997.
	\newblock \href{https://arxiv.org/abs/hep-th/9705131}{{\tt hep-th/9705131}}.
	
	\bibitem{Ohta:1996hq}
	Y.~Ohta, \emph{{Prepotential of $N=2$ $SU(2)$ Yang-Mills Gauge Theory Coupled
			with a Massive Matter Multiplet}},
	\href{http://dx.doi.org/10.1063/1.531764}{\emph{J. Math. Phys.} {\bf 37}
		(1996) 6074--6085}, [\href{https://arxiv.org/abs/hep-th/9604051}{{\tt
			hep-th/9604051}}].
	
	\bibitem{DHoker_1997}
	E.~D'Hoker, I.~Krichever and D.~Phong, \emph{The effective prepotential of n =
		2 supersymmetric su(nc) gauge theories},
	\href{http://dx.doi.org/10.1016/s0550-3213(97)00035-7}{\emph{Nuclear Physics
			B} {\bf 489} (Mar, 1997) 179--210}.
	
	\bibitem{Ne}
	N.~A. Nekrasov, \emph{{Seiberg-Witten prepotential from instanton counting}},
	\href{http://dx.doi.org/10.4310/ATMP.2003.v7.n5.a4}{\emph{Adv. Theor. Math.
			Phys.} {\bf 7} (2003) 831--864},
	[\href{https://arxiv.org/abs/hep-th/0206161}{{\tt hep-th/0206161}}].
	
	\bibitem{ohta1999}
	Y.~Ohta, \emph{Non-perturbative solutions to n=2 supersymmetric yang-mills
		theories -progress and perspective-},  1999.
	
	\bibitem{NekOk}
	N.~Nekrasov and A.~Okounkov, \emph{{Seiberg-Witten theory and random
			partitions}}, \href{http://dx.doi.org/10.1007/0-8176-4467-9_15}{\emph{Prog.
			Math.} {\bf 244} (2006) 525--596},
	[\href{https://arxiv.org/abs/hep-th/0306238}{{\tt hep-th/0306238}}].
	
	\bibitem{Sonnenschein_1996}
	J.~Sonnenschein, S.~Theisen and S.~Yankielowicz, \emph{On the relation between
		the holomorphic prepotential and the quantum moduli in susy gauge theories},
	\href{http://dx.doi.org/10.1016/0370-2693(95)01399-7}{\emph{Physics Letters
			B} {\bf 367} (Jan, 1996) 145--150}.
	
	\bibitem{Eguchi:1995jh}
	T.~Eguchi and S.-K. Yang, \emph{{Prepotentials of N=2 supersymmetric gauge
			theories and soliton equations}},
	\href{http://dx.doi.org/10.1142/S0217732396000151}{\emph{Mod. Phys. Lett.}
		{\bf A11} (1996) 131--138}, [\href{https://arxiv.org/abs/hep-th/9510183}{{\tt
			hep-th/9510183}}].
	
	\bibitem{DHoker:1996yyu}
	E.~D'Hoker, I.~Krichever and D.~Phong, \emph{{The Renormalization group
			equation in N=2 supersymmetric gauge theories}},
	\href{http://dx.doi.org/10.1016/S0550-3213(97)00156-9}{\emph{Nucl. Phys. B}
		{\bf 494} (1997) 89--104}, [\href{https://arxiv.org/abs/hep-th/9610156}{{\tt
			hep-th/9610156}}].
	
	\bibitem{Malmendier:2008yj}
	A.~Malmendier, \emph{{The Signature of the Seiberg-Witten surface}},
	\href{http://dx.doi.org/10.4310/SDG.2010.v15.n1.a8}{\emph{Surveys Diff.
			Geom.} {\bf 15} (2010) 255--278},
	[\href{https://arxiv.org/abs/0802.1363}{{\tt 0802.1363}}].
	
	\bibitem{Caorsi:2018ahl}
	M.~Caorsi and S.~Cecotti, \emph{{Special Arithmetic of Flavor}},
	\href{http://dx.doi.org/10.1007/JHEP08(2018)057}{\emph{JHEP} {\bf 08} (2018)
		057}, [\href{https://arxiv.org/abs/1803.00531}{{\tt 1803.00531}}].
	
	\bibitem{Closset:2021lhd}
	C.~Closset and H.~Magureanu, \emph{{The $U$-plane of rank-one 4d
			$\mathcal{N}=2$ KK theories}},
	\href{http://dx.doi.org/10.21468/SciPostPhys.12.2.065}{\emph{SciPost Phys.}
		{\bf 12} (2022) 065}, [\href{https://arxiv.org/abs/2107.03509}{{\tt
			2107.03509}}].
	
	\bibitem{shioda1972}
	T.~Shioda, \emph{On elliptic modular surfaces},
	\href{http://dx.doi.org/10.2969/jmsj/02410020}{\emph{J. Math. Soc. Japan}
		{\bf 24} (01, 1972) 20--59}.
	
	\bibitem{schuett2009}
	M.~Schuett and T.~Shioda, \emph{Elliptic surfaces},  2009.
	
	\bibitem{maier2006}
	R.~S. {Maier}, \emph{{On Rationally Parametrized Modular Equations}},
	{\emph{arXiv Mathematics e-prints} (Nov., 2006) math/0611041},
	[\href{https://arxiv.org/abs/math/0611041}{{\tt math/0611041}}].
	
	\bibitem{miranda1995}
	R.~Miranda, \emph{An overview of algebraic surfaces}, {\emph{Lecture Notes in
			Pure and Appl. Math.} (1997) 157--217}.
	
	\bibitem{Gaiotto:2009hg}
	D.~Gaiotto, G.~W. Moore and A.~Neitzke, \emph{{Wall-crossing, Hitchin Systems,
			and the WKB Approximation}},  \href{https://arxiv.org/abs/0907.3987}{{\tt
			0907.3987}}.
	
	\bibitem{Eguchi1999}
	T.~Eguchi, \emph{Seiberg-Witten Theory and S-Duality}, pp.~103--120.
	\newblock Springer Netherlands, Dordrecht, 1999.
	
	\bibitem{EGUCHI1996430}
	T.~Eguchi, K.~Hori, K.~Ito and S.-K. Yang, \emph{{Study of N = 2 Superconformal
			Field Theories in 4 Dimensions}},
	\href{http://dx.doi.org/https://doi.org/10.1016/0550-3213(96)00188-5}{\emph{Nuclear
			Physics B} {\bf 471} (1996) 430 -- 442}.
	
	\bibitem{Persson:1990}
	U.~Persson, \emph{{Configurations of Kodaira fibers on rational elliptic
			surfaces}}, {\emph{Mathematische Zeitschrift} {\bf 205} (1990) 1--47}.
	
	\bibitem{Miranda:1990}
	R.~Miranda, \emph{Persson's list of singular fibers for a rational elliptic
		surface}, {\emph{Mathematische Zeitschrift} {\bf 205} (1990) 191--211}.
	
	\bibitem{Nahm:1996di}
	W.~Nahm, \emph{{On the Seiberg-Witten approach to electric - magnetic
			duality}},  \href{https://arxiv.org/abs/hep-th/9608121}{{\tt
			hep-th/9608121}}.
	
	\bibitem{Brandhuber:1996ng}
	A.~Brandhuber and S.~Stieberger, \emph{{Periods, Coupling Constants and Modular
			Functions in N=2 SU(2) SYM with Massive Matter}},
	\href{http://dx.doi.org/10.1142/S0217751X98000627}{\emph{Int.\ J.\ Mod.\
			Phys.\ A} {\bf 13} (1998) 1329--1344},
	[\href{https://arxiv.org/abs/hep-th/9609130}{{\tt hep-th/9609130}}].
	
	\bibitem{Huang:2009md}
	M.-x. Huang and A.~Klemm, \emph{{Holomorphicity and Modularity in
			Seiberg-Witten Theories with Matter}},
	\href{http://dx.doi.org/10.1007/JHEP07(2010)083}{\emph{JHEP} {\bf 07} (2010)
		083}, [\href{https://arxiv.org/abs/0902.1325}{{\tt 0902.1325}}].
	
	\bibitem{Matone:1995rx}
	M.~Matone, \emph{{Instantons and recursion relations in N=2 SUSY gauge
			theory}}, \href{http://dx.doi.org/10.1016/0370-2693(95)00920-G}{\emph{Phys.
			Lett.} {\bf B357} (1995) 342--348},
	[\href{https://arxiv.org/abs/hep-th/9506102}{{\tt hep-th/9506102}}].
	
	\bibitem{stiller1981}
	P.~F. Stiller, \emph{{Differential equations associated with elliptic
			surfaces}}, \href{http://dx.doi.org/10.2969/jmsj/03320203}{\emph{Journal of
			the Mathematical Society of Japan} {\bf 33} (1981) 203 -- 233}.
	
	\bibitem{matone1996}
	M.~Matone, \emph{Koebe 1/4 theorem and inequalities in n=2 supersymmetric qcd},
	\href{http://dx.doi.org/10.1103/PhysRevD.53.7354}{\emph{Phys. Rev. D} {\bf
			53} (Jun, 1996) 7354--7358}.
	
	\bibitem{Klemm:2012sx}
	A.~Klemm, J.~Manschot and T.~Wotschke, \emph{{Quantum geometry of elliptic
			Calabi-Yau manifolds}},  \href{https://arxiv.org/abs/1205.1795}{{\tt
			1205.1795}}.
	
	\bibitem{Magureanu:2022qym}
	H.~Magureanu, \emph{{Seiberg-Witten geometry, modular rational elliptic
			surfaces and BPS quivers}},
	\href{http://dx.doi.org/10.1007/JHEP05(2022)163}{\emph{JHEP} {\bf 05} (2022)
		163}, [\href{https://arxiv.org/abs/2203.03755}{{\tt 2203.03755}}].
	
	\bibitem{Donaldson90}
	S.~K. Donaldson and P.~B. Kronheimer, \emph{{The Geometry of Four-Manifolds}}.
	\newblock Clarendon Press ; Oxford University Press Oxford : New York, 1990.
	
	\bibitem{Aspman:2023ate}
	J.~Aspman, E.~Furrer and J.~Manschot, \emph{{Topological twists of massive
			SQCD, Part II}},  \href{https://arxiv.org/abs/2312.11616}{{\tt 2312.11616}}.
	
	\bibitem{Moore:1997dj}
	G.~W. Moore, N.~Nekrasov and S.~Shatashvili, \emph{{Integrating over Higgs
			branches}}, \href{http://dx.doi.org/10.1007/PL00005525}{\emph{Commun. Math.
			Phys.} {\bf 209} (2000) 97--121},
	[\href{https://arxiv.org/abs/hep-th/9712241}{{\tt hep-th/9712241}}].
	
	\bibitem{Mathai:1986tc}
	V.~Mathai and D.~G. Quillen, \emph{{Superconnections, Thom classes and
			equivariant differential forms}},
	\href{http://dx.doi.org/10.1016/0040-9383(86)90007-8}{\emph{Topology} {\bf
			25} (1986) 85--110}.
	
	\bibitem{Atiyah:1990tm}
	M.~Atiyah and L.~Jeffrey, \emph{{Topological Lagrangians and cohomology}},
	\href{http://dx.doi.org/10.1016/0393-0440(90)90023-V}{\emph{J.\ Geom.\ Phys.}
		{\bf 7} (1990) 119--136}.
	
	\bibitem{losev1998}
	A.~Losev, N.~Nekrasov and S.~Shatashvili, \emph{Testing seiberg-witten
		solution},  1998.
	
	\bibitem{Cordes:1994fc}
	S.~Cordes, G.~W. Moore and S.~Ramgoolam, \emph{{Lectures on 2-d Yang-Mills
			theory, equivariant cohomology and topological field theories}},
	\href{http://dx.doi.org/10.1016/0920-5632(95)00434-B}{\emph{Nucl. Phys. B
			Proc. Suppl.} {\bf 41} (1995) 184--244},
	[\href{https://arxiv.org/abs/hep-th/9411210}{{\tt hep-th/9411210}}].
	
	\bibitem{Vafa:1994tf}
	C.~Vafa and E.~Witten, \emph{{A Strong coupling test of S duality}},
	\href{http://dx.doi.org/10.1016/0550-3213(94)90097-3}{\emph{Nucl. Phys.} {\bf
			B431} (1994) 3--77}, [\href{https://arxiv.org/abs/hep-th/9408074}{{\tt
			hep-th/9408074}}].
	
	\bibitem{Witten:1995gf}
	E.~Witten, \emph{{On S duality in Abelian gauge theory}},
	\href{http://dx.doi.org/10.1007/BF01671570}{\emph{Selecta Math.} {\bf 1}
		(1995) 383}, [\href{https://arxiv.org/abs/hep-th/9505186}{{\tt
			hep-th/9505186}}].
	
	\bibitem{Nelson:1993nf}
	A.~E. Nelson and N.~Seiberg, \emph{{R symmetry breaking versus supersymmetry
			breaking}}, \href{http://dx.doi.org/10.1016/0550-3213(94)90577-0}{\emph{Nucl.
			Phys. B} {\bf 416} (1994) 46--62},
	[\href{https://arxiv.org/abs/hep-ph/9309299}{{\tt hep-ph/9309299}}].
	
	\bibitem{Marino:1998bm}
	M.~Marino and G.~W. Moore, \emph{{The Donaldson-Witten function for gauge
			groups of rank larger than one}},
	\href{http://dx.doi.org/10.1007/s002200050494}{\emph{Commun. Math. Phys.}
		{\bf 199} (1998) 25--69}, [\href{https://arxiv.org/abs/hep-th/9802185}{{\tt
			hep-th/9802185}}].
	
	\bibitem{Cordova:2018acb}
	C.~C\'ordova and T.~T. Dumitrescu, \emph{{Candidate Phases for SU(2) Adjoint
			QCD$_4$ with Two Flavors from $\mathcal{N}=2$ Supersymmetric Yang-Mills
			Theory}},  \href{https://arxiv.org/abs/1806.09592}{{\tt 1806.09592}}.
	
	\bibitem{Manschot:2019pog}
	J.~Manschot, G.~W. Moore and X.~Zhang, \emph{{Effective gravitational couplings
			of four-dimensional $ \mathcal{N} $ = 2 supersymmetric gauge theories}},
	\href{http://dx.doi.org/10.1007/JHEP06(2020)150}{\emph{JHEP} {\bf 06} (2020)
		150}, [\href{https://arxiv.org/abs/1912.04091}{{\tt 1912.04091}}].
	
	\bibitem{Mari_o_1999}
	M.~Mari{\~n}o, \emph{The uses of whitham hierarchies},
	\href{http://dx.doi.org/10.1143/ptps.135.29}{\emph{Progress of Theoretical
			Physics Supplement} {\bf 135} (1999) 29--52}.
	
	\bibitem{Labastida:1998sk}
	J.~M.~F. Labastida and C.~Lozano, \emph{{Duality in twisted N=4 supersymmetric
			gauge theories in four-dimensions}},
	\href{http://dx.doi.org/10.1016/S0550-3213(98)00653-1}{\emph{Nucl. Phys.}
		{\bf B537} (1999) 203--242},
	[\href{https://arxiv.org/abs/hep-th/9806032}{{\tt hep-th/9806032}}].
	
	\bibitem{Marino:1998ru}
	M.~Marino and F.~Zamora, \emph{{Duality symmetry in softly broken N=2 gauge
			theories}},
	\href{http://dx.doi.org/10.1016/S0550-3213(98)00490-8}{\emph{Nucl. Phys. B}
		{\bf 533} (1998) 373--405}, [\href{https://arxiv.org/abs/hep-th/9804038}{{\tt
			hep-th/9804038}}].
	
	\bibitem{TAKASAKI_1999}
	K.~Takasaki, \emph{Integrable hierarchies and contact terms in u-plane
		integrals of topologically twisted supersymmetric gauge theories},
	\href{http://dx.doi.org/10.1142/s0217751x9900049x}{\emph{International
			Journal of Modern Physics A} {\bf 14} (Mar, 1999) 1001--1013}.
	
	\bibitem{Borcherds:1996uda}
	R.~E. Borcherds, \emph{Automorphic forms with singularities on
		{G}rassmannians}, {\emph{Invent.\ Math.\ {\textbf {132}} (1998) 491} (1998)
	}, [\href{https://arxiv.org/abs/alg-geom/9609022}{{\tt alg-geom/9609022}}].
	
	\bibitem{bringmann2017regularized}
	K.~Bringmann, N.~Diamantis and S.~Ehlen, \emph{Regularized inner products and
		errors of modularity}, {\emph{International Mathematics Research Notices}
		{\bf 2017} (2017) 7420--7458}.
	
	\bibitem{bruinier2004two}
	J.~H. Bruinier and J.~Funke, \emph{On two geometric theta lifts}, {\emph{Duke
			Mathematical Journal} {\bf 125} (2004) 45--90}.
	
	\bibitem{Lerche:1988np}
	W.~Lerche, A.~N. Schellekens and N.~P. Warner, \emph{{Lattices and Strings}},
	\href{http://dx.doi.org/10.1016/0370-1573(89)90077-X}{\emph{Phys. Rept.} {\bf
			177} (1989) 1}.
	
	\bibitem{Dixon:1990pc}
	L.~J. Dixon, V.~Kaplunovsky and J.~Louis, \emph{{Moduli dependence of string
			loop corrections to gauge coupling constants}},
	\href{http://dx.doi.org/10.1016/0550-3213(91)90490-O}{\emph{Nucl. Phys.} {\bf
			B355} (1991) 649--688}.
	
	\bibitem{Harvey:1995fq}
	J.~A. Harvey and G.~W. Moore, \emph{{Algebras, BPS states, and strings}},
	\href{http://dx.doi.org/10.1016/0550-3213(95)00605-2}{\emph{Nucl. Phys.} {\bf
			B463} (1996) 315--368}, [\href{https://arxiv.org/abs/hep-th/9510182}{{\tt
			hep-th/9510182}}].
	
	\bibitem{Bruinier08}
	G.~H. J.H.~Bruinier, G. van der~Geer and D.~Zagier, \emph{The 1-2-3 of Modular
		Forms}.
	\newblock Springer-Verlag Berlin Heidelberg, 2008,
	\href{http://dx.doi.org/10.1007/978-3-540-74119-0}{10.1007/978-3-540-74119-0}.
	
	\bibitem{Petersson1950}
	H.~Petersson, \emph{{Konstruktion der Modulformen und der zu gewissen
			Grenzkreisgruppen geh\"origen automorphen Formen von positiver reeller
			Dimension und die vollst\"andige Bestimmung ihrer Fourierkoeffizienten}},
	{\emph{S.-B. Heidelberger Akad. Wiss. Math.-Nat. Kl.} (1950) 417--494}.
	
	\bibitem{1603.03056}
	K.~Bringmann, N.~Diamantis and S.~Ehlen, \emph{Regularized inner products and
		errors of modularity},
	\href{http://dx.doi.org/10.1093/imrn/rnw225}{\emph{International Mathematics
			Research Notices} {\bf 2017} (2017) 7420--7458}.
	
	\bibitem{Diamond}
	F.~Diamond and J.~Shurman, \emph{A First Course in Modular Forms}, vol.~228 of
	\emph{Graduate Texts in Mathematics}.
	\newblock Springer-Verlag New York, 1~ed., 2005.
	
	\bibitem{ono2004}
	K.~Ono, \emph{The Web of Modularity: Arithmetic of the Coefficients of Modular
		Forms and q-series}, vol.~102.
	\newblock American Mathematical Society, cbms regional conference series in
	mathematics~ed., 2004.
	
	\bibitem{Zagier92}
	D.~Zagier, \emph{Introduction to modular forms; From Number Theory to Physics}.
	\newblock Springer, Berlin (1992), pp. 238-291, 1992.
	
	\bibitem{schultz2015}
	D.~Schultz, ``Notes on modular forms.'' URL:
	\url{https://faculty.math.illinois.edu/~schult25/ModFormNotes.pdf}.
	
	\bibitem{koblitz1993}
	N.~Koblitz, \emph{Introduction to Elliptic Curves and Modular Forms}.
	\newblock Graduate Texts in Mathematics. Springer New York, 1993.
	
\end{thebibliography}
\end{document}